\documentclass[11pt]{article}
\usepackage{amssymb}
\usepackage{amsmath}
\usepackage{latexsym}
\usepackage{graphics}
\usepackage{color}
\catcode`@=11
\renewcommand{\section}{\@startsection{section}{1}{0pt}{\medskipamount}
{\medskipamount}{\large\bf}}
\numberwithin{equation}{section}
\catcode`@=12

\def\th{\theta}
\def\a{\alpha}

\def\de{\delta}

\def\vp{\varphi}

\def\vt{\vartheta}
\def\p{\phi}

\def\sfrac#1#2{{\textstyle\frac{#1}{#2}}}
\def\m{\mu}
\def\n{\nu}
\def\pa{\partial}

\def\beq{\begin{equation}}
\def\eeq{\end{equation}}
\def\bea{\begin{eqnarray}}
\def\eea{\end{eqnarray}}

\newcommand{\id}{{1\!\!1}}
\newcommand{\im}{\,\mathrm{i}\,}
\newcommand{\diff}{\mathrm{d}}

\newcommand{\rts}{{\mathbb{R}^{2n}_\theta{\times}S^2}}
\newcommand{\rt}{{\mathbb{R}^{2n}_\theta}}

\newcommand{\R}{{\mathbb{R}}}
\newcommand{\N}{{\mathbb{N}}}
\newcommand{\C}{{\mathbb{C}}}
\newcommand{\Z}{{\mathbb{Z}}}
\newcommand{\Q}{{\mathbb{Q}}}

\newcommand{\Idd}{\mathbf{1}}
\newcommand{\Hcal}{{\cal H}}

\newcommand{\Ncal}{{\cal N}}
\newcommand{\fh}{\hat{f}}

\newcommand{\xh}{\hat{x}}
\newcommand{\zh}{\hat{z}}
\newcommand{\zbh}{\hat{\bar{z}}}
\newcommand{\yb}{{\bar{y}}}
\newcommand{\zb}{{\bar{z}}}
\newcommand{\ab}{{\bar{a}}}
\newcommand{\bb}{{\bar{b}}}
\newcommand{\cb}{{\bar{c}}}
\newcommand{\ca}{{\cal{A}}}
\newcommand{\cf}{{\cal{F}}}

\newcommand{\man}{{\cal M}}
\newcommand{\cliff}{{{\rm C}\ell}}
\newcommand{\K}{{\rm K}}
\newcommand{\HQ}{{\rm H}}

\newcommand{\rep}{{\rm R}}
\newcommand{\mon}{{\cal O}}
\newcommand{\Lcal}{{\cal L}}
\newcommand{\Tr}{{\rm Tr}}
\newcommand{\STr}{{\rm STr}}
\newcommand{\tr}{{\rm tr}}
\newcommand{\str}{{\rm str}}
\newcommand{\su}{{{\rm SU}(2)}}
\newcommand{\uo}{{{\rm U}(1)}}

\newcommand{\slc}{{{\rm SL}(2,\C)}}
\newcommand{\slcL}{{{\rm sl}(2,\C)}}
\newcommand{\spin}{{\rm Spin}}
\newcommand{\Pt}{{\rm P}}
\newcommand{\quiverm}{{{{\sf A}}_{m+1}}}

\newcommand{\quiver}{{{\sf A}}_{2}}
\newcommand{\path}{{\sf P}}
\newcommand{\pathalg}{{\C\,\quiverm}}
\newcommand{\pathmod}{{\underline{\path}}}
\newcommand{\simple}{{\underline{\sf L}}}
\newcommand{\mphi}{{{\mbf\phi}^{~}_{(m)}}}
\newcommand{\mT}{{{\mbf T}^{~}_{(m)}}}
\newcommand{\mcalT}{{{\mbf{\cal T}}^{~}_{(m)}}}
\newcommand{\mA}{{\mbf A^{(m)}}}
\newcommand{\ma}{{\mbf a^{(m)}}}
\newcommand{\mF}{{\mbf F^{(m)}}}
\newcommand{\mup}{{{\mbf\Upsilon}^{~}_{(m)}}}
\newcommand{\mbf}[1]{{\boldsymbol {#1} }}

\def\Dirac{{D\!\!\!\!/\,}} 
\def\chern{{\rm ch}}
\def\Hom{{\rm Hom}}
\def\Ext{{\rm Ext}}

\def\>{\rangle}
\def\<{\langle}
\def\+{\dagger}
\def\={\ =\ }

\begin{document}
\begin{titlepage}
\setcounter{page}{0}
\begin{flushright}
hep-th/0504025\\
ITP--UH--05/05\\
HWM--05--03\\
EMPG--05--04\\
\end{flushright}

\vskip 1.5cm

\begin{center}

{\Large\bf Quiver Gauge Theory of Nonabelian Vortices \\[6pt] and
  Noncommutative Instantons in Higher Dimensions}

\vspace{12mm}

{\large Alexander D. Popov}
\\[2mm]
\noindent {\em Institut f\"ur Theoretische Physik,
Universit\"at Hannover \\
Appelstra\ss{}e 2, 30167 Hannover, Germany }
\\ and \\
\noindent {\em Bogoliubov Laboratory of Theoretical Physics, JINR\\
141980 Dubna, Moscow Region, Russia}\\ Email:
{\tt popov@itp.uni-hannover.de}
\\[10mm]
{\large Richard J. Szabo}\\[2mm]
\noindent {\em Department of Mathematics, Heriot-Watt University\\
Colin Maclaurin Building, Riccarton, Edinburgh EH14 4AS, U.K.}
\\ Email: {\tt R.J.Szabo@ma.hw.ac.uk}

\vspace{15mm}

\begin{abstract}
\noindent
We construct explicit BPS and non-BPS solutions of the Yang-Mills
equations on the noncommutative space $\rts$ which have manifest
spherical symmetry. Using $\su$-equivariant dimensional reduction
techniques, we show that the solutions imply an equivalence between
instantons on $\rts$ and nonabelian vortices on $\rt$, which can be
interpreted as a blowing-up of a chain of D0-branes on $\rt$ into a chain
of spherical D2-branes on $\rts$. The low-energy dynamics of these
configurations is described by a quiver gauge theory which can be
formulated in terms of new geometrical objects generalizing
superconnections. This formalism enables the explicit assignment of
D0-brane charges in equivariant K-theory to the instanton solutions.

\end{abstract}

\end{center}
\end{titlepage}

\section{Introduction and summary }

One of the most basic questions that arises in trying to understand
the nonperturbative structure of string theory concerns the classification of
vector bundles over real and complex manifolds. In the presence of
D-branes one encounters gauge theories in spacetime dimensionalities
up to ten. Already more than 20 years ago, BPS-type equations in higher
dimensions were proposed~\cite{Corrigan2, Ward} as a generalization of
the self-duality equations in four dimensions. For nonabelian gauge theory
on a K\"ahler manifold the most natural BPS~condition lies in the
Donaldson-Uhlenbeck-Yau equations~\cite{Donaldson}, which arise, for
instance, in compactifications down to four-dimensional Minkowski
spacetime as the condition for at least one unbroken supersymmetry.

While the criteria for solvability of these BPS equations are by now
very well understood, in practice it is usually quite difficult to
write down explicit solutions of them. One recent line of attack has
been to consider noncommutative deformations of these field
theories~\cite{CDS1}--\cite{Harvey}. In certain instances, D-branes can be
realized as noncommutative solitons~\cite{DMR1}, which is a
consequence~\cite{Matsuo1,HM1} of the relationship between D-branes and
K-theory~\cite{MM1}--\cite{Manjarin}. All celebrated BPS
configurations in field theories, such as instantons~\cite{Belavin},
monopoles~\cite{Hooft} and vortices~\cite{Abrikos}, have been
generalized to the noncommutative case, originally in~\cite{NS},
in~\cite{GN} and in~\cite{vortex}, respectively (see~\cite{Haman} for
reviews and further references). Solution generating techniques such as the
ADHM construction~\cite{ADHM}, splitting~\cite{Ward1} and
dressing~\cite{ZS} methods have also been generalized to the
noncommutative setting in~\cite{NS,CKT} and in~\cite{LP1}. Solutions of
the generalized self-duality equations~\cite{Corrigan2, Ward} were
investigated in~\cite{Ward, Fairlie}, for example. Noncommutative
instantons in higher dimensions and their
interpretations as D-branes in string theory have been considered
in~\cite{Mih}--\cite{LPS}. In all of these constructions the usual
worldvolume description of D-branes emerges from the equivalence between
analytic and topological formulations of K-homology.

In this paper we will complete the construction initiated
in~\cite{IL,LPS} of multi-instanton solutions of the Yang-Mills
equations on the manifold which is the product of noncommutative
euclidean space $\R^{2n}_\theta$ with an ordinary two-sphere $S^2$. We consider
both BPS and non-BPS solutions, and extend previous solutions to those
which are explicitly $\su$-equivariant for any value of the
Dirac monopole charge characterizing the gauge field components along the
$S^2$ directions. Dimensional reduction techniques are used to
establish an equivalence between multi-instantons on $\rts$ and
nonabelian vortices on $\rt$. The configurations can be interpreted in
Type~IIA superstring theory as {\it chains} of branes and antibranes
with Higgs-like open string excitations between neighbouring sets of
D-branes. The equivalence between instantons and vortices may then be
attributed to the decay of an unstable configuration of D$(2n)$-branes
into a state of D0-branes (There are no higher brane charges
induced because $\R^{2n}$ is equivariantly contractible). The D0-brane
charges are classified by $\su$-equivariant K-theory and the
low-energy dynamics may be succinctly encoded into a simple quiver
gauge theory. Unlike the standard brane-antibrane systems, the
effective action cannot be recast using the formalism of
superconnections~\cite{AIO1} but requires a more general formulation
in terms of new geometrical entities that we call ``graded
connections''. This formalism makes manifest the interplay between the
assignment of K-theory classes to the explicit instanton solutions and
their realization in terms of a quiver gauge theory.

The organisation of this paper is as follows. The material is
naturally divided into two parts. Sections~2--5 deal with
{\it ordinary} gauge theory on a generic K\"ahler manifold of the form
$M_{2n}\times\C P^1$ in order to highlight the geometric structures
that arise due to dimensional reduction and which play a prominent
role throughout the paper. Sections~6--10 are then concerned with the
noncommutative deformation $\R^{2n}\times\C
P^1\to\R^{2n}_\theta\times\C P^1$ and they construct explicit solutions of
the dimensionally reduced Yang-Mills equations, emphasizing their
interpretations in the context of equivariant K-theory, quiver
gauge theory, and ultimately as states of D-branes. In Section~2 we
introduce basic definitions and set some
of our notation, and present the field equations that are to be
solved. In Section~3 we write down an explicit ansatz for the gauge
field which is used in the $\su$-equivariant dimensional reduction. In
Section~4 we describe three different interpretations of the ansatz as
configurations of D-branes, as charges in equivariant K-theory, and as
field configurations in a quiver gauge theory (Later on these three
descriptions are shown to be equivalent). In Section~5 the dimensional
reduction mechanism is explained in detail in the new language of
graded connections and the resulting nonabelian vortex equations,
arising from reduction of the Donaldson-Uhlenbeck-Yau equations, are
written down. In Section~6 we introduce the noncommutative
deformations of all these structures. In Section~7 we find explicit
BPS and non-BPS solutions of the noncommutative Yang-Mills equations
and show how they naturally realize representations of the pertinent
quiver. In Section~8 we develop an $\su$-equivariant generalization of
the (noncommutative) Atiyah-Bott-Shapiro construction, which provides
an explicit and convenient representation of our solution in terms of
K-homology classes. In Section~9 we compute the topological charge of
our instanton solutions directly in the noncommutative gauge theory,
and show that it coincides with the corresponding K-theory charge,
which then allows us to assign D0-brane charges to the solutions from a
worldvolume perspective. Finally, in Section~10 we construct some
novel BPS solutions in the vacuum sectors of the noncommutative field
theory and describe their relation to stable states of brane-antibrane
systems.

\bigskip

\section{Yang-Mills equations\label{Fieldeqs}}

In this section we will introduce the basic definitions and notation
that will be used throughout this paper, as well as the pertinent
field equations that we will solve.

\bigskip

\noindent
{\bf The manifold $\mbf{\man_q\times S^2}$.\ } Let $\man_q$ be a real
$q$-dimensional lorentzian manifold with nondegenerate metric of signature $(-
+\cdots +)$, and $S^2\cong\C P^1$ the standard two-sphere of constant radius
$R$. We shall consider the manifold $\man_q\times S^2$ with local real
coordinates $x'=(x^{\m '}\,)\in\R^q$ on $\man_q$ and coordinates
$\vartheta\in[0,\pi]$,
$\vp\in[0,2\pi]$ on $S^2$. In these coordinates the metric on
$\man_q\times S^2$ reads
\begin{equation}\label{metric1}
\diff \hat s^2\= g_{\hat\mu\hat\nu}~\diff x^{\hat\mu}~\diff x^{\hat\nu}\=
g_{\m'\n'}~\diff x^{\m'}~\diff x^{\n'}
+ R^2\,\left(\diff \vt^2 + \sin^2\vt~\diff\vp^2\right) \ ,
\end{equation}
where hatted indices $\hat\mu , \hat\nu, \ldots$ run over $0,1,\ldots
, q+1$ while primed indices $\m', \n', \ldots$ run through
$0, 1, \ldots , q-1$. We use the Einstein summation convention for
repeated spacetime indices.

\bigskip

\noindent
{\bf The K\"ahler manifold $\mbf{M_{2n}\times \C P^1}$.\ }
As a special instance of the manifold $\man_q$ we shall consider the product
$\man_q=\R^1\times M_{2n}$ of dimension $q=2n+1$ with metric
\begin{equation}\label{metric2}
g_{\m'\n'}~\diff x^{\m'}~\diff x^{\n'}= - \left(\diff x^0\right)^2
+ g_{\m\n}~\diff x^{\m}~\diff x^{\n}
\ .
\end{equation}
Here $M_{2n}$ is a K\"ahler manifold of real dimension $2n$ with local
real coordinates $x=(x^\m )\in\R^{2n}$, where the indices $\m ,\n
,\ldots$ run through $1,\ldots , 2n$. The cartesian product $M_{2n}\times
\C P^1$ is also a K\"ahler manifold with local complex coordinates
$(z^1,\ldots , z^n,y)\in\C^{n+1}$ and their complex conjugates, where
\begin{equation}\label{zz}
z^a\=x^{2a-1}-\im\,x^{2a} \qquad\textrm{and}\qquad
\zb^{\ab}\=x^{2a-1}+\im\,x^{2a} \qquad\textrm{with}\quad
a\=1,\ldots,n
\end{equation}
while
\begin{equation}\label{zn1}
y\=\frac{R\,\sin\vt}{1+\cos\vt}\,\exp{(-\im\vp)} \qquad\mbox{and}\qquad
\yb\=\frac{R\,\sin\vt}{1+\cos\vt}\,\exp{(\im\vp)}
\end{equation}
are stereographic coordinates on the northern hemisphere of $S^2$. In
these coordinates the metric on $M_{2n}\times \C P^1$ takes the form
\begin{equation}\label{metric3}
\diff s^2\= g_{\m\n}~\diff x^{\m}~\diff x^{\n}
+ R^2\,\left(\diff \vt^2 + \sin^2\vt\ \diff\vp^2\right)
\=2\,g_{a\bb}\ \diff z^a~\diff \zb^{\bb} +
\frac{4\,R^4}{\left(R^2+y\yb\right)^2}\ \diff y~\diff \yb\ ,
\end{equation}
while the K\"ahler two-form $\omega $ is given by
\begin{equation}\label{kahler}
\omega\=\sfrac{1}{2}\,\omega_{\m\n}~\diff x^\m\wedge\diff x^\n +
R^2\,\sin\vt~\diff\vt\wedge\diff\vp\=-
2\im g_{a\bb}\ \diff z^a\wedge\diff \zb^{\bb}-
\frac{4\im R^4}{\left(R^2+y\yb\right)^2}\ \diff y\wedge\diff \yb \ .
\end{equation}

\bigskip

\noindent
{\bf Yang-Mills equations.\ } Consider a rank $k$ hermitean vector bundle
${\cal E}\to\man_q\times S^2$ with gauge connection
$\ca$ of curvature $\cf =\diff\ca + \ca\wedge\ca$. In local
coordinates, wherein $\ca=\ca_{\hat\mu}~\diff x^{\hat\mu}$, the
two-form $\cf$ has components
$\cf_{\hat\mu\hat\nu}=\pa_{\hat\mu}\ca_{\hat\nu} -
\pa_{\hat\nu}\ca_{\hat\mu} + [\ca_{\hat\mu}, \ca_{\hat\nu}]$, where
$\pa_{\hat\mu}:=\pa /\pa x^{\hat\mu}$. Both $\ca_{\hat\mu}$ and
$\cf_{\hat\mu\hat\nu}$
take values in the Lie algebra ${\rm u}(k)$. For the usual Yang-Mills
lagrangian\footnote{The Yang-Mills coupling constant $g_{\rm YM}^{~}$
can be introduced via the redefinition $\ca\mapsto g_{\rm YM}^{~}\,\ca$.}
\begin{equation}\label{lagr}
L^{~}_{\rm YM}=-\sfrac{1}{4}\,\sqrt{g}~\tr^{~}_{k\times k}\
\cf_{\hat{\m}\hat{\n}}\,\cf^{\hat{\m} \hat{\n}}
\end{equation}
the equations of motion are
\begin{equation}\label{YM}
\frac{1}{\sqrt{g}}\, \pa_{\hat\m}\bigl(\sqrt{g}~\cf^{\hat{\m}\hat{\n}}\bigr) +
\bigl[\ca_{\hat\m}\,,\, \cf^{\hat\m\hat\n}\bigr]=0 \ ,
\end{equation}
where $g={|\det (g_{\hat\mu\hat\nu})|}$. The curvature two-form can be
written in local coordinates on $\man_q\times\C P^1$ as
\begin{equation}
{\cf}=\sfrac{1}{2}\,{\cf}_{\m'\n'}~\diff x^{\m'}\wedge\diff x^{\n'} +
{\cf}_{\m' y}~\diff x^{\m'}\wedge\diff y +
{\cf}_{\m'\yb}~\diff x^{\m'}\wedge\diff\yb
+ {\cf}_{y\yb}~\diff y\wedge\diff\yb
\label{curvprod}\end{equation}
and the Yang-Mills lagrangian becomes
\bea
L^{~}_{\rm YM}&=&-\sfrac{1}{4}\,\sqrt{g}~\tr^{~}_{k\times k}
\left[\cf_{\mu'\nu'}\,\cf^{\mu'\nu'}+
\frac{\left(R^2+y\bar y\right)^2}{R^4}\,g^{\mu'\nu'}\,
\left(\cf_{\mu'y}\,\cf_{\nu'\bar y}+\cf_{\mu'\yb}\,\cf_{\nu'y}
\right)\right.\nonumber\\ && \qquad\qquad\qquad\quad
-\Biggl.\frac12\,\left(\frac{\left(R^2+y\yb\right)^2}{R^4}\,
\cf_{y\yb}\right)^2\,\Biggr] \ .
\label{lagrprod}\eea

\bigskip

\noindent
{\bf Donaldson-Uhlenbeck-Yau equations.\ }
For static field configurations in the temporal gauge $\ca_0=0$, the
Yang-Mills equations (\ref{YM}) on $\R^1\times M_{2n}\times \C P^1$
reduce to equations on  $M_{2n}\times \C P^1$.
Their stable solutions are provided by solutions of the
Donaldson-Uhlenbeck-Yau (DUY) equations which can be formulated on any
K\"ahler manifold~\cite{Donaldson}. The importance of these
equations derives from the fact that they yield the BPS solutions of
the full Yang-Mills equations.

The DUY equations on $M_{2n}\times\C P^1$ are
\begin{equation}\label{DUY}
*\omega\wedge {\cf}\ =\ 0 \qquad\textrm{and}\qquad
{\cf}^{0,2}\=0\ ,
\end{equation}
where $*$ is the Hodge duality operator and
$\cf =\cf^{2,0}+\cf^{1,1}+\cf^{0,2}$ is the K\"ahler decomposition of
the gauge field strength. In the local complex coordinates $(z^a, y)$
these equations take the form
\begin{eqnarray}\label{DUY1}
g^{a\bb}\,{\cf}_{z^a\zb^{\bb}}+\frac{\left(R^2+y\yb\right)^2}{2\,R^4}
\,{\cf}_{y\yb}&=&0 \ ,
\\[4pt]\label{DUY2}
{\cf}_{\zb^{\ab}\zb^{\bb}}&=&0~=~\cf_{z^az^b} \ ,
\\[4pt]\label{DUY3}
{\cf}_{\zb^{\ab}\yb}&=&0~=~\cf_{z^ay} \ ,
\end{eqnarray}
where the indices $a,b,\ldots$ run through $1,\ldots
,n$. Eq.~(\ref{DUY1}) is a hermitean condition on the gauge field
strength tensor, while eqs.~(\ref{DUY2}) and (\ref{DUY3}) are
integrability conditions implying that the bundle $\cal E$ endowed
with a connection $\ca$ is holomorphic. It is easy
to show that any solution of these $n(n{+}1){+}1$ equations also
satisfies the full Yang-Mills equations.

\bigskip

\section{Invariant gauge fields\label{Invgauge}}

In this section we shall write down the fundamental form of the gauge
potential $\ca$ on $\man_q\times\C P^1$ that will be used later on to
dimensionally reduce the Yang-Mills equations for $\ca$ to equations on
$\man_q$. This will be achieved by prescribing a specific $\C P^1$
dependence for $\ca$, which we proceed to describe first.

\bigskip

\noindent
{\bf Monopole bundles.\ } Consider the hermitean line bundle
$\Lcal^m\to \C P^1$ over the sphere with $\Lcal^m:=(\Lcal)^{\otimes m}$
and unique $\su$-invariant unitary connection $a_m$ having, in the
local complex coordinate $y$ on $\C P^1$, the form
\begin{equation}\label{f1}
a_m=\frac{m}{2\left(R^2 +y\yb\right)}\ \left(\yb~\diff y - y~\diff\yb
\right) \ ,
\end{equation}
where $m$ is an integer. The curvature of this connection is
\begin{equation}\label{f2}
f_m\=\diff a_m\= - \frac{m\,R^2}{\left(R^2 +y\yb\right)^2}~
\diff y\wedge \diff \yb \ .
\end{equation}
The topological charge of this gauge field configuration
is given by the first Chern number (equivalently the degree) of the
associated complex line bundle as
\begin{equation}\label{f3}
{\rm deg}~\Lcal^m\=\frac{\im}{2\pi}\,\int_{\C P^1}f_m \= m
\ .
\end{equation}
In terms of the spherical coordinates $(\vt,\vp)$ the configuration
(\ref{f1},\ref{f2}) has the form
\begin{equation}\label{am}
a_m\=-\frac{\im m}{2}\,(1-\cos\vt )~\diff\vp
\quad\mbox{and}\quad
f_m\=\diff a_m \= - \frac{\im m}{2}\,\sin\vt\ \diff\vt\wedge\diff\vp
\ .
\end{equation}
It describes $|m|$ Dirac monopoles or antimonopoles sitting on top of
each other.

The $m$-monopole bundle is classified by the Hopf fibration
$S^1\hookrightarrow S^3\to S^2$. For each $m\in\Z$ there is a
one-dimensional representation ${\mbf\nu}_m=({\mbf\nu}_1)^{\otimes m}$ of the
circle group ${\rm U}(1)\cong S^1$ defined by
\beq
{\mbf\nu}_m~: \quad v~\longmapsto~\zeta\cdot v\=\zeta^m\,v
\qquad\textrm{with}\qquad \zeta~\in~S^1 \quad\mbox{and}\quad v~\in~\C \ .
\label{U1irreps}\eeq
We denote this irreducible ${\rm U}(1)$-module by
$\underline{S}_{\,m}\cong\C$. Regarding the sphere as the
homogeneous space $\C P^1\cong\su/\uo$, the
$\su$-equivariant line bundle $\Lcal^m\to\C P^1$ corresponds to the
representation ${\mbf\nu}_m$ in the sense that it can be expressed as
\beq
\Lcal^m=\su\times_\uo\,\underline{S}_{\,m} \ ,
\label{monmexpl}\eeq
where the quotient on $\su\times\underline{S}_{\,m}$ is by the $\uo$
action $\zeta\cdot(g,v)=(g\,\zeta^{-1},\zeta^m\,v)$ for
$g\in\su$, $v\in\underline{S}_{\,m}$ and $\zeta\in\uo$. The action of
$\su$ on $\su\times\underline{S}_{\,m}$ given by
$g'\cdot(g,v)=(g'\,g,v)$ descends to an action on
(\ref{monmexpl}). Any $\su$-equivariant hermitean vector bundle over
the sphere is a Whitney sum of bundles (\ref{monmexpl}).

There is an alternative description in terms of
the holomorphic line bundle $\mon(m)\to\C P^1$ defined as the
$m$-th power of the tautological bundle over the complex projective
line. The universal complexification of the Lie group $\su$ is
$\slc$, and we may regard the sphere as a projective variety through
the natural diffeomorphism $\C P^1\cong\su/\uo\cong\slc/\Pt$, where
$\Pt$ is the parabolic subgroup of lower triangular matrices in
$\slc$. The $\su$ action on (\ref{monmexpl}) lifts to a smooth $\slc$
action, and the complexification of (\ref{monmexpl}) is realized as
the $\slc$-equivariant line bundle
\beq
\mon(m)=\slc\times_\Pt\,\underline{S}_{\,m}
\label{monmholexpl}\eeq
over $\C P^1$. Only the Cartan subgroup $\C^\times\subset\Pt$ of
non-zero complex numbers acts non-trivially on the modules
$\underline{S}_{\,m}$, with the $\C^\times$ action
defined analogously to (\ref{U1irreps}). The two descriptions are
equivalent after the introduction of a hermitean
metric on the fibres of $\mon(m)$. This holomorphic line bundle has
transition function $y^m$ transforming sections from the northern
hemisphere to the southern hemisphere of $S^2$. However, the
monopole connection (\ref{f1}) is transformed on the intersection of
the two patches covering $\C P^1$ via the transition function
$(y/\yb)^{m/2}$, which is the unitary reduction of the holomorphic
transition function $y^m$. Thus the bundle $\mon(m)$ regarded as a
hermitean line bundle has transition function $(y/\yb)^{m/2}$ and can
be substituted for the monopole bundle $\Lcal^m$.

\bigskip

\noindent
{\bf SU(2)-invariant gauge potential.\ } The form of our ansatz for the gauge
connection on $\man_q\times\C P^1$ is fixed by imposing invariance
under the $\su$ isometry group of $\C P^1$ acting through rigid
rotations of the sphere. Let ${\cal E}\to\man_q\times
\C P^1$ be an $\su$-equivariant ${\rm U}(k)$-bundle, with the group
$\su$ acting trivially on $\man_q$ and in the standard way on $\C
P^1=\su/\uo$. Let $\ca$ be a connection on ${\cal E}$.
Imposing the condition of $\su$-equivariance means that we should look
for representations of the group $\su$ inside the ${\rm U}(k)$
structure group, i.e. for homomorphisms $\rho:\su\to{\rm
  U}(k)$. The ansatz for $\ca$ is thus given by $k$-dimensional
representations of $\su$. Up to isomorphism, for each positive integer
$d$ there is a unique irreducible $\su$-module
$\underline{V}_{\,d}\cong\C^d$ of dimension $d$. Therefore, for each
positive integer $m$, the module
\beq
\underline{\cal V}\=\bigoplus_{i=0}^m\,
\underline{V}_{\,k_i} \qquad\textrm{with}\qquad
\sum_{i=0}^mk_i\=k
\label{genrepSU2Uk}\eeq
gives a representation $\rho$ of $\su$ inside ${\rm
  U}(k)$. The total number of such homomorphisms is the number of
  partitions of the positive integer ${\rm rank}({\cal E})=k$ into
  $\leq(m+1)$ components. The original ${\rm U}(k)$ gauge symmetry is
  then broken down to the centralizer subgroup of $\rho(\su)$ in ${\rm
    U}(k)$ as
\beq
{\rm U}(k)~\longrightarrow~\prod_{i=0}^m{\rm U}(k_i) \ .
\label{gaugebroken}\eeq

It is natural to allow for gauge transformations
to accompany the $\su$ action~\cite{Forgacs}, and so some
``twisting'' can occur in the reduction of the connection $\ca$ on
$\man_q\times\C P^1$. The $\C P^1$ dependence in this case is uniquely
determined by the above $\su$-invariant Dirac monopole
configurations~\cite{Prada, BGP}. The ${\rm u}(k)$-valued gauge
potential $\ca$ thus splits into $k_i\times k_j$ blocks $\ca^{ij}$,
\begin{equation}\label{f4}
\ca\=\left(\ca^{ij}\right) \qquad\mbox{with}\qquad
\ca^{ij}~\in~\mbox{Hom}\bigl(\,\underline{V}_{\,k_j}\,,\,
\underline{V}_{\,k_i}\bigr) \ ,
\end{equation}
where the indices $i,j,\ldots$ run over $0,1,\ldots,m$,
$k_0+k_1+\ldots +k_m=k$ and
\bea\label{f5}
\ca^{ii}&=&A^i(x'\,)\otimes1 + \Idd_{k_i}\otimes a_{m-2i}(y) \ , \\[4pt]
\label{f6}
\ca^{i\,i+1}~=:~\Phi^{~}_{i+1}&=&
\phi^{~}_{i+1}(x'\,)\otimes\bar{\beta}(y) \ , \\[4pt]
\label{f7}
\ca^{i+1\,i}~=~-\left(\ca^{i\,i+1}\right)^\+~=~-\bigl(\Phi^{~}_{i+1}
\bigr)^\+&=&-\phi_{i+1}^\+ (x'\,)\otimes\beta(y) \ , \\[4pt]
\label{f9}
\ca^{i\,i+l}&=&0~=~\ca^{i+l\,i}\quad \mbox{for} \quad l~\ge~2 \ .
\eea
Here
\begin{equation}\label{f8}
\beta\= \frac{R\ \diff y}{R^2 +y\yb}
\quad\mbox{and}\quad
\bar{\beta} \= \frac{R\ \diff\yb}{R^2 +y\yb}
\end{equation}
are the unique covariantly constant, ${\rm SU}(2)$-invariant
forms of type $(1,0)$ and $(0,1)$ such that the K\"ahler
$(1,1)$-form on $\C P^1$ is $4\,R^2\,\beta\wedge\bar\beta$. They
respectively take values in the bundles $\Lcal^2$ and $\Lcal^{-2}$.

It is easy to see that the gauge potential $\ca$ given by
(\ref{f5})--(\ref{f9}) is anti-hermitean and $\su$-invariant. Note that
we do not use the Einstein summation convention for the repeated indices
$i$ labelling the components of the irreducible representation
$\underline{V}_{\,m+1}\cong\C^{m+1}$ of the group $\su$. Thus the
gauge potential ${\ca}\in {\rm u}(k)$ decomposes into
gauge potentials $A^i\in {\rm u}(k_i)$ with $i=0,1,\ldots,m$
and a multiplet of scalar fields $\p_{i+1}$ with
$i=0,1,\ldots, m-1$ transforming in the bi-fundamental
representations
$\underline{V}^{~}_{\,k_i}\otimes\underline{V}_{\,k_{i+1}}^\vee$ of
the subgroup ${\rm U}(k_i)\times {\rm U}(k_{i+1})$ of the original
${\rm U}(k)$ gauge group. All fields
$(A^i, \p_{i+1})$ depend only on the coordinates $x'\in\man_q$. Every
$\su$-invariant unitary connection $\ca$ on $\man_q\times\C P^1$ is of
the form given in (\ref{f4})--(\ref{f9})~\cite{BGP}, which follow from
the fact that the complexified cotangent bundle of $\C P^1$ is
$\Lcal^2\oplus\Lcal^{-2}$. This ansatz amounts to
an equivariant decomposition of the original rank~$k$
$\su$-equivariant bundle ${\cal E}\to\man_q\times\C P^1$ in the form
\beq
{\cal E}\=\bigoplus_{i=0}^m\,{\cal E}_i \qquad\textrm{with}\qquad
{\cal E}_i\=E_{k_i}\otimes\Lcal^{m-2i} \ ,
\label{calEansatz}\eeq
where $E_{k_i}\to\man_q$ is a hermitean vector bundle of rank $k_i$
with typical fibre the module $\underline{V}_{\,k_i}$, and ${\cal
  E}_i\to\man_q\times\C P^1$ is the bundle with fibres $({\cal
  E}_i)_{(x',y,\yb)}=(E_{k_i})_{x'}\otimes(\Lcal^{m-2i})_{(y,\yb)}$.
By regarding $\Phi_i\in{\rm Hom}
({\cal E}^{~}_i,{\cal E}^{~}_{i-1})\cong{\rm H}^0(\man_q\times \C
P^1;{\cal E}_{i-1}^{~}\otimes{\cal E}_{i}^\vee\,)$ for $i=1,\dots,m$
and defining $\Phi_0:=0=:\Phi_{m+1}$, we can summarize our ansatz
  through the following chain of bundles:
\begin{equation}
\begin{picture}(0,0)%
\includegraphics{vortex1.pstex}%
\end{picture}%
\setlength{\unitlength}{3947sp}%
\begingroup\makeatletter\ifx\SetFigFont\undefined%
\gdef\SetFigFont#1#2#3#4#5{%
  \reset@font\fontsize{#1}{#2pt}%
  \fontfamily{#3}\fontseries{#4}\fontshape{#5}%
  \selectfont}%
\fi\endgroup%
\begin{picture}(6825,600)(2551,-4261)
\put(6376,-4036){\makebox(0,0)[lb]{\smash{\SetFigFont{11}{13.2}{\familydefault}{\mddefault}{\updefault}{\color[rgb]{0,0,0}$\cdots$}%
}}}
\put(9376,-4036){\makebox(0,0)[lb]{\smash{\SetFigFont{12}{14.4}{\familydefault}{\mddefault}{\updefault}{\color[rgb]{0,0,0}.}%
}}}
\put(9001,-4036){\makebox(0,0)[lb]{\smash{\SetFigFont{11}{13.2}{\familydefault}{\mddefault}{\updefault}{\color[rgb]{0,0,0}$0$}%
}}}
\put(8101,-4261){\makebox(0,0)[lb]{\smash{\SetFigFont{11}{13.2}{\familydefault}{\mddefault}{\updefault}{\color[rgb]{0,0,0}${\cal E}_0$}%
}}}
\put(6976,-4261){\makebox(0,0)[lb]{\smash{\SetFigFont{11}{13.2}{\familydefault}{\mddefault}{\updefault}{\color[rgb]{0,0,0}${\cal E}_1$}%
}}}
\put(7501,-4186){\makebox(0,0)[lb]{\smash{\SetFigFont{9}{10.8}{\familydefault}{\mddefault}{\updefault}{\color[rgb]{0,0,0}$\Phi_1^\dag$}%
}}}
\put(7501,-3811){\makebox(0,0)[lb]{\smash{\SetFigFont{9}{10.8}{\familydefault}{\mddefault}{\updefault}{\color[rgb]{0,0,0}$\Phi_1^{~}$}%
}}}
\put(4426,-4261){\makebox(0,0)[lb]{\smash{\SetFigFont{11}{13.2}{\familydefault}{\mddefault}{\updefault}{\color[rgb]{0,0,0}${\cal E}_{m-1}$}%
}}}
\put(5026,-4186){\makebox(0,0)[lb]{\smash{\SetFigFont{9}{10.8}{\familydefault}{\mddefault}{\updefault}{\color[rgb]{0,0,0}$\Phi_{m-1}^\dag$}%
}}}
\put(5026,-3811){\makebox(0,0)[lb]{\smash{\SetFigFont{9}{10.8}{\familydefault}{\mddefault}{\updefault}{\color[rgb]{0,0,0}$\Phi_{m-1}^{~}$}%
}}}
\put(3301,-4261){\makebox(0,0)[lb]{\smash{\SetFigFont{11}{13.2}{\familydefault}{\mddefault}{\updefault}{\color[rgb]{0,0,0}${\cal E}_m$}%
}}}
\put(3826,-4186){\makebox(0,0)[lb]{\smash{\SetFigFont{9}{10.8}{\familydefault}{\mddefault}{\updefault}{\color[rgb]{0,0,0}$\Phi_m^\dag$}%
}}}
\put(3826,-3811){\makebox(0,0)[lb]{\smash{\SetFigFont{9}{10.8}{\familydefault}{\mddefault}{\updefault}{\color[rgb]{0,0,0}$\Phi_m^{~}$}%
}}}
\put(5701,-4261){\makebox(0,0)[lb]{\smash{\SetFigFont{11}{13.2}{\familydefault}{\mddefault}{\updefault}{\color[rgb]{0,0,0}${\cal E}_{m-2}$}%
}}}
\put(2551,-4036){\makebox(0,0)[lb]{\smash{\SetFigFont{11}{13.2}{\familydefault}{\mddefault}{\updefault}{\color[rgb]{0,0,0}$0$}%
}}}
\end{picture}

\label{bundlechain}\end{equation}

\bigskip

\noindent
{\bf Field strength tensor.\ } The calculation of the curvature
(\ref{curvprod}) for $\ca$ of the form (\ref{f4})--(\ref{f9}) yields
\begin{equation}\label{f10}
\cf\=\left(\cf^{ij}\right) \qquad\mbox{with}\qquad
\cf^{ij} \= {\diff}\ca^{ij} +
\sum_{l=0}^{m}\ca^{il}\wedge \ca^{lj}
\ ,
\end{equation}
where
\bea\label{f11}
\cf^{ii}&=&F^i+ f_{m-2i}+\bigl(
\phi^{~}_{i+1}\,\phi^\+_{i+1} - \phi_i^\+\,\phi^{~}_i\bigr)\
\beta\wedge\bar\beta \ , \\[4pt]
\label{f12}
\cf^{i\,i+1}&=&D \phi_{i+1}\wedge\bar\beta \ , \\[4pt]
\label{f13}
\cf^{i+1\,i}~=~-\left( \cf^{i\,i+1}\right)^\+&=&-
\bigl(D \phi_{i+1}\bigr)^\+ \wedge \beta \ , \\[4pt]
\label{f14}
\cf^{i\,i+l}&=&0~=~\cf^{i+l\,i}\quad \mbox{for} \quad l~\ge~2
\ .
\eea
Here we have defined $F^i:=\diff A^i+A^i\wedge
A^i=\frac12\,F_{\mu'\nu'}^i(x'\,)~\diff x^{\mu'}\wedge\diff x^{\nu'}$
and introduced the bi-fundamental covariant derivatives
\begin{equation}\label{der}
D \phi_{i+1}:= \diff\phi_{i+1} + A^i\,\phi_{i+1} - \phi_{i+1}\,A^{i+1}\ .
\end{equation}
{}From (\ref{f11})--(\ref{f14}) we find the non-vanishing field
strength components
\bea\label{f15}
\cf^{ii}_{{\mu}'{\nu}'}&=&F_{{\mu}'{\nu}'}^i \ , \\[4pt]
\label{f16}
\cf^{i\,i+1}_{{\mu}'\yb}&=&\frac{R}{R^2 +y{\yb}}\,
D_{{\mu}'} \phi_{i+1}~=~-\bigl(\cf^{i+1\,i}_{{\mu}'y}\bigr)^\+ \ , \\[4pt]
\label{Fyyb}
\cf^{ii}_{y\yb}&=& - \frac{R^2}{(R^2 +y{\yb})^2}\,
\left(m-2i+\phi_i^\+\,\phi^{~}_i - \phi^{~}_{i+1}\,\phi^\+_{i+1}\right)
\ .
\eea

\bigskip

\section{Description of the ansatz\label{Ansatzdescr}}

In this section we shall clarify some features of the ansatz
constructed in the previous section from three different points of
view. To set the stage for the string theory interpretations of the
solutions that we will construct later on, we begin by indicating how
the ansatz can be interpreted in terms of configurations of D-branes
in Type~II superstring theory. This leads into a discussion
of how the ansatz is realized in topological K-theory, which
classifies the Ramond-Ramond charges of these brane systems, and we
will derive the decomposition (\ref{calEansatz}) directly within the
framework of $\su$-equivariant
K-theory. We will then explain how seeking explicit realizations of
our ansatz is equivalent to finding representations of the ${\rm A}_{m+1}$
quiver. One of the goals of the subsequent sections will be to
establish the precise link between these three descriptions,
showing that they are all equivalent.

\bigskip

\noindent
{\bf Physical interpretation.\ }
Before entering into the formal mathematical characterizations of the
ansatz of the previous section, let us first explain the physical situation
which they will describe. Our ansatz implies an equivalence between
brane-antibrane systems on $\man_q$ and wrapped branes on
$\man_q\times\C P^1$. In the standard D-brane interpretation, our
initial rank $k$ hermitean vector bundle ${\cal E}\to\man_q\times \C P^1$
corresponds to $k$ coincident D($q$+1)-branes wrapping the
worldvolume manifold $\man_q\times \C P^1$. The condition
of $\su$-equivariance imposed on this bundle fixes the
dependence on the coordinates of $\C P^1$ and breaks the
gauge group U($k$) as in (\ref{gaugebroken}). The rank $k_i$ sub-bundle
$E_{k_i}\to\man_q$ of this bundle is twisted by the
Dirac multi-monopole bundle $\Lcal^{m-2i}\to\C P^1$. The system of $k$
coincident D($q$+1)-branes thereby splits into blocks of
$k_0+k_1+\ldots+k_m = k$ coincident D($q$+1)-branes, associated to
irreducible representations of $\su$ and wrapping a common
sphere $\C P^1$ with the monopole fields. This system is equivalent to
a system of $k_0+k_1+\ldots+k_m =k\ $\  D($q{-}$1)-branes carrying
different magnetic fluxes on their common worldvolume $\man_q$. The
D($q{-}$1)-branes which carry negative magnetic flux have opposite
orientation with respect to the D($q{-}$1)-branes with positive magnetic flux,
i.e. they are antibranes. This will become evident from the K-theory
formalism, which will eventually lead to an explicit worldvolume
construction, and also from the explicit calculation of the topological
charges of the instanton solutions. In addition to the usual
Chan-Paton gauge field degrees of freedom $A^i\in{\rm End}(E_{k_i})$
living on each block of branes, the field content on the brane
configuration contains bi-fundamental scalar fields $\p_{i+1}\in{\rm
  Hom}(E_{k_{i+1}},E_{k_{i}})$ corresponding to massless open string
excitations between neighbouring blocks of $k_i$ and $k_{i+1}$ \
D($q{-}$1)-branes. Other excitations are suppressed by the condition
of $\su$-equivariance.

However, as we shall see explicitly in the following, the fields $\phi_{i+1}$
should not be regarded as tachyon fields, but rather only as
(holomorphic) Higgs fields responsible for the symmetry breaking
(\ref{gaugebroken}). Only the brane-antibrane pairs whose constituents
carry equal and opposite monopole charges are neutral and can thus
annihilate to the vacuum, which carries no monopole charge (although
it can carry a K-theory charge from the virtual Chan-Paton bundles
over $\man_q$). Other brane pairs are stable because their overall
non-vanishing Chern number over $\C P^1$ is an obstruction to decay,
and the monopole bundles thereby act as a source of flux stabilization
for such brane pairs by giving them a conserved topological
charge. In particular, neighbouring blocks of D$(q-1)$-branes are
marginally bound by the massless open strings stretching between
them. In this sense, the $\su$-invariant reduction of D-branes on
$\man_q\times\C P^1$ induces brane-antibrane systems on $\man_q$. Note
that while the system on $\man_q$ is generically unstable, the original brane
configuration on $\man_q\times\C P^1$ can be nonetheless stable.

\bigskip

\noindent
{\bf K-theory charges.\ } Given that the charges of configurations of D-branes
in string theory are classified topologically by
K-theory~\cite{MM1,OS1,Manjarin}, let us now
seek the K-theory representation of the above physical
situation. The one-monopole bundle $\Lcal$ is a
crucial object in establishing the Bott periodicity isomorphism
\beq
\K\left(\man_q\times \C P^1\right)=\K\bigl(\man_q\bigr)
\label{Bottper}\eeq
in topological K-theory. The isomorphism is generated by taking the
K-theory product of the tachyon field $\phi_1:E_{k_1}\to E_{k_0}$ of
a virtual bundle $[E_{k_0},E_{k_1};\phi_1]\in\K(\man_q)$ with that of the
class of the line bundle $\Lcal$ which represents the Bott generator of
$\widetilde{\K}(\C P^1)=\Z$~\cite{OS1}. The topological equivalence
(\ref{Bottper}) then implies the equivalence of brane-antibrane
systems on $\man_q\times\C P^1$ and $\man_q$, with the brane and
antibrane systems each carrying a single unit of monopole charge. When
they carry $m>1$ units of charge, the isomorphism breaks down, and it
is necessary to introduce the notion of ``D-operations'' to establish the
relationship~\cite{LPS}. While these operations are natural, they are
not isomorphisms and they reflect the fact that the explicit solutions
in this setting are not ${\rm SU}(2)$-invariant, so that the
equivalence breaks down due to spurious moduli dependences of the
system of branes on the $\C P^1$ factor. In what follows
we will derive a modification of the relation (\ref{Bottper}) in {\it
  equivariant} K-theory which will naturally give the desired
isomorphism, reflecting the equivalence of the brane-antibrane systems
for arbitrary monopole charge, and bypass the need for introducing
D-operations. This is only possible by augmenting the basic
brane-antibrane system to a {\it chain} of $(m+1)$ branes and antibranes
with varying units of monopole charge as described above, and we will thereby
arrive at an independent purely K-theoretic derivation of our ansatz.

The representation ring $\rep_G$ of a group $G$~\cite{Segal} is
the Grothendieck ring of the category of finite dimensional
representations of $G$, with addition induced by direct sum of vector
spaces, $[\,\underline{V}\,]+[\,\underline{V'}\,]:=
[\,\underline{V}\oplus\underline{V'}\,]$, and multiplication induced
by tensor product of modules,
$[\,\underline{V}\,]\cdot[\,\underline{V'}\,]:=
[\,\underline{V}\otimes\underline{V'}\,]$. As an abelian group it is
generated by the irreducible representations of $G$. Alternatively,
since the isomorphism class of a $G$-module $\underline{V}$ is
completely determined by its character
$\chi^{~}_{\underline{V}}:G\to\C$, the map
$\underline{V}\mapsto\chi^{~}_{\underline{V}}$ identifies $\rep_G$ as
a subring of the ring of $G$-invariant functions on $G$.
If $\man_q$ is a $G$-space, then the Grothendieck
group of $G$-equivariant bundles over $\man_q$ is called the
$G$-equivariant K-theory group $\K_G(\man_q)$. This group unifies
ordinary K-theory with group representation theory, in the sense that
for the trivial space $\K_G({\rm pt})=\rep_G$ is the representation ring
of $G$, while for the trivial group $\K_{\rm id}(\man_q)=\K(\man_q)$ is the
ordinary K-theory of $\man_q$. The former property implies that
$\K_G(\man_q)$ is an $\rep_G$-module and the coefficient ring in
equivariant K-theory is $\rep_G$, rather than just $\Z$ as in the
ordinary case. If the $G$-action on $\man_q$ is trivial, then any
$G$-equivariant bundle $E\to \man_q$ may be decomposed as a finite
Whitney sum
\beq
E=\bigoplus_{\underline{V}
\in{\rm Rep}(G)}\,{\rm Hom}^{~}_G\bigl(\id^{~}_{\underline{V}}\,,\,E
\bigr)\otimes\id^{~}_{\underline{V}}
\label{trivialEdecomp}\eeq
where $\id^{~}_{\underline{V}}=\man_q\times\underline{V}$ is the
trivial bundle over $\man_q$ with fibre the irreducible $G$-module
$\underline{V}$. It follows that for trivial $G$-actions the
equivariant K-theory takes the simple form
\beq
\K_G(\man_q)=\K(\man_q)\otimes\rep_G \ .
\label{KGXqtrivialG}\eeq
The $\K_G$-functor behaves analogously to the ordinary $\K$-functor,
and in addition $\K_G$ is functorial with respect to group
homomorphisms. A useful computational tool is the equivariant excision
theorem. If $F$ is a closed subgroup of $G$ and $\man_q$ is an $F$-space,
then the inclusion $\imath:F\hookrightarrow G$ induces an
isomorphism~\cite{Segal}
\beq
\imath^*\,:\,\K_G(G\times_F\man_q)~\stackrel{\approx}{\longrightarrow}~
\K_F(\man_q) \ ,
\label{excision}\eeq
where the quotient on $G\times \man_q$ is by the $F$-action
$f\cdot(g,x'\,)=(g\,f^{-1},f\cdot x'\,)$ for $g\in G$,
$x'\in\man_q$ and $f\in F$. The $G$-action on $G\times_F\man_q$ descends from
that on $G\times \man_q$ given by $g'\cdot (g,x'\,)=(g'\,g,x'\,)$.

Let us specialize to our case of interest by taking $G=\su$,
$F=\uo$ and the trivial action of $\su$ on the space $\man_q$. Using
(\ref{KGXqtrivialG}) and (\ref{excision}) we may then compute
\bea
\K_\su\bigl(\man_q\times\C P^1\bigr)&=&\K_\su\bigl(\su\times_\uo\man_q
\bigr)\nonumber\\[4pt]&=&\K_\uo\bigl(\man_q\bigr)
{}~=~\K\bigl(\man_q\bigr)\otimes\rep_\uo \ .
\label{excisionslc}\eea
This K-theoretic equality asserts a one-to-one correspondence between
classes of $\su$-equivariant bundles over $\man_q\times\C P^1$ and classes
of $\uo$-equivariant bundles over $\man_q$ with $\uo$ acting trivially on
$\man_q$. The isomorphism (\ref{excisionslc}) of equivariant K-theory
groups is constructed explicitly as follows~\cite{Segal}. Given an
$\su$-equivariant bundle ${\cal E}\to\man_q\times\C P^1$, we can
induce a $\uo$-equivariant bundle $E=\imath^*{\cal E}\to\man_q$ by
restriction to the slice $\man_q\cong
\man_q\times\uo/\uo\stackrel{\imath}{\hookrightarrow}
\man_q\times\su/\uo$. Conversely, if $E\to \man_q$ is a
$\uo$-equivariant bundle, then ${\cal E}=\su\times_\uo E\to
\man_q\times\C P^1$ is an $\su$-equivariant bundle, where the quotient
on $\su\times E$ is by the action of $\uo$ on both factors,
$\zeta\cdot(g,e)=(g\,\zeta^{-1},\zeta\cdot e)$ for $g\in\su,e\in
E,\zeta\in\uo$, and the action of $g'\in\su$ on $\su\times_\uo E$
descends from that on $\su\times E$ given by
$g'\cdot(g,e)=(g'\,g,e)$. This construction defines equivalence
functors between the categories of $\su$-equivariant vector bundles over
$\man_q\times\C P^1$ and $\uo$-equivariant vector bundles over $\man_q$, and
hence the corresponding Grothendieck groups coincide, as in
(\ref{excisionslc}).

The role of the representation ring $\rep_\uo$ is unveiled by setting
$\man_q={\rm pt}$ in (\ref{excisionslc}) to get
\beq
\K_\su\left(\C P^1\right)=\rep_\uo \ ,
\label{KslcCP1RP}\eeq
which establishes a one-to-one correspondence between classes of
homogeneous vector bundles over the sphere $\C P^1$ and classes of
finite-dimensional representations of $\uo$. Since the
corresponding irreducible representations are the ${\mbf\nu}_m$ given by
(\ref{U1irreps}), the representation ring of $\uo$ is
the ring of formal Laurent polynomials in the variable ${\mbf\nu}_1$,
$\rep_\uo=\Z[{\mbf\nu}_1^{~},{\mbf\nu}_1^{-1}]$. Using
(\ref{monmexpl}) we can associate the monopole bundle $\Lcal$ to the
generator ${\mbf\nu}_1$, and thereby identify (\ref{KslcCP1RP}) as the
Laurent polynomial ring
\beq
\K_\su\left(\C P^1\right)=\Z\left[\Lcal\,,\,\Lcal^\vee\,\right] \ .
\label{KslcCP1mon}\eeq
In particular, the relationship (\ref{excisionslc}) can be expressed
as
\beq
\K_\su\left(\man_q\times\C P^1\right)=\K\bigl(\man_q
\bigr)\otimes\Z\left[\Lcal\,,\,\Lcal^\vee\,\right] \ .
\label{Botteq}\eeq
This is the appropriate modification of the Bott periodicity
isomorphism (\ref{Bottper}) to the present setting. The crucial
difference now is that virtual bundles over $\man_q$ are multiplied by
arbitrary powers of the one-monopole bundle, allowing us to extend the
equivalence to arbitrary monopole charges $m\in\Z$. In the equivariant
setting, there is no need to use external twists of the monopole
bundle, nor the ensuing K-theory product as done in~\cite{LPS}. The monopole
fluxes are now naturally incorporated by the coefficient ring
$\rep_\uo$ of the $\uo$-equivariant K-theory, superseding the need for
introducing D-operations.

It is instructive to see precisely how the correspondence
(\ref{Botteq}) works. For this, it is convenient to work instead in
the category of holomorphic $\slc$-equivariant bundles~\cite{BGP}. If ${\cal
  E}$ is an $\su$-equivariant vector bundle over $\man_q\times\C
P^1$, then the action of $\su$ can be extended to an $\slc$
action. Everything we have said above carries through by replacing
the group $\su$ with its complexification $\slc$ and the Cartan torus
$\uo\subset\su$ with the subgroup $\Pt\subset\slc$ of lower triangular
matrices. We are then interested in $\Pt$-equivariant bundles over
$\man_q$ with $\Pt$ acting trivially on $\man_q$. The Lie algebra
$\slcL$ is generated by the three Pauli matrices
\beq
\sigma_3\=\begin{pmatrix}1&0\\0&-1\end{pmatrix} \ , \quad
\sigma_+\=\begin{pmatrix}0&1\\0&0\end{pmatrix} \quad\mbox{and}\quad
\sigma_-\=\begin{pmatrix}0&0\\1&0\end{pmatrix}
\label{sl2cmatrices}\eeq
with the commutation relations
\beq
\left[\sigma_3\,,\,\sigma_\pm\right]\=\pm\,2\,\sigma_\pm
\qquad\textrm{and}\qquad \left[\sigma_+\,,\,\sigma_-\right]\=\sigma_3 \ .
\label{sl2cLie}\eeq
The Lie algebra of the subgroup $\Pt$ is generated by the elements
$\sigma_3$ and $\sigma_-$, while the Cartan subgroup
$\C^\times\subset\Pt$ is generated by the element $\sigma_3$ with the
corresponding irreducible representations being the ${\mbf\nu}_m$
given by (\ref{U1irreps}).

Since the manifold $\man_q$ carries a trivial action of the subgroup
$\C^\times$, any $\C^\times$-equivariant bundle
$E'\to \man_q$ can be written using (\ref{trivialEdecomp}) as a finite
Whitney sum
\beq
E'=\bigoplus_{l\in\triangle(E'\,)}\,E_l'\otimes\underline{S}^{~}_{\,l} \ ,
\label{EWhitneySl}\eeq
where $\triangle(E'\,)\subset\Z$ is the set of eigenvalues for the
$\C^\times$-action on $E'$ and $E'_l\to \man_q^{~}$ are bundles carrying
the trivial $\C^\times$-action. The rest of the $\Pt$-equivariant
structure is determined by the generator $\sigma_-$. Since
$[\sigma_3,\sigma_-]=-2\,\sigma_-$, the action of $\sigma_-$ on
$E_l'\otimes\underline{S}^{~}_{\,l}$ corresponds to holomorphic bundle
morphisms $E_l'\to E_{l-2}'$ and the trivial $\sigma_-$-action on the
irreducible $\C^\times$-modules $\underline{S}_{\,l}$. Thus every
indecomposable
$\Pt$-equivariant bundle $E'\to \man_q$ has weight set of the form
$\triangle(E'\,)=\{m_0,m_0+2,\dots,m_1-2,m_1\}$ for some
$m_0,m_1\in\Z$ with $m_0\leq m_1$. After an appropriate twist by a
$\C^\times$-module and a relabelling, the $\sigma_3$-action is given by the
$\C^\times$-equivariant decomposition
\beq
E=\bigoplus_{i=0}^m\,E_{k_i}\otimes\underline{S}_{\,m-2i}
\label{HactionE}\eeq
while the $\sigma_-$-action is determined by a {\it chain}
\beq
0~\longrightarrow~E_{k_m}~\stackrel{\phi_m}{\longrightarrow}~E_{k_{m-1}}~
\stackrel{\phi_{m-1}}{\longrightarrow}~\cdots~
\stackrel{\phi_2}{\longrightarrow}~E_{k_1}~
\stackrel{\phi_1}{\longrightarrow}~E_{k_0}~\longrightarrow~0
\label{holchain}\eeq
of holomorphic bundle maps between consecutive $E_{k_i}$'s. We can now
consider the underlying $\uo$-equivariant hermitean vector bundle
defined by the unitary ${\rm U}(k)$ reduction of the ${\rm GL}(k,\C)$
structure group of the holomorphic bundle (\ref{HactionE}), after
introducing a hermitean metric on its fibres. Then the
corresponding bundle ${\cal E}\to \man_q\times\C P^1$ is given by
\beq
{\cal E}=\su\times_\uo E \ .
\label{calEinduction}\eeq
Using (\ref{monmexpl}) one finds that (\ref{calEinduction}) coincides
with the original equivariant decomposition
(\ref{calEansatz}). Conversely, given an $\su$-equivariant bundle
${\cal E}\to \man_q\times\C P^1$, its restriction $E=\imath^*{\cal E}$ defines
a $\uo$-equivariant bundle over $\man_q$ which thereby admits an isotopical
decomposition of the form (\ref{HactionE}) and $\cal E$ may be
recovered from (\ref{calEinduction}).

\bigskip

\noindent
{\bf Quiver gauge theory.\ }
The ansatz for the gauge potential on $\man_q\times\C P^1$, represented
symbolically by the bundle chain (\ref{bundlechain}), corresponds to
the disjoint union of two copies of the quiver
\beq
\begin{picture}(0,0)%
\includegraphics{vortex2.pstex}%
\end{picture}%
\setlength{\unitlength}{3947sp}%
\begingroup\makeatletter\ifx\SetFigFont\undefined%
\gdef\SetFigFont#1#2#3#4#5{%
  \reset@font\fontsize{#1}{#2pt}%
  \fontfamily{#3}\fontseries{#4}\fontshape{#5}%
  \selectfont}%
\fi\endgroup%
\begin{picture}(6012,345)(1501,-3961)
\put(6826,-3736){\makebox(0,0)[lb]{\smash{\SetFigFont{9}{10.8}{\familydefault}{\mddefault}{\updefault}{\color[rgb]{0,0,0}$\phi_1$}%
}}}
\put(4351,-3736){\makebox(0,0)[lb]{\smash{\SetFigFont{9}{10.8}{\familydefault}{\mddefault}{\updefault}{\color[rgb]{0,0,0}$\phi_{m-1}$}%
}}}
\put(3301,-3736){\makebox(0,0)[lb]{\smash{\SetFigFont{9}{10.8}{\familydefault}{\mddefault}{\updefault}{\color[rgb]{0,0,0}$\phi_m$}%
}}}
\put(4801,-3961){\makebox(0,0)[lb]{\smash{\SetFigFont{9}{10.8}{\familydefault}{\mddefault}{\updefault}{\color[rgb]{0,0,0}$-m+4$}%
}}}
\put(3676,-3961){\makebox(0,0)[lb]{\smash{\SetFigFont{9}{10.8}{\familydefault}{\mddefault}{\updefault}{\color[rgb]{0,0,0}$-m+2$}%
}}}
\put(1501,-3886){\makebox(0,0)[lb]{\smash{\SetFigFont{11}{13.2}{\familydefault}{\mddefault}{\updefault}{\color[rgb]{0,0,0}$\quiverm~:$}%
}}}
\put(7426,-3961){\makebox(0,0)[lb]{\smash{\SetFigFont{9}{10.8}{\familydefault}{\mddefault}{\updefault}{\color[rgb]{0,0,0}$m$}%
}}}
\put(6076,-3961){\makebox(0,0)[lb]{\smash{\SetFigFont{9}{10.8}{\familydefault}{\mddefault}{\updefault}{\color[rgb]{0,0,0}$m-2$}%
}}}
\put(2626,-3961){\makebox(0,0)[lb]{\smash{\SetFigFont{9}{10.8}{\familydefault}{\mddefault}{\updefault}{\color[rgb]{0,0,0}$-m$}%
}}}
\put(5551,-3886){\makebox(0,0)[lb]{\smash{\SetFigFont{11}{13.2}{\familydefault}{\mddefault}{\updefault}{\color[rgb]{0,0,0}$\cdots$}%
}}}
\end{picture}

\label{Aquiver}\eeq
with the second copy obtained from (\ref{Aquiver}) by reversing the
directions of the arrows and replacing $\phi_i^{~}$ with
$\phi_i^\dag$ for each $i=1,\dots,m$. The vertices of the quiver are
labelled by the degrees of the monopole bundles $\Lcal^{m-2i}$, while the
arrows correspond to module morphisms
$\phi_i:\underline{V}_{\,k_i}\to\underline{V}_{\,k_{i-1}}$ (locally at
each point $x'\in \man_q$). Equivalently, the vertices may be labelled
by irreducible chiral representations of the group $\Pt$. Thus our
ansatz determines a representation of the quiver $\quiverm$ in the
category of complex vector bundles over the manifold
$\man_q$~\cite{quiverbun}. Such a representation is called an
$\quiverm$-bundle. Many properties of the explicit solutions that we
construct later on find their most natural explanation in the context
of such a quiver gauge theory, which provides a more refined
description of the brane configurations than just their K-theory
charges. This framework encompasses the algebraic and representation
theoretic aspects of the problem~\cite{quiverbooks}.

The quiver graph (\ref{Aquiver}) is identical to the Dynkin diagram of
the Lie algebra ${\rm A}_{m+1}$. The adjacency matrix of the quiver
has matrix elements specifying the number of links between each pair
of vertices $m-2i,m-2j$, and in the case (\ref{Aquiver}) it is given by ${\rm
  Adj}(\quiverm)=(\delta_{i,j-1})_{i,j=0,1,\dots,m}$. The matrix
elements $C_{ij}=2\,\delta_{ij}-{\rm Adj}(\quiverm)_{ij}$ are then identical to
those of the Cartan matrix $C_{ij}=\vec e_i\cdot\vec e_j$, where $\vec
e_i$, $i=0,1,\dots,m$  are the simple roots of
${\rm A}_{m+1}$. Corresponding to the gauge symmetry breaking
(\ref{gaugebroken}), the dimension vector $\vec
k^{~}_{\underline{\cal V}}:=(k_0,k_1,\dots,k_m)$ can be regarded as a positive
root of ${\rm A}_{m+1}$ associated with the Cartan matrix $C=(C_{ij})$
by writing it as
\beq
\vec k^{~}_{\underline{\cal V}}\=\sum_{i=0}^mk_i~\vec e_i
\qquad\textrm{with}\qquad
\bigl|\,\vec k^{~}_{\underline{\cal V}}\,\bigr|~:=~\sum_{i=0}^mk_i\=k \ .
\label{veckroots}\eeq
By Kac's theorem~\cite{quiverbooks}, there is a one-to-one
correspondence between the isomorphism classes of indecomposable
representations of the quiver $\quiverm$ and the set of positive roots
of the Lie algebra ${\rm A}_{m+1}$. This property is a consequence of
the $\su$-invariance of our ansatz.

Let us focus for a while on the case $\man_q={\rm pt}$. In this case
eq.~(\ref{calEansatz}), with the $m$-monopole bundles $\Lcal^m$
substituted everywhere by the holomorphic line bundles
(\ref{monmholexpl}), gives a relation between
the categories of homogeneous holomorphic vector bundles over $\C
P^1=\slc/\Pt$ and of finite-dimensional chiral representations of $\Pt$, while
the quiver representation further gives a relation with the abelian
category of finite-dimensional representations of
$\quiverm$~\cite{quiverbun}. To describe this latter category, it is
convenient to introduce the notion of a path $\path$ in $\quiverm$,
which is generally defined as a sequence of arrows of the quiver which
compose. In the present case any path is of the form
\beq
\begin{picture}(0,0)%
\includegraphics{vortex3.pstex}%
\end{picture}%
\setlength{\unitlength}{3947sp}%
\begingroup\makeatletter\ifx\SetFigFont\undefined%
\gdef\SetFigFont#1#2#3#4#5{%
  \reset@font\fontsize{#1}{#2pt}%
  \fontfamily{#3}\fontseries{#4}\fontshape{#5}%
  \selectfont}%
\fi\endgroup%
\begin{picture}(5862,420)(1651,-3961)
\put(4201,-3661){\makebox(0,0)[lb]{\smash{\SetFigFont{9}{10.8}{\familydefault}{\mddefault}{\updefault}{\color[rgb]{0,0,0}$\phi_{\frac{m-m_0}2+1}$}%
}}}
\put(6676,-3661){\makebox(0,0)[lb]{\smash{\SetFigFont{9}{10.8}{\familydefault}{\mddefault}{\updefault}{\color[rgb]{0,0,0}$\phi_{\frac{m-m_1}2+1}$}%
}}}
\put(3151,-3661){\makebox(0,0)[lb]{\smash{\SetFigFont{9}{10.8}{\familydefault}{\mddefault}{\updefault}{\color[rgb]{0,0,0}$\phi_{\frac{m-m_0}2}$}%
}}}
\put(1651,-3886){\makebox(0,0)[lb]{\smash{\SetFigFont{11}{13.2}{\familydefault}{\mddefault}{\updefault}{\color[rgb]{0,0,0}$\path~:$}%
}}}
\put(7351,-3961){\makebox(0,0)[lb]{\smash{\SetFigFont{9}{10.8}{\familydefault}{\mddefault}{\updefault}{\color[rgb]{0,0,0}$m_1$}%
}}}
\put(6001,-3961){\makebox(0,0)[lb]{\smash{\SetFigFont{9}{10.8}{\familydefault}{\mddefault}{\updefault}{\color[rgb]{0,0,0}$m_1-2$}%
}}}
\put(2626,-3961){\makebox(0,0)[lb]{\smash{\SetFigFont{9}{10.8}{\familydefault}{\mddefault}{\updefault}{\color[rgb]{0,0,0}$m_0$}%
}}}
\put(4801,-3961){\makebox(0,0)[lb]{\smash{\SetFigFont{9}{10.8}{\familydefault}{\mddefault}{\updefault}{\color[rgb]{0,0,0}$m_0+4$}%
}}}
\put(3676,-3961){\makebox(0,0)[lb]{\smash{\SetFigFont{9}{10.8}{\familydefault}{\mddefault}{\updefault}{\color[rgb]{0,0,0}$m_0+2$}%
}}}
\put(5551,-3886){\makebox(0,0)[lb]{\smash{\SetFigFont{11}{13.2}{\familydefault}{\mddefault}{\updefault}{\color[rgb]{0,0,0}$\cdots$}%
}}}
\end{picture}

\label{path}\eeq
with $-m\leq m_0\leq m_1\leq m$. We will denote it by the formal vector
$|m_0,\dots,m_1)$. The non-negative integer
$|\path\,|:=\frac12\,(m_1-m_0)$ is the length of the
path~(\ref{path}). The trivial path of length~$0$ based at a
single vertex $m_0$ is denoted $|m_0)$. The path algebra
$\C\,\quiverm$ of the quiver (\ref{Aquiver}) is then defined as the
algebra generated by all paths $\path$ of $\quiverm$, i.e. as the
vector space
\beq
\C\,\quiverm=\bigoplus_{\stackrel{\scriptstyle m_0,m_1=-m}
{\scriptstyle m_0\leq m_1}}^m\,\C|m_0,\dots,m_1)
\label{pathalg}\eeq
together with the $\C$-linear multiplication induced by (left)
concatenation of paths where possible,
\beq
|m_0,\dots,m_1)\cdot|n_0,\dots,n_1)=
\delta_{m_1n_0}~|m_0,\dots,n_1) \ .
\label{pathalgmult}\eeq
This makes $\C\,\quiverm$ into a finite-dimensional quasi-free
algebra. The path algebra has a natural $\Z_{m+1}$-grading by path
length,
\beq
\C\,\quiverm\=\bigoplus_{i=0}^m\,(\C\,\quiverm)_i
\qquad\mbox{with}\qquad (\C\,\quiverm)_i\=\bigoplus_{m_0=-m}^{m-2i}
\C|m_0,\dots,m_0+2i) \ ,
\label{pathgrading}\eeq
and can thereby be alternatively described as the tensor algebra over
the ring
\beq
C_0\=\bigoplus_{i=0}^m\,\C|m-2i)~\cong~\C^{m+1}
\label{C0ring}\eeq
of the $C_0$-bimodule
\beq
C_1=\bigoplus_{i=0}^m\,\C|m-2i,m-2i+2) \ .
\label{C1bimodule}\eeq

The importance of the path algebra stems from the fact that the
category of representations of the quiver $\quiverm$ is equivalent to
the category of (left)
$\C\,\quiverm$-modules~\cite{quiverbooks}. Given a representation
$\underline{W}_{\,m-2i}\stackrel{\eta_i}{\longrightarrow}
\underline{W}_{\,m-2i+2}$, $i=1,\dots,m$, of $\quiverm$, the
associated $\C\,\quiverm$-module $\underline{\cal W}$ is
\beq
\underline{\cal W}=\bigoplus_{i=0}^m\,\underline{W}_{\,m-2i}
\label{calWdef}\eeq
with multiplication extended $\C$-linearly from the definitions
\beq
|m-2i)\cdot w_j\=\delta_{ij}~w_j \qquad\textrm{and}\qquad
|m-2i,m-2i+2)\cdot w_j\=\delta_{i,j+1}~\eta_j(w_j)
\label{quivalgmod}\eeq
for $w_j\in\underline{W}_{\,m-2j}$. Conversely, given a left
$\C\,\quiverm$-module $\underline{\cal W}$, we can set
$\underline{W}_{\,m-2i}:=|m-2i)\cdot\underline{\cal W}$
for $i=0,1,\dots,m$ and define
$\eta_i:\underline{W}_{\,m-2i}\to\underline{W}_{\,m-2i+2}$ for
$i=1,\dots,m$ by
\beq
\eta_i(w_i)=|m-2i,m-2i+2)\cdot w_i \ .
\label{etaiwi}\eeq
One can further show that morphisms of representations of $\quiverm$
correspond to $\C\,\quiverm$-module
homomorphisms~\cite{quiverbooks}. Thus, the problem of determining
finite-dimensional representations of the quiver $\quiverm$, or
equivalently homogeneous vector bundles over $\C P^1$, is equivalent
to finding representations of its path algebra.

As an example, consider the ${\rm A}_2$ quiver
\beq
\begin{picture}(0,0)%
\includegraphics{vortex4.pstex}%
\end{picture}%
\setlength{\unitlength}{3947sp}%
\begingroup\makeatletter\ifx\SetFigFont\undefined%
\gdef\SetFigFont#1#2#3#4#5{%
  \reset@font\fontsize{#1}{#2pt}%
  \fontfamily{#3}\fontseries{#4}\fontshape{#5}%
  \selectfont}%
\fi\endgroup%
\begin{picture}(2550,345)(1651,-3961)
\put(1651,-3886){\makebox(0,0)[lb]{\smash{\SetFigFont{11}{13.2}{\familydefault}{\mddefault}{\updefault}{\color[rgb]{0,0,0}$\quiver~:$}%
}}}
\put(4201,-3811){\makebox(0,0)[lb]{\smash{\SetFigFont{9}{10.8}{\familydefault}{\mddefault}{\updefault}{\color[rgb]{0,0,0}.}%
}}}
\put(2701,-3961){\makebox(0,0)[lb]{\smash{\SetFigFont{9}{10.8}{\familydefault}{\mddefault}{\updefault}{\color[rgb]{0,0,0}$-1$}%
}}}
\put(3901,-3961){\makebox(0,0)[lb]{\smash{\SetFigFont{9}{10.8}{\familydefault}{\mddefault}{\updefault}{\color[rgb]{0,0,0}$+1$}%
}}}
\put(3301,-3736){\makebox(0,0)[lb]{\smash{\SetFigFont{9}{10.8}{\familydefault}{\mddefault}{\updefault}{\color[rgb]{0,0,0}$\phi_1$}%
}}}
\end{picture}

\label{A2quiver}\eeq
It represents the standard brane-antibrane system, and as expected
$\su$-equivariance implies that it can only carry $m=1$ unit of
monopole charge~\cite{LPS}. The corresponding path algebra is
\beq
\C\,\quiver\=\C|-1)~\oplus~\C|+1)~\oplus~\C|-1,+1)\=
\begin{pmatrix}\C&\C\\0&\C\end{pmatrix} \ .
\label{A2pathalg}\eeq
Representations of this algebra yield the standard superconnections
characterizing the low-energy field content on the worldvolume of a
brane-antibrane system~\cite{AIO1}. In the next section we will show how to
generalize the superconnection formalism to account for
representations of generic path algebras (\ref{pathalg}). Later on we
shall write down explicit solutions with generic monopole charge
$m\in\Z$ that also correspond to the basic brane-antibrane system.

Our technique for generating D-branes from a quiver gauge theory on
$\man_q$ arises via a quotient with respect to a generalized
$\su$-action on Chan-Paton bundles over $\man_q\times\C P^1$. This new
construction is rather different from the well-known quiver gauge
theories that arise from orbifolds with respect to the action of a
{\it discrete} group $G$~\cite{quiverG}. In the latter case the nodes of a
quiver represent the irreducible representation fractional branes into
which a regular representation D-brane decays into when it is taken to
an orbifold point of $\man_q\,/\,G$, and they can be thought of in
terms of a projection of branes sitting on the leaves of the covering
space $\man_q$. While our quiver gauge theory is fundamentally
different, it shares many of the physical features of orbifold
theories of D-branes. For instance, the blowing up of vortices on
$\man_q$ into instantons on $\man_q\times\C P^1$ is reminescent
of the blowing up of fractional D$(q-1)$-branes into D$(q+1)$-branes
wrapping a non-contractible $\C P^1$ that is used to resolve the
orbifold singularity in $\man_q\,/\,G$. Our solutions provide explicit
realizations of this blowing up phenomenon, but in a completely smooth
setting.

\bigskip

\section{Dimensional reduction\label{Dimred}}

The condition of $\su$-equivariance uniquely prescribes a specific
$\C P^1$ dependence for the gauge potential $\ca$ and reduces
the Yang-Mills equations (\ref{YM}) on $\man_q\times \C P^1$ to
equations on $\man_q$. In this section we will formulate this
reduction in detail and relate it to representations of the path
algebra (\ref{pathalg}). This will be done by developing a new
formalism of $\Z_{m+1}$-graded connections which describes the field
content corresponding to the bundle chains (\ref{bundlechain}) and
(\ref{holchain}), and which generalizes the standard superconnection
field theories on the worldvolumes of brane-antibrane
systems~\cite{AIO1}. This formalism will be the crux to merging
together the three interpretations of the previous section.

\bigskip

\noindent
{\bf Reduction of the Yang-Mills functional.\ }
The dimensional reduction of the Yang-Mills equations can be seen at
the level of the Yang-Mills lagrangian (\ref{lagr}). Substituting
(\ref{f15})--(\ref{Fyyb}) into (\ref{lagrprod}) and performing the integral
over $\C P^1$ we arrive at the action\footnote{A set of Yang-Mills coupling
constants $g_{\rm YM}^i$, $i=0,1,\dots,m$ can be introduced via the
redefinitions $A^i\mapsto g_{\rm YM}^i\,A^i$.}
\bea
S^{~}_{\rm YM}&:=&\int_{{\man_q}\times {\C P^1}}
{\diff^{q+2}}x~L_{\rm YM}^{~}\nonumber\\[4pt] &=&
\pi\,R^2\,\int_{\man_q}\diff^{q}x'~\sqrt{g'}~
\sum_{i=0}^m\,\tr^{~}_{k_i\times k_i}\left[
\bigl(F_{{\mu}'{\nu}'}^i\bigr)^\+\,\bigl(F^{i\,{\mu}'{\nu}'}\,\bigr)
+ \frac{1}{R^2}\,\bigl(D_{{\mu}'} \phi_{i+1}\bigr)\,\bigl(D^{{\mu}'} \phi_{i+1}
\bigr)^\+\right.\nonumber\\ && +\left.
\frac{1}{R^2}\,\bigl(D_{{\mu}'} \phi_{i}
\bigr)^\+\,\bigl(D^{{\mu}'} \phi_{i}\bigr)+
\frac{1}{2\,R^4}\,\left(m-2i+\phi_i^\+\,\phi^{~}_i -
\phi^{~}_{i+1}\,\phi^\+_{i+1}\right)^2\,\right] \ ,
\label{SYMred}\eea
where $g'=|\det (g_{{\mu}'{\nu}'})|$. In the remainder of this paper
we shall only consider static field configurations
on $\man_q=\R^1\times M_{2n}$ in the temporal gauge
$\ca_0=0$. In this case one can introduce the corresponding energy
functional
\bea\label{EF}
E^{~}_{\rm YM}&=&\pi\,R^2\ \int_{M_{2n}}\diff^{2n}x~\sqrt{g_{n}}\ \sum_{i=0}^m
\,\tr^{~}_{k_i\times k_i}\left[\bigl(F_{{\mu}{\nu}}^i\bigr)^\+\,
\bigl(F^{i\,{\mu}{\nu}}\bigr)
+ \frac{1}{R^2}\,\bigl(D_{{\mu}} \phi_{i+1}\bigr)
\,\bigl(D^{{\mu}} \phi_{i+1}\bigr)^\+
\right.\nonumber\\ &&+\left.
 \frac{1}{R^2}\,\bigl(D_{{\mu}} \phi_{i}\bigr)^\+\,
\bigl(D^{{\mu}}\phi_{i}\bigr)+
\frac{1}{2\,R^4}\,\left(m-2i+\phi_i^\+\,\phi^{~}_i - \phi^{~}_{i+1}
\,\phi^\+_{i+1}\right)^2\,\right] \ ,
\eea
where $g_{n}=\det (g_{\m\n})$. The functional (\ref{EF}) is
non-negative.

\bigskip

\noindent
{\bf Graded connections.\ }
The energy functional (\ref{EF}) is analysed most efficiently by
introducing a framework specific to connections on the rank $k\ $
$\Z_{m+1}$-graded vector bundle
\beq
E:=\bigoplus_{i=0}^m\,E_{k_i}
\label{gradedbundle}\eeq
over $M_{2n}$ whose typical fibre is the module
(\ref{genrepSU2Uk}). The endomorphism algebra bundle corresponding to
(\ref{gradedbundle}) is given by the direct sum decomposition
\beq
{\rm End}(E)=\bigoplus_{i=0}^m\,{\rm End}(E_{k_i})~\oplus~
\bigoplus_{\stackrel{\scriptstyle i,j=0}{\scriptstyle i\neq j}}^m
\,{\rm Hom}(E_{k_i},E_{k_j}) \ .
\label{Endgraded}\eeq
We may naturally associate to (\ref{Endgraded}) a distinguished
representation of the ${\rm A}_{m+1}$ quiver. For this, we note that
the path algebra $\C\,\quiverm$ is itself a $\C\,\quiverm$-module, and
that the elements $|m-2i)\in\C\,\quiverm$ define a complete set of
orthogonal projectors of the path algebra,
i.e. $|m-2i)\cdot|m-2j)=\delta_{ij}~|m-2i)$ for $i,j=0,1,\dots,m$ with
$\sum_{i=0}^m|m-2i)=1$. Analogously to the construction of
(\ref{calWdef})--(\ref{etaiwi}), we may thereby define a {\it projective}
$\C\,\quiverm$-module $\pathmod_{\,i}:=|m-2i)\cdot\pathalg$ for each
$i=0,1,\dots,m$~\cite{quiverbooks}, which is the subspace of
$\pathalg$ generated by all paths which start at the $i$-th vertex of
the quiver $\quiverm$. Then $(\,\pathmod_{\,i})_{m-2j}\cong\C$ is the
vector space generated by the path from the $i$-th vertex to the
$j$-th vertex, and the corresponding dimension vector is
\beq
\vec k^{~}_{\pathmod_{\,i}}=\sum_{j=i}^m\,\vec e_j \ .
\label{Pidimvec}\eeq
The modules $\pathmod_{\,i}$, $i=0,1,\dots,m$ are exactly the
set of all indecomposable projective representations of the ${\rm
  A}_{m+1}$ quiver~\cite{quiverbooks}, with
\beq
\pathalg=\bigoplus_{i=0}^m\,\pathmod_{\,i} \ .
\label{pathalgmoddecomp}\eeq

The importance of this path algebra representation stems from the fact
that, for any quiver representation (\ref{genrepSU2Uk}), there is a
natural isomorphism~\cite{quiverbooks}
\beq
{\rm Hom}(\,\pathmod_{\,i}\,,\,\underline{\cal V}\,)
\cong\underline{V}_{\,k_i} \ .
\label{Hompathnatural}\eeq
We may thereby identify ${\rm Hom}(\,\underline{V}_{\,k_j}\,,\,
\underline{V}_{\,k_i}\,)$ in terms of appropriate combinations of the
spaces
\beq
{\rm Hom}(\,\pathmod_{\,j}\,,\,\pathmod_{\,i})~\cong~
|m-2j)\cdot\pathalg\cdot|m-2i)~\cong~\C \ .
\label{HomPij}\eeq
This is the vector space generated by the path from
the $i$-th vertex to the $j$-th vertex of $\quiverm$. A natural
representation of this path is by a matrix of dimension
$(m+1)\times(m+1)$ with $1$ in its $(ij)$-th entry and $0$'s
everywhere else. The path algebra (\ref{pathalgmoddecomp}) is thereby
identified with the algebra of upper triangular $(m+1)\times(m+1)$
complex matrices~\cite{quiverbooks}. For a given quiver representation
(\ref{genrepSU2Uk}), this algebra may be represented by assembling the
chiral Higgs fields $\phi_1,\dots,\phi_m$ into the $k\times k$ matrix
\beq
\mphi:=\begin{pmatrix}0&\phi_1&0&\dots&0\\0&0&\phi_2&\dots&0\\
\vdots&\vdots&\ddots&\ddots&\vdots\\0&0&0&\dots&\phi_m\\
0&0&0&\dots&0\end{pmatrix}
\label{mgradedphidef}\eeq
with respect to the decomposition (\ref{gradedbundle}). This object
generates a representation of the path algebra in the category of
complex vector bundles over $M_{2n}$, corresponding to the
off-diagonal $i<j$ components of the decomposition (\ref{Endgraded}). The
finite dimensionality of $\pathalg$ is reflected in the property that
generically
\beq
\mphi,\bigl(\mphi\bigr)^2,\dots,\bigl(\mphi\bigr)^m~\neq~0
\qquad\mbox{but}\qquad \bigl(\mphi\bigr)^{m+1}\=0 \ .
\label{mphi0s}\eeq
The field configuration (\ref{mgradedphidef}) generates the basic
zero-form component of a geometric object that we shall refer to as a
``$\Z_{m+1}$-graded connection'' on $M_{2n}$. For $m=1$ it corresponds
to a standard superconnection~\cite{Quillen1}, while for $m>1$ it is
the appropriate entity that constructs representations corresponding to the
enlargement of the path algebra $\pathalg$. Its matrix form is similar
to (\ref{f4})--(\ref{f9}), but without the one-forms on $\C P^1$.

To formulate the definition precisely, we note
that the algebra $\Omega(M_{2n},E)$ of differential forms on $M_{2n}$
with values in the bundle (\ref{gradedbundle}) has a natural
$\Z\times\Z_{m+1}$ grading, where the $\Z$-grading is by form
degree. We can thereby induce a total $\Z_{m+1}$-grading by the
decomposition
\beq
\Omega_\bullet(M_{2n},E)\=\bigoplus_{p=0}^m\,\Omega_{(p)}(M_{2n},E)
\qquad\mbox{with}\qquad \Omega_{(p)}(M_{2n},E)\=
\bigoplus_{i+j\equiv^{~}_{m+1}\,p}\,\Omega^i(M_{2n},E_{k_j}) \ ,
\label{Omegagrading}\eeq
where $\equiv^{~}_{m+1}$ denotes congruence modulo $(m+1)$. By using
(\ref{Endgraded}) and the usual tensor product grading, this induces a
$\Z_{m+1}$-grading on the corresponding endomorphism algebra as
\beq
\Omega_\bullet(M_{2n},{\rm End}~E)=\bigoplus_{p=0}^m\,\Omega_{(p)}(M_{2n},
{\rm End}~E)
\label{OmegaEnddecomp}\eeq
with
\beq
\Omega_{(p)}(M_{2n},{\rm End}~E)=
\bigoplus_{i=0}^m~\bigoplus_{a=0}^p~\bigoplus_{i_a\equiv^{~}_{m+1}\,(p-a)}
{}~\Omega^{i_a}(M_{2n})\otimes{\rm Hom}(E_{k_i},E_{k_{i+a}}) \ .
\label{OmegaEndgrading}\eeq
A {\it graded connection} on (\ref{Omegagrading}) is defined to be a linear
operator $\Omega_\bullet(M_{2n},E)\to\Omega_{\bullet+1}(M_{2n},E)$ which
shifts the total $\Z_{m+1}$-grading by $1$ modulo $(m+1)$, i.e. an
element of
\bea
\Omega_{(1)}(M_{2n},{\rm End}~E)&=&\bigoplus_{i=0}^m~\biggl(\,
\bigoplus_{i_1\equiv^{~}_{m+1}\,1}\,\Omega^{i_1}(M_{2n})\otimes
{\rm End}(E_{k_i})\biggr.\nonumber\\ && \qquad\quad
{}~\biggl.\oplus~\bigoplus_{i_0\equiv^{~}_{m+1}\,0}\,
\Omega^{i_0}(M_{2n})\otimes{\rm Hom}(E_{k_i},E_{k_{i+1}})
\biggr) \ ,
\label{Omega1conn}\eea
and which satisfies the usual Leibniz rule on $\Omega(M_{2n})$. As in
the standard cases, the $\Z_{m+1}$-graded connections form an affine
space modelled on a set of local operators.

In our case we retain only the $i_0=0$ and $i_1=1$ components of
(\ref{Omega1conn}) corresponding to the lowest lying massless degrees
of freedom on the given configuration of D-branes. From the Leibniz
rule it follows that the pertinent graded connections are then of the
form $(\diff+\mA+(\mphi)+(\mphi)^\dag)$, where
\beq
\mA:=\sum_{i=0}^mA^i\otimes\Pi_i
\label{mAdef}\eeq
and $\Pi_i:E\to E_{k_i}$ are the canonical orthogonal projections of
rank~$1$,
\beq
\Pi_i\,\Pi_j=\delta_{ij}~\Pi_i \ ,
\label{Piortho}\eeq
which may be represented, with respect to the decomposition
(\ref{gradedbundle}), by diagonal matrices
$\Pi_i=(\delta_{ji}\,\delta_{li})_{j,l=0,1,\dots,m}$ of unit trace. In this
geometric
framework all $\phi_i$ are assumed to anticommute with a given local
basis $\diff x^\mu$ of the cotangent bundle of the K\"ahler manifold
$M_{2n}$, as if they were $m$ basic odd complex elements of a
superalgebra. This requisite property may be explicitly realized
by extending the graded connection formalism to $M_{2n}\times\C
P^1$. For this, we rewrite the ansatz (\ref{f4})--(\ref{f8}) in terms
of the above field configurations as
\bea
\ca_\mu&=&\bigl(\mA\bigr)_\mu\otimes1 \ , \label{calAgradedmu}\\[4pt]
\ca_y&=&\Idd_k\otimes\bigl(\ma\bigr)_y-\bigl(\mphi\bigr)^\dag\otimes
\beta_y \ , \label{calAgradedy}\\[4pt]
\ca_\yb&=&\Idd_k\otimes\bigl(\ma\bigr)_\yb+\bigl(\mphi\bigr
)\otimes\bar\beta_\yb \ ,
\label{calAgraded}\eea
where
\beq
\ma:=\sum_{i=0}^ma_{m-2i}\otimes\Pi_i
\label{madef}\eeq
and $\Pi_i:{\cal E}\to{\cal E}_i$ are the canonical projections on
(\ref{calEansatz}). The coupling of $\mphi$ to $\diff\yb$ in
(\ref{calAgraded}) yields the desired anticommutativity with $\diff
x^\mu$.

Alternatively, we may use the canonical isomorphism
$\Omega(M_{2n}\times\C P^1)\cong\cliff(M_{2n}\times\C P^1)$ to map the
cotangent basis $\diff x^{\hat\mu}\mapsto\Gamma^{\hat\mu}$ onto the
generators of the Clifford algebra
\beq
\Gamma^{\hat\mu}\,\Gamma^{\hat\nu}+\Gamma^{\hat\nu}\,\Gamma^{\hat\mu}\=
-2\,g^{\hat\mu\hat\nu}~\Idd_{2^{n+1}}
\qquad\mbox{with}\qquad \hat\mu,\hat\nu\=1,\dots,2n+2 \ .
\label{2n2Cliffalg}\eeq
The gamma-matrices in (\ref{2n2Cliffalg}) may be decomposed as
\beq
\bigl\{\Gamma^{\hat\mu}\bigr\}\=\bigl\{\Gamma^\mu,\Gamma^y,
\Gamma^\yb\bigr\} \qquad\mbox{with}\qquad \Gamma^\mu\=\gamma^\mu
\otimes\Idd_2 \ , ~~ \Gamma^y\=\gamma\otimes\gamma^y
\quad\mbox{and}\quad \Gamma^\yb\=\gamma\otimes\gamma^\yb \ ,
\label{gamma2n2decomp}\eeq
where the $2^n\times2^n$ matrices $\gamma^\mu=-(\gamma^\mu)^\dag$
act on the spinor module $\underline{\Delta}\,(M_{2n})$ over the
Clifford algebra $\cliff(M_{2n})$,
\beq
\gamma^\mu\,\gamma^\nu+\gamma^\nu\,\gamma^\mu\=-2\,g^{\mu\nu}~\Idd_{2^n}
\qquad\mbox{with}\qquad \mu,\nu\=1,\dots,2n \ ,
\label{2nCliffalg}\eeq
while
\beq
\gamma\=\frac{\im^n}{(2n)!~\sqrt{g_n}}~\epsilon_{\mu_1\cdots\mu_{2n}}\,
\gamma^{\mu_1}\cdots\gamma^{\mu_{2n}} \qquad\mbox{with}\qquad
(\gamma)^2\=\Idd_{2^n} \quad\mbox{and}\quad
\gamma\,\gamma^\mu\=-\gamma^\mu\,\gamma
\label{chiralityop}\eeq
is the corresponding chirality operator. Here
$\epsilon_{\mu_1\dots\mu_{2n}}$ is the Levi-Civita symbol with
$\epsilon_{12\cdots2n}=+\,1$. The action of the Clifford
algebra $\cliff(\C P^1)$ on the spinor module $\underline{\Delta}\,(\C
P^1)$ is generated by
\beq
\gamma^y\=\frac1{R^2}\,\left(R^2+y\yb\right)\,\sigma^y \quad\mbox{and}\quad
\gamma^\yb\=\frac1{R^2}\,\left(R^2+y\yb\right)\,\sigma^\yb
\label{CP1Cliffalg}\eeq
with constant $2\times2$ Pauli matrices $\sigma^\yb=\sigma_-$ and
$\sigma^y=-\sigma_+$ obeying $[\sigma^y,\sigma^\yb\,]=-\sigma_3$. The gauge
potential (\ref{f4})--(\ref{f9}) may then be written in an algebraic
form as
\bea
\hat\ca&:=&\Gamma^{\hat\mu}\,\ca_{\hat\mu} \nonumber\\[4pt]
&=&\gamma^\mu\,\bigl(\mA\bigr)_\mu\otimes\Idd_2+\bigl(\mphi\bigr)
\,\gamma\otimes\gamma^\yb\,\bar\beta_\yb-\bigl(\mphi\bigr)^\dag\,
\gamma\otimes\gamma^y\,\beta_y \nonumber\\ &&+\,\gamma\otimes\left(\gamma^y\,
\bigl(\ma\bigr)_y+\gamma^\yb\,\bigl(\ma\bigr)_\yb\right) \ ,
\label{calAgammas}\eea
and the coupling of (\ref{mgradedphidef}) with the chirality operator
(\ref{chiralityop}) realizes the desired anticommutativity with the
one-form representatives $\gamma^\mu$. Note that the products
\beq
\bigl(\mphi\bigr)\,\gamma\otimes\gamma^\yb\,\bar\beta_\yb\=
\frac1R\,\bigl(\mphi\bigr)\,\gamma\otimes\sigma^\yb \quad\mbox{and}\quad
\bigl(\mphi\bigr)^\dag\,\gamma\otimes\gamma^y\,\beta_y\=
\frac1R\,\bigl(\mphi\bigr)^\dag\,\gamma\otimes\sigma^y
\label{CP1indepprods}\eeq
are independent of the coordinates $(y,\yb)\in\C P^1$.

The curvature
$(\diff+\mA+(\mphi)+(\mphi)^\dag)^2\in\Omega_{(2)}(M_{2n},{\rm End}~E)$ of the
graded connection is also most elegantly expressed through dimensional
reduction from $M_{2n}\times\C P^1$. From (\ref{f10})--(\ref{Fyyb}) it
is given by
\bea
\hat\cf&:=&\mbox{$\frac14$}\,\bigl[\Gamma^{\hat\mu}\,,\,
\Gamma^{\hat\nu}\bigr]\,\cf_{\hat\mu\hat\nu}\nonumber\\[4pt] &=&
\mbox{$\frac14$}\,\bigl[\gamma^\mu\,,\,\gamma^\nu\bigr]\,
\bigl(\mF\bigr)_{\mu\nu}\otimes\Idd_2-\frac1R\,\gamma\,\bigl(\gamma^\mu\,
D_\mu\mphi\bigr)^\dag\otimes\sigma^y-\frac1R\,\gamma\,\bigl(\gamma^\mu\,
D_\mu\mphi\bigr)\otimes\sigma^\yb\nonumber\\ && +\,\frac1{2\,R^2}\,
\left(\mup+\bigl(\mphi\bigr)^\dag\,\bigl(\mphi\bigr)-\bigl
(\mphi\bigr)\,\bigl(\mphi\bigr)^\dag\right)\,
\Idd_{2^n}\otimes\sigma_3
\label{gradedcurv}\eea
where $\mF:=\diff\mA+\mA\wedge\mA$ and
\beq
\mup:=\sum_{i=0}^m\,(m-2i)~\Pi_i \ .
\label{mupdef}\eeq
The contribution (\ref{mupdef}) is generated by the monopole
connection on $\C P^1$ in (\ref{calAgammas}), while the Higgs
potentials in (\ref{gradedcurv}) are produced by
(\ref{CP1indepprods}). The graded curvature is independent of $(y,\bar
y)\in\C P^1$, and the standard gamma-matrix trace formulas
\bea
\Tr^{~}_{\C^{2^{n+1}}}
\Bigl(\gamma^\mu\,\gamma^\nu\otimes\Idd_2\Bigr)&=&-2^{n+1}
\,g^{\mu\nu} \ , \label{Trgammaid1} \\[4pt] \Tr^{~}_{\C^{2^{n+1}}}
\left(\gamma^\mu\,\gamma^\nu\,
\gamma^\lambda\,\gamma^\rho\otimes\Idd_2\right)&=&2^{n+1}\,
\left(g^{\mu\nu}\,g^{\lambda\rho}+g^{\mu\rho}\,g^{\nu\lambda}-
g^{\mu\lambda}\,g^{\nu\rho}\right) \ , \label{Trgammaid2} \\[4pt]
\Tr^{~}_{\C^{2^{n+1}}}\left(\bigl[\gamma^\mu\,,\,\gamma^\nu\bigr]\,
\bigl[\gamma^\lambda\,,\,\gamma^\rho\bigr]\otimes
\Idd_2\right)&=&2^{n+3}\,\left(g^{\mu\rho}\,g^{\nu\lambda}-g^{\mu\lambda}
\,g^{\nu\rho}\right) \ , \label{Trgammaid3} \\[4pt] \Tr^{~}_{\C^{2^{n+1}}}
\Bigl(\gamma^\mu\,
\gamma\,\gamma^\nu\,\gamma\otimes\sigma^\yb\,\sigma^y\Bigr)
&=&-2^n\,g^{\mu\nu}~=~\Tr^{~}_{\C^{2^{n+1}}}\Bigl(\gamma^\mu\,
\gamma\,\gamma^\nu\,\gamma\otimes\sigma^y\,\sigma^\yb\Bigr)
\label{Trgammaid4}\eea
imply that the energy functional (\ref{EF}) can be compactly written
in terms of (\ref{gradedcurv}) as
\beq
E^{~}_{\rm YM}=\frac{\pi\,R^2}{2^n}\,\int_{M_{2n}}\diff^{2n}x~
\sqrt{g_n}~\tr_{k\times k}^{~}~\Tr^{~}_{\C^{2^{n+1}}}~\hat\cf^2 \ .
\label{EFgraded}\eeq

\bigskip

\noindent
{\bf Nonabelian coupled vortex equations.\ } Let us now examine the
reduction of the DUY equations on $M_{2n}\times \C P^1$
for a gauge potential of the form proposed in Section 3 (with static
configurations in the gauge $\ca_0 =0$). Substituting
(\ref{f11})--(\ref{f14}) into (\ref{DUY1})--(\ref{DUY3}), we obtain
\bea\label{f24}
g^{a\bb}\,F^i_{a{\bb}}&=&\frac{1}{2\,R^2}\,\left(m-2i+\phi_i^\+\,
\phi^{~}_i -\phi^{~}_{i+1}\,\phi^\+_{i+1}\right) \ , \\[4pt]
\label{f240} F_{\ab\bb}^i&=&0~=~F_{ab}^i \ , \\[4pt] \label{f25}
\bar\pa^{~}_{\bar a}\phi^{~}_{i+1} + A^i_{\bar a}\,\phi^{~}_{i+1} -
\phi^{~}_{i+1}\,A^{i+1}_{\bar a}&=&0
\eea
for each $i=0,1,\dots,m$, where $\phi_0:=0=:\phi_{m+1}$. Recall that
there is no summation over $i$ in these equations. We have
abbreviated $F_{ab}^i:=F_{z^az^b}^i$ etc., and defined the derivatives
$\partial_a:=\partial_{z^a}=\frac12\,(\partial_{2a-1}+\im\partial_{2a})$
and
$\partial_\ab:=\partial_{\zb^\ab}=\frac12\,(\partial_{2a-1}-\im\partial_{2a})$
with $a,b=1,\dots,n$. We shall call (\ref{f24})--(\ref{f25}) the
nonabelian coupled vortex equations.

Eq.~(\ref{f240}) implies that the vector bundles $E_{k_i}\to M_{2n}$
are holomorphic, while eq.~(\ref{f25}) implies that the Higgs fields
$\phi_{i+1}:E_{k_{i+1}}\to E_{k_i}$ are holomorphic maps. By using a
Bogomolny-type transformation~\cite{Prada} one can show that
solutions to these equations realize absolute minima of the energy
functional (\ref{EF}). These field configurations describe
supersymmetric BPS states of D-branes.

\bigskip

\noindent
{\bf Seiberg-Witten monopole equations.\ } For
$n=2$, $m=1$ and $k_0=k_1=1$ (so that $k=k_0+k_1=2$), the
equations (\ref{f24})--(\ref{f25}) coincide with the perturbed
abelian Seiberg-Witten monopole equations on a K\"ahler
four-manifold $M_4$~\cite{Witten}. In this case we have
\begin{equation}
A^0\=-A^1~=:~A~\in~{\rm u}(1) \ , ~~ F^0\=-F^1~=:~F \quad\mbox{and}
\quad \phi_1~=:~\phi~\in~\C \ ,
\end{equation}
and the equations (\ref{f24})--(\ref{f25}) reduce to
\bea\label{f26}
g^{a\bb}\,F_{a{\bb}}&=&\frac{1}{2\,R^2}\,\left(1-\phi\,\bar\phi\,\right)
\ , \\[4pt] F_{\ab\bb}&=&0~=~F_{ab} \ , \\[4pt] \label{f27}
\bar\pa_{\bar a}\phi + 2\,A_{\bar a}\,\phi&=&0 \ .
\eea
The perturbation, i.e. the term $\frac{1}{2\,R^2}$ in (\ref{f26}), is
needed whenever $M_4$ has non-negative scalar curvature in order to
produce a non-trivial and non-singular moduli
space of finite energy ${\rm L}^2$-solutions. It is usually introduced
into the Seiberg-Witten equations by hand. In the present context, it
arises automatically from the extra space $\C P^1$ and the reduction
from $M_4\times \C P^1$ to $M_4$.

\bigskip

\section{Noncommutative gauge theory\label{NCgauge}}

To build further on the interpretation of our ansatz in terms of
configurations of D-branes as described in Section~\ref{Ansatzdescr},
we should now proceed to construct explicit solutions of the reduced
Yang-Mills equations on $M_{2n}$. Unfortunately, even solutions of the
vortex equations (\ref{f24})--(\ref{f25}) are difficult to come by and
there is no known general method for explicitly constructing the
appropriate field configurations. As we will demonstrate in the
following, explicit realizations of these D-brane states are
possible in the context of {\it noncommutative} gauge
theory, which can be mapped afterwards onto commutative worldvolume
configurations. For this, we will now specialize the K\"ahler manifold
$M_{2n}\times \C P^1$ to be $\R^{2n}\times \C P^1$ with metric tensor
$g_{\m\n}=\de_{\m\n}$ on $\R^{2n}$ and pass to a noncommutative
deformation of the flat part of the space, i.e. $\R^{2n}\times \C
P^1 \to \R^{2n}_\th\times \C P^1$. Note that the $\C P^1$ factor
remains a commutative space throughout this paper. Then we will deform
the Yang-Mills, DUY and nonabelian coupled vortex
equations, and in the subsequent sections construct various solutions
of them.

\bigskip

\noindent
{\bf Noncommutative deformation.\ }
Field theory on $\R^{2n}_\th$ may be realized in an operator formalism
which turns Schwartz functions~$f$ on $\R^{2n}$ into compact
operators~$\fh$ acting on the $n$-harmonic oscillator Fock
space~$\Hcal$~\cite{Harvey}. The noncommutative space~$\R^{2n}_\th$ is
then defined by declaring its coordinate functions
$\xh^1,\ldots,\xh^{2n}$ to obey the Heisenberg algebra relations
\begin{equation}
[ \xh^\mu\,,\,\xh^\nu\,] = \im\th^{\mu\nu}
\end{equation}
with a constant real antisymmetric tensor~$\th^{\mu\nu}$.
Via an orthogonal transformation of the coordinates, the matrix
$\theta=(\th^{\m\n})$ can be rotated into its canonical block-diagonal
form with non-vanishing components
\begin{equation}\label{tha}
\th^{{2a-1}\ {2a}} \= -\th^{{2a}\ {2a-1}} \ =:\ \th^a
\end{equation}
for $a=1,\dots,n$. We will assume for definiteness that all
$\th^a>0$. The noncommutative version of the complex coordinates
(\ref{zz}) has the non-vanishing commutators
\begin{equation}\label{zzb}
\big[\zh^a\,,\,\zbh^{\bb}\,\big] \= -2\,\de^{a\bb}\,\th^a \
=:\ \th^{a\bb} \= -\th^{\bb a}\ \le\ 0 \ .
\end{equation}
Taking the product of $\R^{2n}_\th$ with the commutative sphere $\C P^1$
means extending the noncommutativity matrix $\th$ by vanishing entries
along the two new directions.

The Fock space~$\Hcal$ may be realized as the linear span
\begin{equation}
\Hcal=\bigoplus_{r_1,\dots,r_n=0}^\infty\,\C|r_1,\ldots,r_n\> \ ,
\label{Fockspace}\eeq
where the orthonormal basis states
\beq
|r_1,\ldots,r_n\>=\prod_{a=1}^{n}\left(2\,
\th^a\,r_a!\right)^{-1/2}\,(\zh^{a})^{r_a}|0,\ldots,0\>
\end{equation}
are connected by the action of creation and annihilation operators
subject to the commutation relations
\begin{equation}
\Bigl[\,\frac{\zbh^{\bb}}{\sqrt{2\,\th^b}}\ ,\ \frac{\zh^a}{\sqrt{2\,\th^a}}\,
\Bigr] = \de^{a\bb} \ .
\end{equation}
In the Weyl operator realization $f\mapsto\fh$, coordinate
derivatives are given by inner derivations of the noncommutative
algebra according to
\begin{equation}\label{pazzf}
\widehat{\pa_{z^a} f}\=\th_{a\bb}\,\big[\zbh^{\bb} \,,\, \fh\,\big]
\ =:\ \pa_{\zh^a} \fh
\qquad\textrm{and}\qquad
\widehat{\pa_{\zb^{\ab}} f}\=\th_{\ab b}\,\big[\zh^b \,,\, \fh\,\big]
\ =:\ \pa_{\zbh^{\,\ab}} \fh\ ,
\end{equation}
where $\th_{a\bb}$ is defined via $\th_{b\bar{c}}\,\th^{\bar{c}a}=\de^a_b$
so that $\th_{a\bb}=-\th_{\bb a}=\frac{\de_{a\bb}}{2\,\th^a}$.
On the other hand, integrals are given by traces over the Fock space
$\Hcal$ as
\begin{equation}\label{intNC}
\int_{\R^{2n}} \diff^{2n} x~f(x)=
\Bigl(\,\prod_{a=1}^n 2\pi\,\th^a \Bigr)~\textrm{Tr}^{~}_\Hcal~\fh\ .
\end{equation}

The transition to the noncommutative Yang-Mills and DUY
equations is trivially achieved by going over to operator-valued objects
everywhere. In particular, vector bundles $E\to\R^{2n}$ whose typical
fibres are complex vector spaces $\underline{V}$ are replaced by the
corresponding (trivial) projective modules $\underline{V}\otimes\Hcal$ over
$\R^{2n}_\theta$. The field strength components along $\rt$
in~(\ref{YM}) and (\ref{DUY1})--(\ref{DUY3}) read
$\hat{\cf}_{\m\n}=\pa_{\xh^{\m}}\hat{\ca}_{\n}-
\pa_{\xh^{\n}}\hat{\ca}_{\m}+[\hat{\ca}_{\m},\hat{\ca}_{\n}]$,
where $\hat{\ca}_{\m}$ are simultaneously ${\rm u}(k)$ and operator valued.
To avoid a cluttered notation, we drop the hats over operators from now on.
Thus all our equations have the same form as previously but are considered
now as operator equations.

\bigskip

\noindent
{\bf Noncommutative coupled vortex equations.\ }
By reducing the noncommutative version of the DUY
equations on $\R^{2n}_\th\times \C P^1$ to $\R^{2n}_\th$ we obtain the
noncommutative nonabelian coupled vortex equations.
Instead of working with the gauge potentials $A^i_\m$
we shall use the operators $X^{i}_\m$ defined by
\begin{equation}\label{X}
X_{a}^{i}\ :=\ A_{a}^{i} + \th_{a\bb}\,\zb^{\bb}
\qquad\textrm{and}\qquad
X_{{\ab}}^{i}\ :=\ A_{{\ab}}^{i} + \th^{~}_{\ab b}\,z^b\ .
\end{equation}
In terms of these operators the field strength tensor reads
\begin{equation}
F_{{a}{\bb}}^{i}\ =\ \big[X_{a}^{i}\,,\,X_{{\bb}}^{i}\,\big] + \th_{a\bb}
\ , ~~ F_{{\ab}{\bb}}^{i}\ =\ \big[X_{{\ab}}^{i}\,,\, X_{\bb}^{i}\,\big]
\quad\textrm{and}\quad F_{{a}{b}}^{i}\ =\ \big[X_{{a}}^{i}\,,\,
X_{b}^{i}\,\big]\ ,
\label{fieldstrengthX}\end{equation}
while the bi-fundamental covariant derivatives become
\beq
D_\ab^{~}\phi^{~}_{i+1}\=X_\ab^i\,\phi^{~}_{i+1}-\phi^{~}_{i+1}
\,X_\ab^{i+1} \quad\mbox{and}\quad D_a^{~}\phi^{~}_{i+1}\=X_a^i\,
\phi^{~}_{i+1}-\phi^{~}_{i+1}\,X_a^{i+1} \ .
\label{bifundX}\eeq

The nonabelian vortex equations (\ref{f24})--(\ref{f25})
can then be rewritten as
\bea\label{ddd1}
\delta^{a\bb}\,\Bigl(\big[X_{a}^{i}\,,\,X_{{\bb}}^{i}\,\big] +
\th_{a\bb}\Bigr)
&=&\frac{1}{4\,R^2}\,\left(m-2i+\phi_i^\+\,\phi^{~}_i -
\phi^{~}_{i+1}\,\phi^\+_{i+1}\right)\ , \\[4pt] \label{ddd}
\big[X_{\ab}^{i}\,,\,X_{{\bb}}^{i}\,\big]&=&0~=~
\big[X_{a}^{i}\,,\,X_{{b}}^{i}\,\big]\ , \\[4pt] \label{ddd2}
X_{\ab}^{i}\,\phi^{~}_{i+1} - \phi^{~}_{i+1}\,X_{{\ab}}^{i+1}&=&0
\eea
for $i=0,1,\dots,m$. Note that for $m=1$ we obtain the equations
\bea\label{f28}
\de^{a\bb}\,F^0_{a{\bb}}&=&\frac{1}{4\,R^2}\,\left(1-\phi^{~}_1
\,\phi_1^\+\,\right)
\quad\mbox{and}\quad\ \ F_{\ab\bb}^0~=~0~=~F_{ab}^0\ , \\[4pt] \label{f29}
\de^{a\bb}\,F_{a{\bb}}^1&=&-\frac{1}{4\,R^2}\,\left(1-\phi_1^\+\,
\phi^{~}_1\right)
\quad\mbox{and}\quad F_{\ab\bb}^1~=~0~=~F_{ab}^1\ , \\[4pt] \label{f30}
\bar\pa^{~}_{\bar a}\phi^{~}_1 + A^0_{\bar a}\,\phi^{~}_1
- \phi^{~}_1\,A^{1}_{\bar a}&=&0
\eea
which are considered in~\cite{IL,LPS}. In particular, for
$n=2$ and $k_0=k_1=1$ the equations (\ref{f28})--(\ref{f30}) coincide
with the perturbed Seiberg-Witten ${\rm U}_+(1)\times{\rm U}_-(1)$
monopole equations on $\R^4_\th$ as considered in~\cite{PSW}.

\bigskip

\section{Explicit solutions of the noncommutative Yang-Mills
  equations\label{NCYMSols}}

We are now ready to construct solutions to the Yang-Mills equations on
$\R_\theta^{2n}\times\C P^1$. We shall first present the generic non-BPS
solutions of the full Yang-Mills equations, and then proceed to solve
the nonabelian coupled vortex equations (\ref{ddd1})--(\ref{ddd2}),
and thus the DUY equations on~$\R_\theta^{2n}\times\C P^1$,
which describe the stable BPS states. Our technique will make use of
appropriate partial isometry operators $T^{~}_{N_i}$ in the
noncommutative space.

\bigskip

\noindent
{\bf Ansatz for explicit solutions. }
Let us fix a monopole charge $m>0$ and an arbitrary integer
$0<r\leq k$. Consider the ansatz
\bea\label{ansatz3}
X_{a}^{i} &=& \th_{a\bb}\,T^{~}_{N_{i}}\,\zb^{\bb}\,T_{N_{i}}^\+
\qquad\mbox{and}\qquad
X_{{\ab}}^{i}~=~\th^{~}_{\ab b}\, T^{~}_{N_{i}}
\,z^{b}\,T_{N_{i}}^\+ \ ,
\\[4pt] \label{ansatz3p}
\p^{~}_{i+1} &=& \a^{~}_{i+1}\,T^{~}_{N_{i}}\,T_{N_{i+1}}^\+
\qquad\mbox{and}\quad
\p^{\+}_{i+1}~=~\bar\a^{~}_{i+1}\,T^{~}_{N_{i+1}}\,T_{N_{i}}^\+
\eea
for $i=0,1,\dots,m$, where $\a_i\in\C$ are some constants with
$\a_0=\a_{m+1}=0$. Denoting by $\Hcal$ the $n$-oscillator Fock space,
the Toeplitz operators
$T^{~}_{N_{i}}:\C^r\otimes\Hcal\to\underline{V}_{\,k_i}\otimes\Hcal$
are partial isometries described by {\it rectangular} $k_i\times r$
matrices (with operator entries acting on $\Hcal$) possessing the
properties
\begin{equation}\label{ansatz4}
T_{N_{i}}^\+\,T^{~}_{N_{i}}\=\Idd_r \qquad\textrm{while}\qquad
T^{~}_{N_{i}}\,T_{N_{i}}^\+\ =\ \Idd_{k_i} - P^{~}_{N_{i}}\ ,
\end{equation}
where $P^{~}_{N_{i}}$ is a hermitean projector of finite rank $N_i$ on
the Fock space $\underline{V}_{\,k_i}\otimes\Hcal$ so that
\beq
P_{N_i}^2\=P^{~}_{N_i}\=P^\dag_{N_i} \qquad\mbox{and}\qquad
\Tr^{~}_{\underline{V}_{\,k_i}\otimes\Hcal}~P^{~}_{N_i}\=N_i \ .
\label{PNiTrace}\eeq
{}From (\ref{ansatz4}) it follows that the operator $T^{~}_{N_i}$ has a
trivial kernel, while the kernel of $T_{N_i}^\dag$ is the
$N_i$-dimensional subspace of $\underline{V}_{\,k_i}\otimes\Hcal$
corresponding to the range of $P^{~}_{N_i}$. Thus
\beq
\dim\ker T^{~}_{N_i}\=0 \qquad\mbox{but}\qquad
\dim\ker T^\dag_{N_i}\=N_i \ .
\label{dimkerTNi}\eeq

Substituted into (\ref{fieldstrengthX}) this ansatz yields the gauge
field strength
\begin{equation}
F_{a{\bb}}^{i} \=  \th_{a\bb}\,P^{~}_{N_{i}}\=\frac1{2\,\theta^a}\,
\delta_{a\bb}\,P^{~}_{N_i} \  \quad\mbox{and}\quad
F_{\ab{\bb}}^{i} \= 0\=F_{ab}^i \ ,
\label{ansatzfieldstrength}\end{equation}
while from (\ref{bifundX}) one finds the covariant derivatives
\beq
D_\ab\phi_{i+1}\=0\=D_a\p_{i+1} \ .
\label{ansatzcovconst}\eeq
Thus our ansatz describes {\it holomorphic} fields, and the projector
$P^{~}_{N_i}$ defines a noncommutative gauge field configuration of
rank $N_i$ and constant curvature in the subspace $\ker
T_{N_i}^\dag\subset\underline{V}_{\,k_i}\otimes\Hcal$. In
particular, the Higgs fields $\phi_{i+1}$ are covariantly constant
with
\beq
\phi^\dag_{i}\,\phi^{~}_{i}\=\left|\a_{i}\right|^2\,
\bigl(\Idd^{~}_{k_{i}}- P^{~}_{N_{i}}\bigr) \quad\mbox{and}\quad
\p^{~}_{i+1}\,\p^{\+}_{i+1}\=\left|\a_{i+1}\right|^{2}\,
\left(\Idd^{~}_{k_i}-P^{~}_{N_{i}}\right) \ .
\label{phiphidags}\eeq

The ranks $N_i$ are generically non-negative integers. If some
$N_i=0$, then we should formally set $P^{~}_{N_i}=0$,
$T_{N_i}^{~}=1$ and $\phi^{~}_{i+1}=\alpha^{~}_{i+1}$ in the $i$-th
component of the ansatz. Then
\beq
X_a^i\=\th_{a\bb}\,\zb^\bb \qquad\mbox{and}\qquad
X^i_\ab\=\th^{~}_{\ab b}\,z^b
\label{XNi0}\eeq
which leads to the vacuum gauge field configuration
\beq
A^i\=0 \qquad\mbox{and}\qquad F^i\=0 \ .
\label{AFNi0}\eeq
These matter fields correspond to open strings with one end on a
D-brane and the other end on the closed string vacuum.

Our ansatz has a natural interpretation in quiver gauge
theory. Consider the module
\beq
\underline{\cal T}~:=~\bigoplus_{i=0}^m\,\ker T_{N_i}^\dag
\qquad\mbox{with}\qquad \vec k^{~}_{\underline{\cal T}}\=
\sum_{i=0}^mN_i~\vec e_i
\label{VNAm}\eeq
over the quiver $\quiverm$, which is a finite-dimensional submodule of
the infinite-dimensional representation $\underline{\cal
  V}\otimes\Hcal$ of $\quiverm$ given by the noncommutative quiver
bundle. Let us fix an integer $0\leq s\leq m$, and take
$N_i\neq0$ for all $i\leq s$ and $N_i=0$ for all $i>s$. The quiver
representation (\ref{VNAm}) is a combination
of the indecomposable projective representations $\pathmod_{\,i}$ of
$\quiverm$ that we encountered in Section~\ref{Dimred}. The
$\pathmod_{\,i}$'s form a complete set of projective representations
in the sense that any quiver representation has a
projective resolution in terms of sums of them~\cite{quiverbooks}. In
particular, the canonical Ringel resolution of (\ref{VNAm}) is given
by the exact sequence
\beq
0~\longrightarrow~\bigoplus_{i=1}^{s}\,\pathmod_{\,i-1}\otimes
\ker T_{N_i}^\dag~\longrightarrow~\bigoplus_{i=0}^s\,
\pathmod_{\,i}\otimes\ker T_{N_i}^\dag~\longrightarrow~
\underline{\cal T}~\longrightarrow~0 \ .
\label{Ringelres}\eeq

\bigskip

\noindent
{\bf Solving the Yang-Mills equations.\ }
We shall now demonstrate that the field configurations
(\ref{ansatz3})--(\ref{ansatz4}) yield solutions of the full Yang-Mills
equations on $\R^{2n}_\theta\times\C P^1$ for any values of $m$,
$N_0,N_1,\dots,N_m$ and $\a_1,\dots,\a_m$. For this, we write the
ansatz in the form
\bea\label{chia}
\ca_a- \th_{a\bb}\,\zb^{\bb} & =&
\sum_{i=0}^m X^i_a\otimes \Pi^{~}_i~=~
\th_{a\bb}\,\sum_{i=0}^m T^{~}_{N_i}\,\zb^{\bb}\,T_{N_i}^\+
\otimes\Pi^{~}_i \ , \\[4pt]
\label{chiab}
\ca_{\ab} - \th_{\ab b}\, z^{b} & =&
\sum_{i=0}^m X^i_{\ab}\otimes \Pi^{~}_i~=~
\th_{{\ab} b}\,\sum_{i=0}^m T^{~}_{N_i}\,z^{b}\,T_{N_i}^{\+}
\otimes\Pi^{~}_i \ .
\eea
We also have
\bea
\ca_y^{ii}&=&\frac{(m-2i)\,\yb}{2\,\left(R^2+y\yb\right)}~\Idd_{k_i}
\ , \label{ayii} \\[4pt]
\ca_\yb^{ii}&=&-\frac{(m-2i)\,y}{2\,\left(R^2+y\yb\right)}~\Idd_{k_i}
\ , \label{aybii} \\[4pt] \label{avt}
{\ca}_{\yb}^{i\, i+1} &=&
\frac{R }{R^2+y{\yb}}\,\phi^{~}_{i+1}~=~\frac{R\ \a^{~}_{i+1}}
{R^2+y{\yb}}\,T^{~}_{N_i}\,T_{N_{i+1}}^\+ \ , \\[4pt] \label{avp}
{\ca}_y^{i+1\, i} & = &-
\frac{R}{R^2+y{\yb}}\,\phi^{\+}_{i+1}~=~
-\frac{R\ \bar\a^{~}_{i+1}}{R^2+y{\yb}}\,T^{~}_{N_{i+1}}\,T_{N_i}^{\+}
\ ,
\eea
with
\beq
{\ca}_{\yb}^{ij}\=0\={\ca}_{y}^{i+1\, j}
\quad\mbox{for}\quad j~\ne~i,i+1 \ .
\label{av0}\eeq
Thus for the ansatz (\ref{ansatz3})--(\ref{ansatz4}) the field
strength tensor is given by
\bea\label{cfabb}
\cf_{a\bb} &=&\th_{a\bb}\, \sum_{i=0}^m P^{~}_{N_i}\otimes
\Pi^{~}_i \ , \\[4pt] \label{cfvtvp}
\cf_{y\yb} &=&- \frac{R^2}{\left(R^2+y\yb\right)^2}\,
\sum_{i=0}^m\Bigl(m-2i+\bigl(|\alpha^{~}_i|^2
-|\alpha^{~}_{i+1}|^2\,\bigr)\,\bigl(\Idd_{k_i}-P^{~}_{N_i}
\bigr)\Bigr)\otimes\Pi^{~}_i \ ,
\eea
with all other components of ${\cf}_{\hat\m\hat\n}$ vanishing.

Let us now insert these expressions into the Yang-Mills equations
(\ref{YM}) (for static configurations with ${\ca}_0=0$).
It is enough to consider the cases $\hat\n =c$ and $\hat\n = \yb$,
since the cases $\hat\n =\bar c$ and $\hat\n = y$ can be obtained by hermitean
conjugation of (\ref{YM}) due to the anti-hermiticity of $\ca_{\hat\m}$ and
$\cf_{\hat\m\hat\n}$. For $\hat\n =c$, eq.~(\ref{YM}) becomes
\begin{equation}\label{rYM1}
\delta^{\cb a}\,\delta^{\bb
c}\,\bigl(\pa_{\cb}\cf_{a\bb}+[\ca_{\cb},{\cf}_{a\bb}]\bigr)=0
\end{equation}
which is equivalent to
\begin{equation}\label{rYM}
\delta^{\cb a}\,\delta^{\bb c}\,\bigl[\ca_{\cb}-\th_{\cb b}\,z^{b}\,,\,
{\cf}_{a\bb}]=0\ .
\end{equation}
Substituting (\ref{chiab}) and (\ref{cfabb}),
we see that (\ref{rYM}) is satisfied due to the identities
(\ref{Piortho}) and
\begin{equation}\label{iden}
T_{N_i}^\+\,P^{~}_{N_i} \= P^{~}_{N_i}\,T^{~}_{N_i}\=0 \ .
\end{equation}
In the case $\hat\n =\yb$, eq.~(\ref{YM}) simplifies to
\begin{equation}\label{rrYM}
\pa_y\bigl(\sqrt{g}~{\cf}^{y\yb}\,\bigr) + \sqrt{g}~\bigl[\ca_y\,,\,
\cf^{y\yb}\,\bigr]=0
\end{equation}
with $\sqrt g=2\,R^4/(R^2+y\yb)^2$. Substituting (\ref{ayii}), (\ref{avp}),
(\ref{av0}) and (\ref{cfvtvp}), we find that (\ref{rrYM}) is also
satisfied due to the identities (\ref{Piortho}) and
(\ref{iden}). Hence, the Yang-Mills equations on
$\R_\theta^{2n}\times\C P^1$ are solved by our choice of ansatz.

\bigskip

\noindent
{\bf Finite-energy solutions.\ }
The arbitrary coefficients $\alpha_i\in\C$ can be fixed (up to a
phase) by demanding that the solution (\ref{ansatz3})--(\ref{ansatz4})
yield finite-energy field configurations. For this, we evaluate the
energy functional (\ref{EF}) using (\ref{intNC}). From
(\ref{ansatzfieldstrength}) we may compute
\beq
\bigl(F_{{\mu}{\nu}}^i\bigr)^\+\,\bigl(F^{i\,{\mu}{\nu}}\bigr)\=
8\,\delta^{a\cb}\,\delta^{d\bb}\,F_{a\bb}^i\,F_{d\cb}^i\=2\,
\Bigl(\,\sum_{a=1}^n\frac1{(\theta^a)^2}\Bigr)~P^{~}_{N_i} \ ,
\label{EFNCfieldstrength}\eeq
and combining this with (\ref{ansatzcovconst}) and (\ref{phiphidags})
we find the noncommutative Yang-Mills energy
\bea
E^{~}_{\rm YM}&=&2\pi\,R^2\,\Bigl(\,\prod_{a=1}^n2\pi\,\theta^a\Bigr)\,
\sum_{i=0}^m\,\Tr^{~}_{\underline{V}_{\,k_i}\otimes\Hcal}\biggl[
\Bigl(\,\sum_{b=1}^n\frac1{(\theta^b)^2}\Bigr)~P^{~}_{N_i}\biggr.
\nonumber\\ && +\biggl.\frac1{4\,R^4}\,\Bigl(m-2i+\bigl(|\alpha^{~}_i|^2
-|\alpha^{~}_{i+1}|^2\,\bigr)\,\bigl(\Idd_{k_i}-P^{~}_{N_i}
\bigr)\Bigr)^2\,\biggr] \ .
\label{EFNCgen}\eea
Because of the trace over the infinite-dimensional Fock space $\Hcal$,
the constant terms in (\ref{EFNCgen}) which are not proportional to
the projectors $P^{~}_{N_i}$ must all vanish in order for the energy
to be finite. This leads to the finite-energy conditions
\begin{equation}\label{cnstr1}
m-2i +|\a_i|^2 - |\a_{i+1}|^2 = 0
\end{equation}
for each $i=0,1,\dots,m$.

With $\a_0=\a_{m+1}=0$, the constraints (\ref{cnstr1}) are solved by
\begin{equation}\label{7.8}
|\a_{i+1}|^2 \= (i+1)\,m - 2\,\sum^i_{j=0}\,j \= (i+1)\,(m-i)
\end{equation}
and the energy (\ref{EFNCgen}) can thereby be written as
\beq
E^{~}_{\rm YM}=2\pi\,R^2\,\Bigl(\,\prod_{a=1}^n2\pi\,\theta^a\Bigr)\,
\sum_{i=0}^{\lfloor\frac m2\rfloor}\,\left(N_i+N_{m-i}\right)\,
\left[\Bigl(\,\sum_{b=1}^n
\frac1{(\theta^b)^2}\Bigr)+\frac{(m-2i)^2}{4\,R^4}\right] \ .
\label{EFNCfinite}\eeq
We have naturally split the sum over nodes $i$ into contributions from
Dirac monopoles and antimonopoles, which for each $i=0,1,\dots,m$ have the same
Yang-Mills energy on the sphere $\C P^1$. Later on we will see that
this splitting corresponds to a $\Z_2$-grading of the chain of
D-branes into brane-antibrane pairs. The monopole independent terms in
(\ref{EFNCfinite}) can be interpreted as the tension of
$\sum_{i=0}^m\,N_i\ $ D0-branes inside a D$(2n)$-brane~\cite{Strom} in
the Seiberg-Witten decoupling limit~\cite{Seiberg}.

\bigskip

\noindent
{\bf BPS solutions.\ }
The solutions we have described generically yield non-BPS solutions
of the full Yang-Mills equations on $\R^{2n}_\theta\times\C P^1$. On
the other hand, the DUY equations on
$\R^{2n}_\theta\times\C P^1$ are BPS conditions for the Yang-Mills
equations. Inserting (\ref{ansatz3})--(\ref{ansatz4}) and
(\ref{ansatzfieldstrength})--(\ref{phiphidags}) into our nonabelian vortex
equations (\ref{ddd1})--(\ref{ddd2}), we find that (\ref{ddd})
and (\ref{ddd2}) are automatically satisfied. The vanishing of the
constant term (not proportional to $P^{~}_{N_i}$) in (\ref{ddd1}) is
precisely the finite-energy constraint (\ref{cnstr1}), whose solution
is given in (\ref{7.8}). Equating the coefficients of $P^{~}_{N_i}$ in
(\ref{ddd1}) for each $N_i\neq0$ leads to the additional constraints
\begin{equation}\label{cnstr2}
\sum^n_{a=1}\,\frac{1}{\th^a}\= \frac{m-2i}{2\,R^2}\qquad
\mbox{with}\qquad i\=0,1,\ldots,s \ .
\end{equation}

For $s>0$ the conditions (\ref{cnstr2}) are incompatible with one
another, implying that the ansatz (\ref{ansatz3})--(\ref{ansatz4})
with $s>0$ does not allow for BPS configurations. For $s=0$, the
equation (\ref{cnstr2}) relates the radius $R$ of the sphere to the
noncommutativity parameters $\th^a$ of $\R^{2n}_\theta$. In this case
we obtain the explicit solutions of the noncommutative vortex and
DUY equations parametrized by the partial isometry
operators $T^{~}_{N_0}$ as
\bea\label{explsl0}
X^0_a&=&{\th}_{a\bb}\,T^{~}_{N_0}\,\zb^{\bb}\, T_{N_0}^\+
\qquad\mbox{and}\qquad \phi^{~}_1\=\alpha^{~}_1\,T^{~}_{N_0} \ ,
\\[4pt] X^i_a&=&\th_{a\bb}\,\zb^{\bb}
\qquad\qquad\quad\ \mbox{and}\qquad\
\phi^{~}_i\=\alpha^{~}_i~\Idd_{k_1} \quad\mbox{for}\quad 0~<~i~\le~m \ .
\label{explsl1}\eea
The BPS conditions (\ref{ddd1})--(\ref{ddd2}) force us to
take $k_1=\dots=k_m$ corresponding to the gauge symmetry breaking
${\rm U}(k)\to{\rm U}(k_0)\times{\rm U}(k_1)^m$, so that $r=k_1$,
$k_0+m\,k_1=k$ with $k_0>0$ and $k_1>0$.
The configurations with $i>0$ correspond to the vacuum
gauge fields (\ref{AFNi0}) with trivial bundle maps $\phi_i$ given as
multiplication by the complex numbers $\alpha_i$ satisfying
(\ref{7.8}). Using (\ref{cnstr2}) and (\ref{EFNCfinite}) we find that
the energies of these BPS states are given by
\beq
E^{~}_{\rm BPS}=2\,(2\pi)^{n+1}\,R^2\,\Bigl(\,\sum_{\stackrel{\scriptstyle
b,c=1}{\scriptstyle b\leq c}}^n~\prod_{\stackrel{\scriptstyle a=1}
{\scriptstyle a\neq b,c}}^n\,\theta^a\Bigr)~N_0 \ .
\label{EBPS}\eeq

These solutions have a natural physical interpretation along the lines
described in Section~\ref{Ansatzdescr}. The original noncommutative
DUY equations are fixed by the positive integers
$n$ and $k$. Our ansatz (\ref{f4})--(\ref{f9}) and
(\ref{ansatz3})--(\ref{ansatz4}) is labelled by the collection of
positive integers $(m, k_i, N_i)$ with $i=0,1,\ldots,s$. According to
the standard identification of D-branes as noncommutative
solitons~\cite{Strom}, the configuration (\ref{explsl0},\ref{explsl1})
with $s=0$ describes a collection of $m\,N_0$ BPS D0-branes as a
stable bound state (i.e. a vortex-like solution on $\R^{2n}_\theta$)
in a system of $k_0+m\,k_1=k\ $  D(2$n$) branes and antibranes. But from the
point of view of the initial branes wrapped on $\R^{2n}_\theta\times\C
P^1$, they are spherical $m\,N_0\ $ D2-branes. This means that
instantons on $\R^{2n}_\theta\times\C P^1$ are the spherical
extensions of vortices which are points in $\rt$. For $s>0$ the
configuration (\ref{ansatz3})--(\ref{ansatz4}) describes an unstable
system of $m\,N_0+|m-2|\,N_1+\ldots+|m-2s|\,N_s\ $ D0-branes
(vortices) in a D(2$n$) brane-antibrane system, because
${\rm deg}\,\Lcal^{m-2i}=m-2i$ for each $i=0,1,\dots,s$. Again they
form a system of spherical D2-branes (i.e. an SU(2)-symmetric
multi-instanton) in the initial brane-antibrane system on
$\R^{2n}_\theta\times\C P^1$. Their orientation depends on the sign of
the magnetic charge $m-2i$ for each $i=0,1,\ldots,s$, which determines
whether we have D2-branes or D2-antibranes. If more than one $N_i\ne
0$ then the ansatz either describes pairs of D0-branes with overall
non-vanishing monopole charges, or both D0-branes and
anti-D0-branes. Such systems cannot be stable, i.e. the corresponding
configuration (\ref{ansatz3})--(\ref{ansatz4}) cannot satisfy the
noncommutative vortex and DUY equations.

The distinction between BPS versus non-BPS solutions is very
natural in quiver gauge theory. The BPS configurations are described
by the simple Schur representations $\simple_{\,i}$, $i=0,1,\dots,m$ of the
${\rm A}_{m+1}$ quiver given by a one-dimensional vector space at vertex
$i$ with all maps equal to~$0$, i.e. the $\quiverm$-module with
$(\,\simple_{\,i})_{m-2j}=\delta_{ij}~\C$ and dimension vector $\vec
k_{\simple_{\,i}}^{~}=\vec e_i$. The BPS states constructed above then
correspond to the quiver representations $(\,\simple_{\,0})^{\oplus
  N_0}$. Together with the projective modules $\pathmod_{\,i}$, the
Schur modules $\simple_{\,i}$ admit the projective resolutions
\bea
0~\longrightarrow~\pathmod_{\,0}~\longrightarrow~\simple_{\,0}&
\longrightarrow&0 \ , \label{PLres0} \\[4pt] 0~\longrightarrow~
\pathmod_{\,i-1}~\longrightarrow~\pathmod_{\,i}~\longrightarrow~
\simple_{\,i}&\longrightarrow&0 \qquad\mbox{for}\qquad i\=1,\dots,s
\label{PLresi}\eea
and satisfy the relations~\cite{quiverbooks}
\beq
{\rm Hom}(\,\simple_{\,i}\,,\,\simple_{\,j})\=\delta_{ij}~\C\=
{\rm Hom}(\,\pathmod_{\,i}\,,\,\simple_{\,j}) \ .
\label{PLHomrels}\eeq

The resolutions (\ref{Ringelres}) and (\ref{PLres0},\ref{PLresi}) exhibit
a sharp homological distinction between BPS and non-BPS solutions. The
constituent D-branes at the vertices of the quiver $\quiverm$ are
associated with the basic representations
$\simple_{\,i}$. Sums $(\,\simple_{\,i})^{\oplus N_i}$ for fixed $i$
correspond to BPS states, associated generally with the symmetry breaking ${\rm
  U}(k)\to{\rm U}(k_i)\times{\rm U}(k_{i+1})^m$, which are constructed
analogously to (\ref{explsl0},\ref{explsl1}) but with the vacuum Higgs
configurations $\phi_j=\alpha_j~\Idd_{k_i}$ for $j<i$ and
$\phi_j=\alpha_j~\Idd_{k_{i+1}}$ for $j>i$. A generic non-BPS state,
associated to the quiver representation (\ref{VNAm}), corresponds to
the decay of the original $\su$-symmetric branes wrapped on
$\R^{2n}_\theta\times\C P^1$ into the collection of constituent branes
$(\,\simple_{\,0})^{\oplus N_0}\oplus(\,\simple_{\,1})^{\oplus
  N_1}\oplus\cdots\oplus(\,\simple_{\,s})^{\oplus N_s}$ in $\R^{2n}_\theta$.
For $s>0$
this collection is unstable. In the quiver gauge theory, we have
thereby arrived at a natural construction of the unstable D-brane
configurations in terms of stable BPS states of D-branes, which may be
succinctly summarized through the sequence of distinguished triangles
of quiver representations
\beq
\begin{matrix}(\,\simple_{\,0})^{\oplus N_0}~=&\underline{\cal T}_{\,0}&
\longrightarrow&\underline{\cal T}_{\,1}&\longrightarrow~\cdots~
\longrightarrow&\underline{\cal T}_{\,m-1}&\longrightarrow&
\underline{\cal T}_{\,m}&=~\underline{\cal T}\\[5pt]
&~~~\nwarrow& &\!\!\!\!\!\!\!\!\!\!\swarrow& &~~~~\nwarrow& &
\!\!\!\!\!\!\!\swarrow& \\[4pt]
& &(\,\simple_{\,1})^{\oplus N_1}& &\cdots&
&(\,\simple_{\,m})^{\oplus N_m}& & \end{matrix}
\label{BPStriangles}\eeq
where $\underline{\cal T}_{\,s}:=\bigoplus_{i=0}^s\,\ker
T_{N_i}^\dag=\ker T_{N_s}^\dag\oplus\underline{\cal T}_{\,s-1}$
and the horizontal maps are the canonical inclusions of
submodules. This exact sequence expresses the fact that, for each
$s=1,\dots,m$, the non-BPS module $\underline{\cal T}_{\,s}$ is an
extension of the BPS module $(\,\simple_{\,s})^{\oplus N_s}$ by the
non-BPS module $\underline{\cal T}_{\,s-1}$.

\bigskip

\section{Generalized Atiyah-Bott-Shapiro construction\label{Explreal}}

In this section we shall construct an explicit realization of the basic
partial isometry operators $T_{N_i}^{~}$ which will be particularly
useful for putting the D-brane interpretation of our noncommutative
multi-instanton solutions on firmer ground. It is based
on an $\su$-equivariant generalization of the (noncommutative)
Atiyah-Bott-Shapiro (ABS) construction of tachyon field
configurations~\cite{HM1}--\cite{OS1}.

\bigskip

\noindent
{\bf Equivariant ABS construction.\ }
If $G$ is a group and $\cliff_{2n}:=\cliff(\R^{2n})$, we denote by
$\rep_{\spin_G(2n)}$ the Grothendieck group of isomorphism
classes of finite-dimensional $\Z_2$-graded $G\times\cliff_{2n}$
modules, i.e. Clifford modules possessing an even ($\Z_2$-degree
preserving) $G$-action which commutes with the
$\cliff_{2n}$-action. More precisely, we consider representations
of $\C[G]\otimes\cliff_{2n}$ with $\C[G]$ the group ring of
$G$. The inclusion $\imath(2n):\cliff_{2n}\hookrightarrow\cliff_{2n+1}$ of
Clifford algebras induces a restriction map
\beq
\imath_G(2n)^*\,:\,\rep_{\spin_G(2n+1)}~\longrightarrow~\rep_{\spin_G(2n)}
\label{restrmapG}\eeq
on equivariant Clifford modules. Following the standard ABS
construction~\cite{ABS1}, we may then obtain the $G$-equivariant K-theory
$\K_G(\R^{2n})$ (with compact support) through the descendent
isomorphism
\beq
\K_G\bigl(\R^{2n}\bigr)~=~{\rm coker}\,\imath_G(2n)^*~=~\rep_{\spin_G(2n)}\,/\,
\imath_G(2n)^*\rep_{\spin_G(2n+1)} \ .
\label{descisoG}\eeq
The image of $\imath_G(2n)^*$ in $\rep_{\spin_G(2n)}$ contains classes
of Clifford modules $[\,\underline{V}\,]$ which admit a
$G\times\cliff_{2n}$-equivariant involution
$\underline{V}\cong\underline{V}^\vee$, where $\underline{V}^\vee$ is the
Clifford module $\underline{V}$ with its $\Z_2$-parity reversed.

In our case, we take $G=\uo\subset\su$ acting trivially on $\R^{2n}$, and
thereby consider $\uo\times\cliff_{2n}$-modules with the $\uo$-action
commuting with the Clifford action. Any such module is a direct sum
of tensor products of a $\uo$-module and a spinor module, and hence
\beq
\rep_{\spin_\uo(2n)}\=\rep_{\spin(2n)}\otimes\rep_\uo
\qquad\textrm{and}\qquad \imath_\uo(2n)^*\=\imath(2n)^*\otimes\Idd  \ .
\label{spinortrivdecomp}\eeq
Since from the standard ABS construction one has the
isomorphism~\cite{ABS1}
\beq
\K\bigl(\R^{2n}\bigr)~=~{\rm coker}\,\imath(2n)^*~=~\rep_{\spin(2n)}\,/\,
\imath(2n)^*\rep_{\spin(2n+1)}
\label{usualABSiso}\eeq
of abelian groups, we can reduce (\ref{descisoG}) for $G=\uo$ to the
isomorphism
\beq
\K_\uo\bigl(\R^{2n}\bigr)=\K\bigl(\R^{2n}\bigr)\otimes\rep_\uo
\label{eqABSiso}\eeq
of $\rep_\uo$-modules, where $\K(\R^{2n})\cong\Z$ (Note that the
isomorphism $\K_\uo(\R^{2n})\cong\R_\uo$ also follows from the fact
that $\R^{2n}$ is equivariantly contractible to a point). We may
describe the isomorphism (\ref{eqABSiso}) along the lines explained in
Section~\ref{Ansatzdescr}. In particular, the spinor module
$\underline{\Delta}_{\,2n}:=\underline{\Delta}\,(\R^{2n})$ admits the
isotopical decomposition
\beq
\underline{\Delta}_{\,2n}\=\bigoplus_{i=0}^m\,\Delta_i\otimes
\underline{S}_{\,m-2i} \qquad\mbox{with}\qquad \Delta_i\={\rm
  Hom}^{~}_{\uo}(\,\underline{S}_{\,m-2i}\,,\,\underline{\Delta}_{\,2n})
\label{spinmoddecomp}\eeq
obtained by restricting $\underline{\Delta}_{\,2n}$ to representations
of $\uo\subset\spin(2n)\subset\cliff_{2n}$. The $\Delta_i$'s in
(\ref{spinmoddecomp}) are the corresponding multiplicity spaces.

The most instructive and useful way to explicitly realize the decomposition
(\ref{spinmoddecomp}) is to use the equivariant excision theorem
(\ref{excisionslc}) directly and consider the $\su$-invariant dimensional
reduction of spinors from $\R^{2n}\times\C P^1$ to $\R^{2n}$. For this,
we introduce the twisted Dirac operator on $\R^{2n}\times\C P^1$ using
the graded connection formalism of Section~\ref{Dimred} to write the
$\Z_{m+1}$-graded Clifford connection
\beq
\hat\Dirac~:=~\Gamma^{\hat\mu}\,D_{\hat\mu}
{}~=~\gamma^\mu\,D_\mu\otimes\Idd_2+\bigl(\mphi\bigr)\,\gamma\otimes
\gamma^\yb\,\beta_\yb-\bigl(\mphi\bigr)^\dag\,\gamma\otimes\gamma^y\,
\beta_y+\gamma\otimes\Dirac_{\C P^1} \ ,
\label{Diracgradeddef}\eeq
where
\beq
\Dirac_{\C P^1}~:=~\gamma^y\,D_y+\gamma^\yb\,D_\yb~=~
\gamma^y\left(\partial_y+\omega_y+\bigl(\ma\bigr)_y\right)+
\gamma^\yb\left(\partial_\yb+\omega_\yb+\bigl(\ma\bigr)_\yb\right)
\label{DiracS2def}\eeq
and $\omega_y,\omega_\yb$ are the components of the Levi-Civita spin
connection on the tangent bundle of $\C P^1$. From
(\ref{Diracgradeddef}) we see that the monopole charges $m-2i$ in the
Yang-Mills energy functional (\ref{EF}) can be understood as
originating from the Dirac operator (\ref{DiracS2def}) on $\C
P^1$. The operator (\ref{Diracgradeddef}) acts on spinors $\Psi$
which are sections of the bundle
\beq
\Psi\=\begin{pmatrix}\Psi^+\\\Psi^-\end{pmatrix}
{}~\in~\bigoplus_{i=0}^m\,\left(E_{k_i}\otimes
\underline{\Delta}_{\,2n}\right)\otimes\begin{pmatrix}\Lcal^{m-2i+1}\\
\Lcal^{m-2i-1}\end{pmatrix}
\label{spinortotgen}\eeq
over $\R^{2n}\times\C P^1$, where $\Lcal^{m-2i+1}\oplus\Lcal^{m-2i-1}$
are the twisted spinor bundles of rank~$2$ over the sphere $\C
P^1$. We are therefore interested in the twisted spinor module
$\underline{\Delta}_{\,\underline{\cal V}}(\R^{2n}\times\C P^1)$ over
the Clifford algebra $\cliff(\R^{2n}\times\C P^1)$ which is the
product of the spinor module
$\underline{\Delta}_{\,2n}\otimes\underline{\Delta}\,(\C P^1)$ with
the fundamental representation (\ref{genrepSU2Uk}) of the gauge group
${\rm U}(k)$ broken as in (\ref{gaugebroken}).

The symmetric fermions on $\R^{2n}$ that we are interested in
correspond to $\su$-invariant spinors on $\R^{2n}\times\C P^1$. They
belong to the kernel of the Dirac operator (\ref{DiracS2def}) on $\C
P^1$ and will be massless on $\R^{2n}$. One can write
\beq
\Dirac_{\C P^1}\=\bigoplus_{i=0}^m\,\Dirac_{m-2i}\=
\bigoplus_{i=0}^m\,\begin{pmatrix}0&\Dirac^-_{m-2i}\\\Dirac^+_{m-2i}&0
\end{pmatrix} \ ,
\label{DiracS2decomp}\eeq
where
\bea
\Dirac^+_{m-2i}&=&\frac1{R^2}\,\left[\left(R^2+y\yb\right)\,
\partial_\yb-\mbox{$\frac12$}\,(m-2i+1)\,y\right] \ , \label{DiracS2p}
\\[4pt] \Dirac^-_{m-2i}&=&-\frac1{R^2}\,\left[\left(R^2+y\yb\right)\,
\partial_y+\mbox{$\frac12$}\,(m-2i-1)\,\yb\right] \ .
\label{DiracS2m}\eea
The operator (\ref{DiracS2decomp}) acts on sections of the bundle
(\ref{spinortotgen}) which we write with respect to this
decomposition as
\beq
\Psi=\bigoplus_{i=0}^m\,\begin{pmatrix}\psi_{(m-2i)}^+\\
\psi_{(m-2i)}^-\end{pmatrix} \ ,
\label{Psidecomp}\eeq
where $\psi^\pm_{(m-2i)}$ are sections of $\Lcal^{m-2i\pm1}$ taking
values in $\underline{\Delta}_{\,2n}\otimes\underline{V}_{\,k_i}$ with
coefficients depending on~$x\in\R^{2n}$.

To describe the kernel of the Dirac operator (\ref{DiracS2decomp}),
we need to solve the differential equations
\beq
\Dirac^+_{m-2i}\psi^+_{(m-2i)}\=0 \qquad\mbox{and}\qquad
\Dirac^-_{m-2i}\psi^-_{(m-2i)}\=0
\label{Diracpkernel}\eeq
for the positive and negative chirality spinors $\psi^+_{(m-2i)}$ and
$\psi^-_{(m-2i)}$ in $\ker\Dirac^+_{m-2i}$ and
$\ker\Dirac^-_{m-2i}$. By recalling the form of the transition functions
for the monopole bundles from Section~\ref{Invgauge}, one easily sees
that the only solutions of these equations which are regular on both
the northern and southern hemispheres of $S^2$ are of the form
\beq
\psi_{(m-2i)}^+\=\frac1{\left(R^2+y\yb\right)^{t_i/2}}\,
\sum_{\ell=0}^{t_i}\psi^+_{(m-2i)\,\ell}(x)~y^\ell \quad\mbox{and}\quad
\psi_{(m-2i)}^-\=0 \quad\mbox{for}\quad m-2i~<~0
\label{solschargeneg}\eeq
and
\beq
\psi_{(m-2i)}^-\=\frac1{\left(R^2+y\yb\right)^{t_i/2}}\,
\sum_{\ell=0}^{t_i}\psi^-_{(m-2i)\,\ell}(x)~\yb^\ell \quad\mbox{and}\quad
\psi_{(m-2i)}^+\=0 \quad\mbox{for}\quad m-2i~>~0 \ .
\label{solschargepos}\eeq
Here $t_i=|m-2i|-1$ and the component functions
$\psi^\pm_{(m-2i)\,\ell}(x)$ on $\R^{2n}$ with $\ell=0,1,\dots,t_i$
form the irreducible representation
$\underline{V}_{\,t_i+1}\cong\C^{|m-2i|}$ of the group $\su$. Thus the
chirality grading is by the sign of the magnetic charges.

This analysis is valid when the monopole charge $m$ is an even or odd
integer. However, when $m$ is even there is precisely one
term in (\ref{spinortotgen}) with $m=2i$ for which the sub-bundle
$E_{k_{\frac m2}}\to\R^{2n}$ is twisted by the ordinary spinor bundle
$\Lcal\oplus\Lcal^\vee\to\C P^1$ of vanishing magnetic charge. This
bundle admits an infinite-dimensional vector space of symmetric ${\rm
  L}^2$-sections comprised of spinor harmonics $\Psi_{lq}\in\C^2$ with
$l\in\N_0+\frac12$, $q\in\{-l,-l+1,\dots,l-1,l\}$ and
$\Dirac_0\Psi_{lq}\neq0$~\cite{GMNRS1}. The spectrum of the
(untwisted) Dirac operator $\Dirac_0$ consists of the eigenvalues
$\pm\,(l+\frac12)$, each of even multiplicity $p+1=2l+1$. After
dimensional reduction, this produces an infinite tower of massive
spinors on $\R^{2n}$, and such fermions of zero magnetic charge have
no immediate interpretation in the present context. However, one has
$\dim\ker\Dirac_0=0$, and this will be enough for our purposes. We
will therefore fix one of these vector spaces, such that after
integration over $\C P^1$ it corresponds to the space
\beq
\underline{H}_{\,p}~\cong~\C^2\otimes\C^{p+1} \qquad\mbox{with}\qquad
p\=1,3,5,\dots \ .
\label{spinorharm}\eeq
All of our subsequent results will be independent of the particular
choice of eigenspace (\ref{spinorharm}).

We have thereby shown that the $\su$-equivariant reduction of the
twisted spinor representation of $\cliff(\R^{2n}\times\C P^1)$
decomposes as a $\Z_2$-graded bundle giving
\beq
\underline{\Delta}_{\,\underline{\cal V}}\bigl(\R^{2n}\times\C P^1
\bigr)^{\su}\=\underline{\Delta}_{\,2n}\otimes\bigl(\,\underline{
\Delta}_{\,\underline{\cal V}}^+~\oplus~\underline{\Delta}_{\,
\underline{\cal V}}^-\,\bigr) \qquad\mbox{for}~~m~~\mbox{odd} \ ,
\label{twistedspingrad}\eeq
where
\beq
\underline{\Delta}_{\,\underline{\cal V}}^+\=\bigoplus_{i=m_+}^m\,
\underline{V}_{\,k_i}\otimes\underline{V}_{\,|m-2i|}
\qquad\mbox{and}\qquad \underline{\Delta}_{\,\underline{\cal V}}^-\=
\bigoplus_{i=0}^{m_-}\,
\underline{V}_{\,k_i}\otimes\underline{V}_{\,m-2i}
\label{twistedspinpm}\eeq
with $m_+=\lfloor\frac{m+1}2\rfloor$ and
$m_-=\lfloor\frac{m-1}2\rfloor$ . When $m$ is an even integer, one
should also couple the eigenspace (\ref{spinorharm}) giving
\beq
\underline{\Delta}_{\,\underline{\cal V}}\bigl(\R^{2n}\times\C P^1
\bigr)^{\su}\=\underline{\Delta}_{\,2n}\otimes\left(\,\underline{
\Delta}_{\,\underline{\cal V}}^+~\oplus~\bigl(\,\underline{V}_{\,k_{\frac
m2}}\otimes\underline{H}_{\,p}\,\bigr)~\oplus~\underline{\Delta}_{\,
\underline{\cal V}}^-\,\right) \qquad\mbox{for}~~m~~\mbox{even}
\label{twistedspingradeven}\eeq
with $m_+=\lfloor\frac{m+3}2\rfloor$ and
$m_-=\lfloor\frac{m-1}2\rfloor$. It remains to work out the
corresponding action of Clifford multiplication
\beq
\mu^{~}_{\underline{\cal V}}\,:\,
\underline{\Delta}_{\,\underline{\cal V}}^-~
\longrightarrow~\underline{\Delta}_{\,\underline{\cal V}}^+ \ .
\label{CliffmultV}\eeq
For this, we recall from Section~\ref{Ansatzdescr} that the action of
the generators of the parabolic subgroup $\Pt\subset\slc$ on the
equivariant decomposition (\ref{twistedspingrad},\ref{twistedspinpm})
is given by $\sigma_3(\,\underline{V}_{\,k_i}\otimes\underline{V}_{\,|m-2i|})
=(m-2i)\,(\,\underline{V}_{\,k_i}\otimes\underline{V}_{\,|m-2i|})$ and
$\sigma_-:\underline{V}_{\,k_i}\otimes\underline{V}_{\,|m-2i|}
\to\underline{V}_{\,k_{i-1}}\otimes\underline{V}_{\,|m-2i|}$. Since the
Clifford action is required to commute with this action, the map
(\ref{CliffmultV}) is thereby uniquely fixed on the isotopical
components in the form
\beq
\mu^{~}_{\underline{\cal V}}\circ\Pi^{~}_i\,:\,\underline{V}_{\,k_i}\otimes
\underline{V}_{\,|m-2i|}~\longrightarrow~
\underline{V}_{\,k_{i+\lfloor\frac m2\rfloor+1}}
\otimes\underline{V}_{\,|m-2i|} \qquad\mbox{for}\qquad
i\=0,1,\dots,m_- \ .
\label{Cliffmultisot}\eeq
Furthermore, since $\sigma_3(\,\underline{H}_{\,p})=0$ for all $p$, the
space of spinor harmonics must lie in the kernel of the Clifford map
and one has
\beq
\mu^{~}_{\underline{\cal V}}\circ\Pi_{\frac m2}\=0  \qquad\mbox{for}~~m~~
\mbox{even} \ .
\label{Cliffordmulteven}\eeq

It is also illuminating to formulate this equivariant dimensional reduction
from a dynamical point of view, as we did for the gauge fields in
Section~\ref{Dimred}. Using the gauged Dirac operator
(\ref{Diracgradeddef}) we may define a fermionic energy functional on
the space of sections of the bundle (\ref{spinortotgen}) by
\beq
E^{~}_{\rm D}:=\int_{\R^{2n}\times\C P^1}\diff^{2n+2}x~\sqrt g~
\Psi^\dag\,\hat\Dirac\Psi \ .
\label{EDdef}\eeq
One has
\beq
\Psi^\dag\,\left(\gamma\,\bigl(\mphi\bigr)\otimes\sigma^\yb-\gamma\,
\bigl(\mphi\bigr)^\dag\otimes\sigma^y\right)\Psi=\begin{pmatrix}
\bigl(\Psi^+\bigr)^\dag&\bigl(\Psi^-\bigr)^\dag\end{pmatrix}\,
\begin{pmatrix}\gamma\,\bigl(\mphi\bigr)^\dag\,\bigl(\Psi^-\bigr) \\
\gamma\,\bigl(\mphi\bigr)\,\bigl(\Psi^+\bigr)\end{pmatrix} \ .
\label{PsiPhiid}\eeq
Substituting (\ref{DiracS2decomp})--(\ref{solschargepos}), we see that
(\ref{PsiPhiid}) vanishes on symmetric spinors and after integration
over $\C P^1$ the energy functional (\ref{EDdef}) for $m$ odd becomes
\bea
E^{~}_{\rm D}&=&4\pi\,R^2\,\int_{\R^{2n}}\diff^{2n}x~\Biggl[~
\sum_{i=m_+}^m~\sum_{\ell=0}^{|m-2i|-1}~\bigl(\psi^+_{(m-2i)\,\ell}
\bigr)^\dag\,\gamma^\mu\,D_\mu\bigl(\psi^+_{(m-2i)\,\ell}\bigr)
\Biggr.\nonumber\\ && \qquad\qquad\qquad\qquad\quad
+\left.\sum_{i=0}^{m_-}~
\sum_{\ell=0}^{m-2i-1}~\bigl(\psi^-_{(m-2i)\,\ell}
\bigr)^\dag\,\gamma^\mu\,D_\mu\bigl(\psi^-_{(m-2i)\,\ell}\bigr)
\right] \ .
\label{EFred}\eea
The symmetric fermion energy functional for $m$ even also contains
mass terms for fermions of vanishing magnetic charge which are
proportional to the multiplicity $(p+1)$ of the spinor harmonics.

\bigskip

\noindent
{\bf Explicit form of the operators $\mbf{T^{~}_{N_{i}}}$.\ }
The operators  $T^{~}_{N_{i}}$ parametrizing the solutions of the previous
section may be realized explicitly by appealing to a noncommutative
version of the above construction. For this, we first note that the
(trivial) action of $\uo\subset\su$ on $\R^{2n}$ induces an action on
functions $f$ on $\R^{2n}$ by $(\zeta\cdot f)(x):=f(\zeta^{-1}\cdot
x)$ for $\zeta\in\uo$. This in turn defines an action of $\uo$ on the
noncommutative space $\R_\theta^{2n}$ through automorphisms $\hat
  f\mapsto\widehat{\zeta\cdot f}$ of the Weyl operator algebra, i.e. a
  representation of $\uo$ in the automorphism group of the algebra. We
  will assume that the Fock space (\ref{Fockspace}) carries a unitary
  representation of $\uo$. We can then decompose it into its
  isotopical components in the usual way as
\beq
\Hcal=\bigoplus_{i=0}^m\,\Hcal_i\otimes\underline{S}_{\,m-2i} \ .
\label{Fockisotop}\eeq
For $\zeta\in\uo$ we denote the corresponding unitary operator
on $\Hcal$ by $\hat\zeta$. If we demand that the representations of
$\uo$ above are covariant with respect to each other~\cite{Mart1},
\beq
\hat\zeta\,\hat f\,\hat\zeta^{-1}=\widehat{\zeta\cdot f} \ ,
\label{covariantreps}\eeq
then they define a representation of the crossed-product of the
algebra of Weyl operators with the group $\uo$. This defines the
(trivial) noncommutative $\uo$-space $\R^{2n}_\theta\rtimes\uo$, and
equivariant field configurations are operators belonging to the
commutant of $\uo$ in the crossed-product algebra. In quiver gauge
theory, the pertinent representation of $\quiverm$ thus labels
isotopical components of the Hilbert space of the noncommutative gauge
theory. Since the $\uo$-action is trivial here, the isotopical
components of the Fock space (\ref{Fockisotop}) are given by
$\Hcal_i\cong\Hcal$ for each $i=0,1,\dots,m$. Note that one has an
isomorphism $(\Hcal)^{\oplus(m+1)}\cong\Hcal$ by the usual Hilbert
hotel argument.

We will now construct a representation on (\ref{Fockisotop}) of the
partial isometry operators $T^{~}_{N_i}$ in
$\R^{2n}_\theta\rtimes\uo$. For this, let us put $r:=2^{n-1}$ and
consider the operators~\cite{HM1}
\begin{equation}\label{TT}
\Sigma\=(\sigma\cdot x)^\+\,\frac{1}{\sqrt{(\sigma\cdot x)\,
(\sigma\cdot x)^\+}}
\quad \mbox{and}\quad
\Sigma^\+\=\frac{1}{\sqrt{(\sigma\cdot x)\,(\sigma\cdot x)^\+}}\,
(\sigma\cdot x) \ ,
\end{equation}
where $\sigma\cdot x := \sigma_{\mu}\,x^{\mu}$, $\m, \n=
1,\ldots,2n$ and the $r\times r$ matrices
${\sigma}_{\mu}$ are subject to the anticommutation relations
\begin{equation}
{\sigma_{\mu}}^\+\,\sigma_{\nu} + {\sigma_{\nu}}^\+\,\sigma_{\mu} \=
2\,\de_{\mu\nu}~\Idd_r \=
\sigma_{\mu}\,{\sigma_{\nu}}^\+ + \sigma_{\nu}\,{\sigma_{\mu}}^\+ \ .
\label{sigmaanti}\end{equation}
Eq.~(\ref{sigmaanti}) implies that the matrices
\begin{equation}\label{G}
\gamma_{\mu}\ = \ \begin{pmatrix}0&{{\sigma_{\mu}}^\+}\,\\
-\sigma_{\mu} & 0\end{pmatrix}
\qquad\textrm{with}\qquad
\gamma_{\mu}\,\gamma_{\nu} + \gamma_{\nu}\,\gamma_{\mu} \=
-2\,\de_{\mu\nu}~\Idd_{2r}
\end{equation}
generate the Clifford algebra $\cliff_{2n}$. Note
that for $n=1$ we have $r=1$, $\sigma_1=1$ and $\sigma_2=\im$, which
yields
\begin{equation}
\Sigma^\+\=\frac{1}{\sqrt{\zb^1 z^1}}\,\zb^1 \=
\sum_{\ell =1}^\infty |\ell{-}1\>\<\ell|
\end{equation}
and we obtain the standard shift operator $(\Sigma)^N$ on the Fock
space $\Hcal$ in this case. Generally, the operators (\ref{TT}) obey
\begin{equation}\label{6}
\bigl(\Sigma^{\+}\bigr)^{N_i}\,\bigl(\Sigma\bigr)^{N_i}\=\Idd_{r}^{~}
\quad\mbox{and}\quad
\bigl(\Sigma\bigr)^{N_i}\,\bigl(\Sigma^{\+}\bigr)^{N_i}\= \Idd^{~}_{r} -
{\cal P}^{~}_{N_i} \ ,
\end{equation}
where ${\cal P}^{~}_{N_i}$ is a projector of rank $N_i$ on the vector
space $\underline{\Delta}^+_{\,2n}\otimes\Hcal$, and
$\underline{\Delta}^\pm_{\,2n}\cong\C^r$ are the irreducible chiral
spinor modules of dimension $r=2^{n-1}$ (with
$\underline{\Delta}^{~}_{\,2n}=\underline{\Delta}^+_{\,2n}\oplus
\underline{\Delta}^-_{\,2n}$) on which the matrices $\sigma_\mu$ act.

The partial isometry operators $(\Sigma)^{N_i}$ in $\R_\theta^{2n}$ do
not act on the isotopical decomposition
(\ref{Fockisotop}) and thus do not properly incorporate the
$\su$-equivariant reduction of the original system of D-branes. The desired
operators $T_{N_i}^{~}$ in $\R_\theta^{2n}\rtimes\uo$ are obtained by
first projecting these partial isometries onto constituent brane
subspaces. With $\Pi_i$ the rank~$1$ projector onto the $i$-th
isotopical component in (\ref{Fockisotop}), we thereby define the
$r\times r$ matrices
\begin{equation}\label{8}
T^{(0)}_{N_i}=\Idd^{~}_{r}\otimes (1- \Pi^{~}_i) +
(\Sigma)^{N_i}\otimes\Pi^{~}_i  \ .
\end{equation}
The operator $T_{N_i}^{(0)}$ acts as the shift operator $(\Sigma)^{N_i}$
on $\Hcal_i$ and as the identity operator $\Idd_r$ on $\Hcal_j$ for
all $j\neq i$. It is easy to see that these matrices satisfy the
equations
\beq
\bigl(T^{(0)}_{N_i}\bigr)^\dag\,\bigl(T^{(0)}_{N_i}\bigr)\=\Idd_r
\quad\mbox{and}\quad \bigl(T^{(0)}_{N_i}\bigr)\,\bigl(T^{(0)}_{N_i}
\bigr)^\dag\=\Idd_r-P_{N_i}^{(0)}
\label{TNi0eqs}\eeq
with
\begin{equation}\label{9}
P^{(0)}_{N_i}={\cal P}^{~}_{N_i}\otimes\Pi^{~}_i
\end{equation}
a projector of rank $N_i$ on the Fock space
$\underline{\Delta}^+_{\,2n}\otimes\Hcal$. They also satisfy the
algebra
\beq
\bigl(T^{(0)}_{N_i}\bigr)^N\=T^{(0)}_{N_i\,N} \qquad\mbox{and}\qquad
T^{(0)}_{N_i}\,T_{N_j}^{(0)}\=T^{(0)}_{N_i}+T_{N_j}^{(0)}-\Idd^{~}_r\=
T^{(0)}_{N_j}\,T_{N_i}^{(0)} \quad\mbox{for}\quad i~\neq~j  \ .
\label{TNiTNj}\eeq
The operator (\ref{8}) may be regarded as a linear map
\begin{equation}
T^{(0)}_{N_i}\,:\
\underline{\Delta}_{\,2n}^-\otimes\Hcal~\longrightarrow~
\underline{\Delta}_{\,2n}^+\otimes\Hcal \ .
\label{TTop}\end{equation}
In particular, the map $(T^{(0)}_1)$ has a trivial kernel, while
$(T^{(0)}_1)^\+$ has a one-dimensional kernel which is spanned by the
vector $|\psi\>\otimes|0,\ldots,0\>$ where $|\psi\>$ denotes the
lowest-weight spinor of SO($2n$).

Finally, the desired {\it rectangular} $k_i\times r$ Toeplitz operators
$T^{~}_{N_i}$ may be realized in terms of the partial isometries
(\ref{8}) by appealing to the Hilbert hotel argument. For this, we
introduce a lexicographic ordering $\N_0^n\sim\N^{~}_0$ on the Fock space
$\Hcal$ so that $|r_1,\dots,r_n\rangle=|q\rangle$ with
$q=0,1,2,\dots$, and fix an orthonormal basis
$\vec\rho^{\,}_0,\vec\rho^{\,}_1,\dots,\vec\rho^{\,}_{r-1}$ of the chiral
spinor representation space
$\underline{\Delta}_{\,2n}^+\cong\C^r$. Then
$\vec\rho^{\,}_{a}\otimes|q\rangle$, $a=0,1,\dots,r-1$ is an orthonormal
basis for $\underline{\Delta}_{\,2n}^+\otimes\Hcal$ and there is a
one-to-one correspondence
$\vec\rho^{\,}_{a}\otimes|q\rangle\leftrightarrow|r\,
q+a\rangle$ of basis states. Similarly, by fixing an
orthonormal basis $\vec\lambda_0^{\,i},\vec\lambda_1^{\,i},\dots,
\vec\lambda_{k_i-1}^{\,i}$ of the $\su$ representation space
$\underline{V}_{\,k_i}\cong\C^{k_i}$, there is a one-to-one
correspondence
$\vec\lambda_{a_i}^{\,i}\otimes|q_i\rangle\leftrightarrow
|k_i\,q_i+a_i\rangle$, $a_i=0,1,\dots,k_i-1$ for the corresponding orthonormal
basis of $\underline{V}_{\,k_i}\otimes\Hcal$. Let us now introduce
unitary isomorphisms
$U^{~}_i:\underline{\Delta}_{\,2n}^+
\otimes\Hcal\to\underline{V}_{\,k_i}\otimes\Hcal$ and
$U_i^\dag:\underline{V}_{\,k_i}\otimes\Hcal\to \underline{\Delta}_{\,2n}^+
\otimes\Hcal$ by the formulas
\bea\label{Uipdef}
U^{~}_i&=&\sum_{a=0}^{r-1}~\sum_{a_i=0}^{k_i-1}~\sum_{\stackrel{
\scriptstyle q,q_i=0}{\scriptstyle r\,q+a=k_i\,q_i+a_i}}^\infty
|k_i\,q_i+a_i\rangle\langle r\,q+a| \nonumber
\\[4pt] &=& \sum_{a=0}^{r-1}~
\sum_{a_i=0}^{k_i-1}~\sum_{\stackrel{
\scriptstyle q,q_i=0}{\scriptstyle r\,q+a=k_i\,q_i+a_i}}^\infty
\vec\lambda^{\,i}_{a_i}\,
{\vec\rho^{\,}_{a}}{}^\dag\otimes|q_i\rangle\langle q| \ , \\[4pt]
U^\dag_i&=&\sum_{a=0}^{r-1}~\sum_{a_i=0}^{k_i-1}~\sum_{\stackrel{
\scriptstyle q,q_i=0}{\scriptstyle r\,q+a=k_i\,q_i+a_i}}^\infty
|r\,q+a\rangle\langle k_i\,q_i+a_i| \nonumber
\\[4pt] &=&\sum_{a=0}^{r-1}~
\sum_{a_i=0}^{k_i-1}~\sum_{\stackrel{
\scriptstyle q,q_i=0}{\scriptstyle r\,q+a=k_i\,q_i+a_i}}^\infty
\vec\rho_{a}^{\,}\,
{\vec\lambda^{\,i}_{a_i}}{}^\dag\otimes|q\rangle\langle q_i| \ .
\label{Uimdef}\eea
By using the shift operators (\ref{8}), we then define the operators
\beq
T^{~}_{N_i}\=U^{~}_i\,\bigl(T_{N_i}^{(0)}\bigr)
\quad\mbox{and}\quad
T_{N_i}^\dag\=\bigl(T_{N_i}^{(0)}\bigr)^\dag\,U^\dag_i
\label{TNiUidef}\eeq
on $\underline{\Delta}_{\,2n}^-\otimes\Hcal\to
\underline{V}_{\,k_i}\otimes\Hcal$ and
$\underline{V}_{\,k_i}\otimes\Hcal
\to\underline{\Delta}_{\,2n}^-\otimes\Hcal$. They satisfy the
requisite equations (\ref{ansatz4}), with the $k_i\times k_i$ matrix
\beq
P^{~}_{N_i}=U_i^{~}\,({\cal P}^{~}_{N_i}\otimes\Pi^{~}_i)\,
U_i^\dag
\label{PNifinal}\eeq
a projector of rank $N_i$ on the Fock space
$\underline{V}_{\,k_i}\otimes\Hcal$.

Notice that the rank $r=2^{n-1}$ used in this construction is an even
integer for $n\ge 2$. To work with odd ranks $r$ one may introduce the
$(2^{n-1}+1)\times (2^{n-1}+1)$ matrices
\begin{equation}\label{10}
T_{N_i}^{(0)\,\prime}=\begin{pmatrix}T^{(0)}_{N_i-1}&0\\
0&\Sigma_1'\end{pmatrix} \ ,
\end{equation}
where $T^{(0)}_{N_i-1}$ is defined as above and
\begin{equation}\label{11}
\Sigma_1'=\sum_{\ell =1}^\infty |0,\dots,0,\ell
\>\<\ell-1,0,\dots,0|
\end{equation}
is a shift operator on the Fock space $\Hcal$. Then the operators
(\ref{10}) satisfy the equations (\ref{TNi0eqs}) with
\begin{equation}\label{12}
P_{N_i}^{(0)\,\prime}=\begin{pmatrix}{\cal P}^{~}_{N_i-1}\otimes\Pi^{~}_i&0\\
0&|0,\ldots,0\>\<0,\ldots,0|\,\end{pmatrix}
\end{equation}
a projector of rank $N_i$ on the Fock space
$(\,\underline{\Delta}_{\,2n}^+\otimes
\Hcal)\oplus\Hcal\cong\C^r\otimes\Hcal$, where $r=2^{n-1} +1$. In this
case the Toeplitz operators $T_{N_i}^{~}$ are obtained by substituting
(\ref{10}) into (\ref{TNiUidef}) with the replacement
$\underline{\Delta}_{\,2n}^\pm\to\underline{\Delta}_{\,2n}^\pm\oplus\C$.

Note also that the partial isometry operator
\bea
T^{(0)}&:=&\prod_{i=0}^m\,T^{(0)}_{N_i} \nonumber\\[4pt]
&=&\Idd_r^{~}+\sum_{i=0}^m
\bigl(T^{(0)}_{N_i}-\Idd_r^{~}\bigr)~=~\Idd_r^{~}+
\sum_{i=0}^m\left((\Sigma)^{N_i}-\Idd_r^{~}\right) \ ,
\label{totalTN}\eea
together with the above representations of the $\uo$ group on the Weyl
operator algebra of $\R^{2n}_\theta$ and on the Fock space $\Hcal$,
defines a cycle in the $\uo$-equivariant analytic K-homology $\K^{\rm
  a}(\R_\theta^{2n}\rtimes\uo)\cong\K_\uo^{\rm a}(\R^{2n})$. After a
twisting appropriate to the inclusion of the pertinent magnetic
monopole bundles, it describes the $\su$-invariant configurations of
D-branes as branes on the (trivial) quotient space $\R^{2n}\,/\,\uo$.
The charge of this class is the same as that of the
cocycle $[\,\underline{\Delta}_{\,\underline{\cal V}}^+\,,\,
\underline{\Delta}_{\,\underline{\cal
    V}}^-\,;\,\mu^{~}_{\underline{\cal V}}\,]$ built
earlier in the topological K-theory (\ref{eqABSiso}) from the
standard ABS brane-antibrane class
$[\,\underline{\Delta}_{\,2n}^+,\underline{\Delta}_{\,2n}^-;\mu]$
which is the generator of
(\ref{usualABSiso})~\cite{HM1}--\cite{OS1},\cite{ABS1}. The
computation of the topological charge, as well as the equivalence between the
commutative (topological) and noncommutative (analytic) K-homology
descriptions of the D-brane configurations, will be presented in the
next section.

\bigskip

\noindent
{\bf Moduli space of solutions.\ }
The realization (\ref{8}) can be generalized in order to introduce
$2n\,\sum_{i=0}^m\,N_i$ real moduli into the solution which specify the
locations of the various noncommutative solitons
in~$\R^{2n}$~\cite{Mart1}. For this, one first has to introduce
``shifted ground states'' centered at $(b^{i\,\m}_{\ell_i})$, $\ell_i
= 1,\dots,N_i$ for each $i=0,1,\dots,m$. The operators (\ref{8}) are
rewritten as
\beq
T^{(0)}_{N_i}=\Idd^{~}_r\otimes\bigl(1-\Pi^{~}_i\bigr)+\bigl
(\Sigma^i_1\,\Sigma^i_2\,
\cdots\Sigma^i_{N_i}\bigr)\otimes\Pi^{~}_i \ ,
\label{TNirewrite}\eeq
where each $\Sigma^i_{\ell_i}$, $\ell_i=1,\dots,N_i$, $i=0,1,\dots,m$ is of
the form of the shift operator $\Sigma$ in (\ref{TT}) but with the
coordinates~$x$ shifted to $x^i_{\ell_i}:=x-b^i_{\ell_i}$. They behave just
like $\Sigma$ except that now the kernel of
$(\Sigma^i_{\ell_i})^\dag$ is spanned by the vector
$|\psi\rangle\otimes|\bb^i_{\ell_i}\rangle$, where $|\psi\rangle$ is the
fermionic ground state and the shifted ground state $|\bb^i_{\ell_i}\rangle$
is a coherent state in the $n$-oscillator Fock space $\Hcal$,
i.e. $\bar z_{\ell_i}^{i\,\ab}|\bb^i_{\ell_i}\rangle=0$. The
states $|\psi\rangle\otimes|\bb^i_1\rangle$ and
$\Sigma^i_1\cdots\Sigma^i_{\ell_i-1}
(|\psi\rangle\otimes|\bb^i_{\ell_i}\rangle)$
for $\ell_i=2,\dots,N_i$ span the kernel of the operator $(T^{(0)}_{N_i})^\dag$
given by (\ref{TNirewrite}), and we find that the equations
(\ref{TNi0eqs}) are obeyed with $P_{N_i}^{(0)}$ the orthogonal
projection onto $\ker(T_{N_i}^{(0)})^\dag$.

The space of partial isometries (\ref{totalTN}) may thereby be
described as the complex manifold $\prod_{i=0}^m\,(\C^n)^{N_i}$. After
a quotient by the appropriate discrete symmetry group, the moduli
space for the full solution consisting of the rectangular Toeplitz
operators (\ref{TNiUidef}) is given by
\beq
{\cal M}\bigl(\,n\,;\,\vec k^{~}_{\underline{\cal V}}\,,\,\vec k^{~}_{
\underline{\cal T}}\,\bigr)={\cal Q}\bigl(\,\vec k^{~}_{
\underline{\cal V}}\,\bigr)\times\prod_{i=0}^m\,{\rm Hilb}^{N_i}
\left(\C^n\right) \ ,
\label{partisospace}\eeq
where ${\rm Hilb}^{N_i}(\C^n)$ is the moduli space of $N_i$
noncommutative solitons on $\rt$~\cite{GHS1} which is given as the
(singular) Hilbert scheme of $N_i$ points in $\C^n$, i.e. the set of
ideals $\cal I$ of codimension $N_i$ in the polynomial ring
$\C[b_1^i,\dots,b_{N_i}^i]$. The factor ${\cal Q}(\,\vec
k^{~}_{\underline{\cal V}}\,)$ is the moduli space of isomorphism
classes of quiver representations (\ref{genrepSU2Uk}) of
dimension~\cite{quiverbooks}
\beq
\dim{\cal Q}\bigl(\,\vec k^{~}_{\underline{\cal V}}\,\bigr)\=
1-\mbox{$\frac12$}\,\vec k^{~}_{\underline{\cal V}}\cdot C
\vec k^{~}_{\underline{\cal V}}\=1+\sum_{i=0}^mk_i\,\left(k_{i+1}-k_i
\right) \ .
\label{dimcalQ}\eeq
Note that real roots (having $\vec k^{~}_{\underline{\cal V}}\cdot C
\vec k^{~}_{\underline{\cal V}}=2$) correspond to rigid
representations of the quiver $\quiverm$ with no moduli, while
imaginary roots (having $\vec k^{~}_{\underline{\cal V}}\cdot C
\vec k^{~}_{\underline{\cal V}}\leq0$) carry moduli associated to the
gauge symmetry breaking (\ref{gaugebroken}). The points of the moduli
space (\ref{partisospace}) label the positions of well-separated
D-branes, and it coincides in the low-energy limit with the moduli
space of the commutative brane description~\cite{Mart1}.

\bigskip

\section{D-brane charges\label{Dbrane}}

In this section we will compute the topological charge of our
multi-instanton solutions in essentially two distinct
ways. The first one is a direct field theoretic calculation of
the $(n+1)$-th Chern number of our gauge field configurations on $\rts$, which
can also be computed using the $\Z_{m+1}$-graded connection formalism
of Section~\ref{Dimred}. The second one is a homological calculation
of the index class of our solutions in K-theory, which is also
equivalent to the Euler-Ringel character of the pertinent
representations of the quiver $\quiverm$. The equivalence of these two
calculations will then lead us directly into a worldvolume description
whereby we can interpret the topological charge in terms of cycles in
topological equivariant K-homology, yielding the claimed D-brane
interpretation of our solutions. The results of this section bridge
together the descriptions presented in Section~\ref{Ansatzdescr} and
justify the brane interpretations that have been given throughout this
paper thus far.

\bigskip

\noindent
{\bf Field theory calculation.\ }
We will first compute the topological charge of the configurations
(\ref{ansatz3})--(\ref{ansatz4}). For
this, it is convenient to parametrize the two-sphere by the angular
coordinates $0\le\vp\le 2\pi$ and $0\le\vartheta<\pi$ defined in
(\ref{zn1}). In these coordinates
\begin{equation}
\label{complfun}
{\cf}_{y\yb}\=\Bigl|\frac{\pa(\vt,\vp)}{\pa(y,\yb)}\Bigr|\,{\cf}_{\vt\vp}
\=\frac{1}{2\im}\,\frac{\sin\vt}{y\yb}\,{\cf}_{\vt\vp}
\=\frac{1}{2\im}\,\frac{(1{+}\cos\vt)^2}{R^2\,\sin\vt}\,{\cf}_{\vt\vp}\ ,
\end{equation}
and we have
\begin{align}
{\cf}^{~}_{2a-1\, 2a} \= & \,2\im\, {\cf}^{~}_{a\ab} \= -\frac{\im}{\th^a}\,
\sum^m_{i=0} P^{~}_{N_i}\otimes\Pi^{~}_i \ ,\\[4pt]
{\cf}^{~}_{\vt\vp} \= &-\im\,\frac{\sin\vt}{2}\,
\sum^m_{i=0}\,(m-2i)~P^{~}_{N_i}\otimes\Pi^{~}_i
\end{align}
giving
\begin{align}
{\cf}^{~}_{12}\,{\cf}^{~}_{34}\cdots{\cf}^{~}_{2n-1\, 2n}\,{\cf}^{~}_{\vt\vp}\=
&\,(-\im)^{n+1}\,\frac{\sin\vt}{2\,\prod\limits^n_{a=1}\th^a}\,
\Bigl(\,\sum^m_{i=0} P^{~}_{N_i}\otimes\Pi^{~}_i \Bigr)^n \,
\Bigl(\,\sum^m_{j=0}\,(m-2j)~P^{~}_{N_j}\otimes\Pi^{~}_j\Bigr)\nonumber\\[4pt]
\=&\,(-\im)^{n+1}\,\frac{\sin\vt}{2\,\prod\limits^n_{a=1}\th^a}\,
\sum^m_{i=0}\,(m-2i)~P^{~}_{N_i}\otimes\Pi^{~}_i \ ,
\end{align}
where we have used the definitions (\ref{PNiTrace}) and
(\ref{Piortho}) of the projectors $P^{~}_{N_i}$ and $\Pi^{~}_i$.

The instanton charge is then given by the $(n+1)$-th Chern number
\begin{align}
Q\ :=\ &\,\frac{1}{(n{+}1)!}\ \Bigl(\frac{\im}{2\pi}\Bigr)^{n+1}\,
\Bigl(\,\prod_{a=1}^n{2\pi\,\th^a}\Bigr)~
\int_{S^2} \Tr^{~}_{\underline{\cal V}\otimes{\cal H}}~
\underbrace{{\cf}\wedge\ldots\wedge {\cf}}_{n+1} \nonumber\\[4pt]
\= &\,\Bigl(\frac{\im}{2\pi}\Bigr)^{n+1}\,
\frac{(-\im)^{n+1}}{2\,\prod\limits^n_{b=1}\th^b}\,
\Bigl(\,\prod_{a=1}^n{2\pi\,\th^a}\Bigr)\,
\sum^m_{i=0}\,(m-2i)\,{N_i}\,
\int_{S^2} \sin\vt~\diff\vt\wedge\diff\vp \ .
\label{Chernnum}\end{align}
After splitting the sum over $i$ into contributions from monopoles and
antimonopoles analogously to (\ref{EFNCfinite}), this becomes
\beq
Q=\sum^{\lfloor\frac m2\rfloor}_{i=0}\,(m-2i)\,
(N_i-N_{m-i}) \ ,
\label{TopchargeYM}\eeq
where we recall that $N_i\ge 0$ for $i=0,1,\dots,m$. The formula
(\ref{TopchargeYM}) clarifies the D-brane interpretation of the
configuration (\ref{ansatz3})--(\ref{ansatz4}). It describes a
collection of $(m-2i)\,N_i\ $ D0-branes for $2i<m$ and $(2i-m)\,N_i\ $
anti-D0-branes for $2i>m$ as a bound state (i.e. a vortex-like configuration
on $\rt$) in a system of $k_0+k_1+\ldots+k_m=k\ $ D(2$n$) branes and
antibranes. However, from the point of view of the initial brane-antibrane
system on $\rts$, they are spherical $|m-2i|\,N_i$ \ D2-branes or
D2-antibranes depending on the sign of the monopole charge
$m-2i$. Note that the vortices with $2i=m$, which always exist
for even $m$, have vanishing instanton charge since they couple with the
trivial line bundle $\Lcal^0=S^2\times\C$. Thus they are not
extended to instantons on $\rts$, but are rather unstable and simply
decay into the vacuum.

The topological charge can be alternatively computed within the graded
connection formalism of Section~\ref{Dimred}. Recalling the
equivariant ABS construction
(\ref{twistedspingrad})--(\ref{twistedspingradeven}), we note that the
$\Z_{m+1}$-graded vector space (\ref{genrepSU2Uk}) (the fibre of the
$\Z_{m+1}$-graded bundle (\ref{gradedbundle})) also has a {\it
  natural} $\Z_2$-grading by the sign of the magnetic charge, i.e. by
the involution $\varepsilon:\underline{\cal V}\to\underline{\cal V}$
defined by $\varepsilon(v_i):={\rm sgn}(m-2i)~v_i$ for
$v_i\in\underline{V}_{\,k_i}$, where throughout we use the convention
${\rm sgn}(0):=0$. The corresponding supertrace is given by
\beq
\str^{~}_{k\times k}~X~:=~\tr_{k\times k}^{~}(\varepsilon\circ X)
{}~=~\sum_{i=0}^m\,{\rm sgn}(m-2i)~\tr_{k_i\times k_i}^{~}~X_i
\label{supertracedef}\eeq
for any linear operator $X\in{\rm End}(\,\underline{\cal V}\,)$ with
block-diagonal components $X_i\in{\rm
  End}(\,\underline{V}_{\,k_i})$. This extends to a supertrace
$\STr^{~}_{\underline{\cal V}\otimes\Hcal}:=\Tr^{~}_{\Hcal}~\str^{~}_{k\times
  k}$ which we may use to express the Chern number in
terms of the graded curvature (\ref{gradedcurv}) as
\beq
Q=\frac{R^2}{2^n\,(n{+}1)!}\ \Bigl(\frac{\im}{2\pi}\Bigr)^{n+1}\,
\Bigl(\,\prod_{a=1}^n{2\pi\,\th^a}\Bigr)~\STr^{~}_{\underline{\cal V}
\otimes{\cal H}}~\Tr^{~}_{\C^{2^{n+1}}}~\bigl(\Gamma\,\hat\cf^{n+1}
\bigr)_{\rm asym} \ ,
\label{Topchargegraded}\eeq
where $\Gamma:=\frac2{\sqrt
  g}\,\Gamma^1\cdots\Gamma^{2n+2}=\gamma\otimes\sigma_3$ and the
antisymmetrized product of gamma-matrices
\beq
\bigl(\Gamma^{\hat\mu_1}\cdots\Gamma^{\hat\mu_q}\bigr)_{\rm asym}
:=\frac1{q!}\,\sum_{\pi\in S_q}\,{\rm
  sgn}(\pi)~\Gamma^{\hat\mu_{\pi(1)}}\cdots\Gamma^{\hat\mu_{\pi(q)}}
\label{asymGamma}\eeq
mimicks the algebraic structure of the exterior product of differential
forms. The formula (\ref{TopchargeYM}) follows from
(\ref{Topchargegraded}) upon repeated application of the Clifford
algebra and the trace identities
(\ref{Trgammaid1})--(\ref{Trgammaid4}), with the supertrace
(\ref{supertracedef}) giving the appropriate sign alternations.

\bigskip

\noindent
{\bf K-theory calculation.\ }
The origin of the topological charge lies in the {\it graded Chern
  character} ${\rm ch}(\,\underline{\cal V}\otimes\Hcal):=\str^{~}_{k\times
  k}\,\exp\hat\cf/2\pi\im$. Standard transgression arguments can be
used to show that the cohomology class defined by this closed
differential form is independent of the choice of graded
connection~\cite{Quillen1}. In particular, we may either compute it by
setting the off-diagonal Higgs fields $\phi_i=0$ or by setting the
diagonal gauge fields $A^i=0$. It is instructive to recall how this
works in the case $m=1$ corresponding to the basic brane-antibrane
system represented by the chain
(\ref{A2quiver})~\cite{Matsuo1,AIO1}. In the former case we would
obtain the difference $\chern(\,\underline{
  V}_{\,k_1}\otimes\Hcal)-\chern(\,\underline{V}_{\,k_0}\otimes\Hcal)$ of
topological charges on the branes and antibranes. In the latter case
we would compute the index of the tachyon field $\phi_1$, or
equivalently the Euler characteristic of the two-term complex
$0\to\underline{
  V}_{\,k_1}\otimes\Hcal\stackrel{\phi_1}{\to}\underline{
  V}_{\,k_0}\otimes\Hcal\to0$. The virtual Euler
class generated by the cohomology of this complex is the
analytic K-homology class $[\phi_1]\in\K^{\rm a}(\R^{2n})$ of the brane
configuration. The equivalence of the two computations is asserted by
the index theorem.

The situation for $m>1$ is more subtle. The action of the graded
connection zero-form (\ref{mgradedphidef}) on the bundle
(\ref{gradedbundle}) produces the
holomorphic chain (\ref{holchain}). In general this is {\it not} a
complex because, according to (\ref{mphi0s}), $(\mphi)^2\neq0$ for
$m>1$, i.e. $\phi_i\,\phi_{i+1}\neq0$. The only physical instance in
which such a chain generates a complex is when it corresponds to an
alternating sequence of branes and antibranes~\cite{Sharpe1}. But if one has a
tachyon field which is a holomorphic map from an antibrane to a brane,
then the adjoint map is antiholomorphic. Recalling
(\ref{f25}), we see that in our chain (\ref{holchain}) all
maps $\phi_i$ are {\it holomorphic} and thus do not represent tachyon
fields between pairs of branes and antibranes. Furthermore, the maps
$\phi_i$ obtained as solutions of the vortex equations, which can be
associated with the ${\rm A}_{m+1}$ quiver and are obtained by
$\su$-invariant reduction, can never satisfy the constraints
$\phi_i\,\phi_{i+1}=0$~\cite{BGP,quiverbun}.

The solution to this problem is to fold the given holomorphic chain
into maps between branes and antibranes. Let us first carry out the
calculation in the case that the monopole Chern number $m$ is an odd
integer. By using the $\Z_2$-grading $\varepsilon:\underline{\cal
  V}\to\underline{\cal V}$ introduced above, we explicitly decompose
(\ref{genrepSU2Uk}) as a $\Z_2$-graded module into the $\pm\,1$
eigenspaces of the involution $\varepsilon $ giving
\beq
\underline{\cal V}\=\underline{\cal V}_{\,+}\oplus\underline{\cal V}_{\,-}
\qquad\mbox{with}\qquad \underline{\cal V}_{\,+}\=\bigoplus_{i=0}^{m_-}\,
\underline{V}_{\,k_i} \quad\mbox{and}\quad\underline{\cal V}_{\,-}\=
\bigoplus_{i=m_+}^m\,\underline{V}_{\,k_i} \ .
\label{calVZ2module}\eeq
Using (\ref{mgradedphidef}) and (\ref{mphi0s}) we now introduce the
operator
\beq
\mbf T^{~}_{(m)}:=\bigl(\mphi\bigr)^{\lfloor\frac m2\rfloor+1} \ .
\label{mTdef}\eeq
With respect to the $\Z_2$-grading (\ref{calVZ2module}), it is an odd
map
\beq
\mT\,:\,\underline{\cal V}_{\,-}\otimes\Hcal~
\longrightarrow~\underline{\cal V}_{\,+}\otimes\Hcal
\qquad\mbox{with}\qquad \bigl(\mT\bigr)^2\=0 \ .
\label{mTmap}\eeq
Thus the triple $[\underline{\cal V}_{\,+}\otimes\Hcal,\underline{\cal
  V}_{\,-}\otimes\Hcal;\mT]$ defines a two-term complex and represents a
brane-antibrane system with tachyon field given in terms of the graded
connection by (\ref{mTdef}). The corresponding index class
$[\mT]\in\K^{\rm a}(\R^{2n})$ is thus
the analytic K-homology class of our configuration of
D-branes. In particular, on isotopical components one has
\beq
\mT\circ\Pi^{~}_{i+1+\lfloor\frac m2\rfloor}
\=\phi^{~}_{i+1}\cdots\phi^{~}_{i+1+\lfloor\frac
m2\rfloor}\=\bigl(\alpha^{~}_{i+1}\cdots\alpha^{~}_{i+1+\lfloor\frac
m2\rfloor}\bigr)~T^{~}_{N_i}\,T^\dag_{N_{i+1+\lfloor\frac
m2\rfloor}}
\label{mTisotopexpl}\eeq
while $\mT\circ\Pi_i^{~}=0$, where $i=0,1,\dots,m_-$. The tachyon field
is thus a holomorphic map between branes of equal and opposite
magnetic charge,
\beq
\mT\circ\Pi^{~}_{i+\lfloor\frac m2\rfloor+1}
\,:\,\underline{V}_{\,k_{i+\lfloor\frac
m2\rfloor+1}}\otimes\Hcal~\longrightarrow~\underline{V}_{\,k_i}
\otimes\Hcal \ ,
\label{mTisotopmap}\eeq
and from (\ref{dimkerTNi}) it follows that it has a finite dimensional
kernel and cokernel with
\beq
\dim\ker\bigl(\mT\circ\Pi^{~}_{i+\lfloor\frac m2\rfloor+1}
\bigr)\=N_{i+\lfloor\frac
m2\rfloor+1} \qquad\mbox{and}\qquad
\dim\ker\bigl(\mT\circ\Pi^{~}_{i+\lfloor\frac m2\rfloor+1}
\bigr)^\dag\=N_i \ .
\label{dimkermT}\eeq

To incorporate the twistings by the magnetic monopole bundles, we use
the ABS construction
(\ref{twistedspingrad})--(\ref{Cliffordmulteven}) to define the
tachyon field
\beq
\mcalT~:=~\mT\otimes\Idd\,:\,\underline{\Delta}_{\underline{\cal
V}}^+\otimes\Hcal~\longrightarrow~\underline{\Delta}_{\underline{\cal
V}}^-\otimes\Hcal  \ .
\label{mcalTdef}\eeq
It behaves like a noncommutative version of Clifford
multiplication $\mu_{\underline{\cal V}}^\dag$ in
(\ref{CliffmultV},\ref{Cliffmultisot}). Since
$\dim\underline{V}_{\,|m-2i|}=|m-2i|$, from (\ref{dimkermT}) it
follows that the index of the tachyon field (\ref{mcalTdef}) is given
by
\bea
{\rm index}~\mcalT&=&\dim\ker\bigl(\mcalT\bigr)-\dim\ker
\bigl(\mcalT\bigr)^\dag \nonumber\\[4pt]
&=&\sum_{i=m_+}^m|m-2i|\,N_i-\sum_{i=0}^{m_-}
|m-2i|\,N_i\=-Q \ .
\label{indexmcalT}\eea
Thus the K-theory charge of the noncommutative soliton configuration
(\ref{ansatz3})--(\ref{ansatz4}) coincides with the Yang-Mills
instanton charge (\ref{Chernnum},\ref{TopchargeYM}) on $\rts$.

When the monopole charge $m$ is even, we introduce the tachyon field
$\mT$ by the same formula (\ref{mTdef}). The only difference now is
that the subspace $\underline{V}_{\,k_{\frac m2}}\otimes\Hcal$ is annihilated
by
both operators $(\mT)$ and $(\mT)^\dag$ so that
\beq
\underline{V}_{\,k_{\frac m2}}\otimes\Hcal\subset\ker\bigl(\mT\bigr)\cap
\ker\bigl(\mT\bigr)^\dag \ .
\label{mTVm20}\eeq
According to (\ref{twistedspingradeven}), this subspace should be
coupled to the eigenspace (\ref{spinorharm}) of spinor harmonics on
$\C P^1$ when defining the extended tachyon field
(\ref{mcalTdef}). Analogously to (\ref{Cliffordmulteven}), one then has
\beq
\ker\bigl(\mcalT\circ\Pi^{~}_{\frac m2}\bigr)\=
\ker\bigl(\mcalT\circ\Pi^{~}_{\frac m2}\bigr)^\dag\=
\underline{V}_{\,k_{\frac m2}}\otimes\underline{H}_{\,p}\otimes\Hcal \
{}.
\label{kermcalTm2}\eeq
With a suitable regularization of the infinite dimensions of the
kernel and cokernel of the operator $\mcalT\circ\Pi^{~}_{\frac m2}$,
these subspaces will make no contribution to the index
(\ref{indexmcalT}). This statement will be justified below by the fact
that ${\rm index}\,\Dirac_0=0$ and that the index class of the
noncommutative tachyon field coincides with that of the twisted
$\su$-invariant Dirac operator on $\R^{2n}\times\C P^1$.

We can give a more detailed picture of how the topological charge of the
system of D-branes arises by relating the index to a homological
computation in the corresponding quiver gauge theory, which shows
precisely how the original brane configuration folds itself into
branes and antibranes. Consider the $\quiverm$-module (\ref{VNAm})
defined by a generic (non-BPS) solution of the Yang-Mills equations on
$\R^{2n}_\theta\times\C P^1$, and let
\beq
\underline{\cal W}\=\bigoplus_{i=0}^m\,\underline{W}_{\,i}
\qquad\mbox{with}\qquad \vec k^{~}_{\underline{\cal W}}\=
\sum_{i=0}^mw_i~\vec e_i
\label{anyWrep}\eeq
be any quiver representation. Applying the functor
$\Hom(\,-\,,\,\underline{\cal W}\,)$ to the projective resolution
(\ref{Ringelres}) gives a complex whose
cohomology in the $p$-th position defines the extension groups
$\Ext^p(\,\underline{\cal T}\,,\,\underline{\cal W}\,)\cong{\rm
  H}^p(\R^{2n}_\theta\,;\,\underline{\cal W}\otimes\underline{\cal
  T}^\vee\,)$, with $\Ext^0=\Hom$ and
$\Ext^1=\Ext$. We may then define the relative Euler character between
these two representations through the corresponding Euler form
\beq
\chi\bigl(\,\underline{\cal T}\,,\,\underline{\cal W}\,
\bigr):=\sum_{p\geq0}(-1)^p~\dim\,\Ext^p(\,
\underline{\cal T}\,,\,\underline{\cal W}\,) \ .
\label{Eulerformgen}\eeq
Since the ${\rm A}_{m+1}$ quiver has no
relations, one has $\Ext^p(\,\underline{\cal
  T}\,,\,\underline{\cal W}\,)=0$ for all $p\geq2$ in the present
case~\cite{quiverbooks}.

By using (\ref{Hompathnatural}), the resolution (\ref{Ringelres})
induces an exact sequence of extension groups given by
\bea
0~\longrightarrow~\Hom\bigl(\,\underline{\cal T}\,,\,
\underline{\cal W}\,\bigr)&\longrightarrow&\bigoplus_{i=0}^s\,
\Hom\bigl(\ker T_{N_i}^\dag\,,\,\underline{W}_{\,i}\bigr)~
\longrightarrow \nonumber\\ &\longrightarrow&
\bigoplus_{i=0}^{s-1}\,\Hom\bigl(
\ker T_{N_{i+1}}^\dag\,,\,\underline{W}_{\,i}\bigr)~
\longrightarrow~\Ext\bigl(\,\underline{\cal T}\,,\,
\underline{\cal W}\,\bigr)~\longrightarrow~0
\label{Extexactseq}\eea
from which we may compute the Euler form (\ref{Eulerformgen})
explicitly to get
\bea
\chi\bigl(\,\underline{\cal T}\,,\,\underline{\cal W}\,
\bigr)&=&\dim\,\Hom\bigl(\,\underline{\cal T}\,,\,
\underline{\cal W}\,\bigr)-\dim\,\Ext\bigl(\,\underline{\cal T}\,,\,
\underline{\cal W}\,\bigr) \nonumber\\[4pt]
&=&\sum_{i=0}^s\dim\,\Hom\bigl(\ker T_{N_i}^\dag\,,\,
\underline{W}_{\,i}\bigr)-\sum_{i=0}^{s-1}\dim\,\Hom\bigl(
\ker T_{N_{i+1}}^\dag\,,\,\underline{W}_{\,i}\bigr) \nonumber\\[4pt]
&=&\sum_{i=0}^mN_i\,w_i-\sum_{i=0}^{m-1}N_{i+1}\,w_i \ .
\label{EulerWgen}\eea
Thus the relative Euler character depends only on the dimension vectors
of the corresponding representations and coincides with the Ringel
form $\langle\,\vec k^{~}_{\underline{\cal T}}\,,\,\vec k^{~}_{\underline{\cal
    W}}\,\rangle$ on the representation ring $\rep_\quiverm$ of the
${\rm A}_{m+1}$ quiver~\cite{quiverbooks}. The map $[\,\underline{\cal
  W}\,]\mapsto\vec k^{~}_{\underline{\cal W}}$ gives a linear map
$\rep_\quiverm\to\Z^{m+1}$ which is an isomorphism of abelian groups
since $\rep_\quiverm$ is generated by the Schur modules
$\simple_{\,i}$, $i=0,1,\dots,m$. By using (\ref{ansatz3p}) and
(\ref{dimkerTNi}) we can write this bilinear pairing in the suggestive
form
\beq
\chi\bigl(\,\underline{\cal T}\,,\,\underline{\cal W}\,
\bigr)=-\sum_{i=0}^mw_i~{\rm index}(\phi_{i+1}) \ .
\label{Eulerindexphi}\eeq

The appropriate representation $\underline{\cal W}$ to couple with in
the present case is dictated by the correct incorporation of magnetic
charges. As before, the fact that the Higgs fields $\phi_{i+1}$ in
(\ref{Eulerindexphi}) themselves are not tachyonic, i.e. do not
generate a complex, means that we have to fold the $\su$
representations $\underline{V}_{\,|m-2i|}$ appearing in the ABS
construction (\ref{twistedspinpm}) appropriately. The correct folding
is expressed by the collection of distinguished triangles
(\ref{BPStriangles}) which shows that we should couple an increasing sequence
$\underline{W}_{\,0}\subset\underline{W}_{\,1}\subset\dots\subset
\underline{W}_{\,m}$ of representations as we move along the chain of
constituent D-branes of the quiver, so that the $\su$-module
$\underline{W}_{\,i}$ gives an extension of the monopole field carried
by the elementary brane state at vertex $i$ by the $\su$-module
$\underline{W}_{\,i-1}$. Thus we take
$\underline{W}_{\,i}\=\bigoplus_{j=0}^i\,\underline{V}_{\,|m-2j|} =
\underline{V}_{\,|m-2i|}\oplus\underline{W}_{\,i-1}$ and embed its
class into the representation ring $\rep_{\quiverm}$ using the
$\Z_2$-grading above as the element
\beq
\bigl[\,\underline{W}_{\,i}\,\bigr]\=\sum_{j=0}^i\,{\rm sgn}(m-2j)~
\bigl[\,\underline{V}_{\,|m-2j|}\,\bigr]\={\rm sgn}(m-2i)~
\bigl[\,\underline{V}_{\,|m-2i|}\,\bigr]+\bigl[\,\underline{W}_{\,i-1}
\,\bigr]
\label{Widef}\eeq
of virtual dimension
\beq
w_i\=\sum_{j=0}^i\,(m-2j)\=(i+1)\,(m-i)
\label{Wiwidef}\eeq
for each $i=0,1,\dots,m$. In this case the Euler-Ringel form
(\ref{EulerWgen}) becomes
\beq
\chi\bigl(\,\underline{\cal T}\,,\,\underline{\cal W}\,
\bigr)\=\sum_{i=0}^m(i+1)\,(m-i)\,(N_i-N_{i+1})\=\sum_{i=0}^m(m-2i)
\,N_i\=Q
\label{Eulertopcharge}\eeq
and it also coincides with the instanton charge of the
gauge field configurations on $\rts$. The equivalence of the relative
Euler character with the index of the tachyon field above
is a consequence of the Grothendieck-Riemann-Roch theorem.

\bigskip

\noindent
{\bf Worldvolume construction.\ }
We can now present a very explicit geometric description of the
equivalence between the brane configurations on $\R^{2n}\times\C P^1$
and on $\R^{2n}$. The crux of the formulation is the well-known map in
K-theory between analytic (noncommutative) and topological
(commutative) descriptions~\cite{HM1,Sz1,LPS,Baum1}. If
$\Dirac:=-\im\sigma\cdot\partial:{\rm
  L}^2(\R^{2n}\,,\,\underline{\Delta}_{\,2n}^-)\to{\rm
  L}^2(\R^{2n}\,,\,\underline{\Delta}_{\,2n}^+)$ is the usual Dirac
operator on $\R^{2n}$, then its index coincides with that of the
noncommutative ABS configuration~(\ref{TT}) giving
\beq
{\rm index}~\Sigma={\rm index}~\Dirac \ .
\label{indexSigmaDirac}\eeq
This coincides with the K-theory charge of the Bott class
$[\,\underline{\Delta}_{\,2n}^+,\underline{\Delta}_{\,2n}^-;\mu]
\in\K(\R^{2n})$ given by the ordinary ABS construction~\cite{ABS1}, where
$\mu_x=\frac{\sigma\cdot
  x}{|x|}:\underline{\Delta}_{\,2n}^-\to\underline{\Delta}_{\,2n}^+$
is Clifford multiplication by $x\in\R^{2n}$. In particular, the Dirac
operator itself can be used to represent the analytic K-homology class
$[\Sigma]=[\Dirac]$ described by the noncommutative ABS field.

Let us represent a system of $k$ Type~IIA D-branes wrapped on $\R^{2n}\times\C
P^1$ with virtual Chan-Paton bundle $\Xi\in\K(\R^{2n}\times\C P^1)$ by the
K-cycle $[\R^{2n}\times\C P^1,\Xi,{\rm id}]$ in the topological
K-homology $\K^{\rm t}(\R^{2n}\times\C P^1)$. Its equivalence class is
invariant under the usual relations of bordism, direct sum and vector
bundle modification~\cite{Sz1,LPS,Baum1}. There is an isomorphism
$\K^{\rm t}(\R^{2n}\times\C
P^1)\cong\K^{\rm a}(\R^{2n}\times\C P^1)$ of abelian groups which
sends this K-cycle to the analytic K-homology class $[\hat\Dirac_\Xi\,]$
defined by the corresponding twisted Dirac operator on
$\R^{2n}\times\C P^1$. Similarly, if $\xi\in\K(\R^{2n})$ and
$\imath:\R^{2n}\hookrightarrow\R^{2n}\times\C P^1$ is the slice
induced by the inclusion $\uo\hookrightarrow\su$ of groups,
then the topological K-cycle $[\R^{2n},\xi,\imath]\in\K^{\rm
  t}(\R^{2n}\times\C P^1)$ corresponds to the analytic K-homology
class $\imath_*[\Dirac_\xi\,]\in\K^{\rm a}(\R^{2n}\times\C P^1)$, where
$\Dirac_\xi$ is the twisted Dirac operator on $\R^{2n}$.

Now consider the $\su$-equivariant reduction of these
cycles. From the construction of the previous section with $\mphi=0$
and the equivariant excision theorem of Section~\ref{Ansatzdescr} we
have the equality
\beq
\bigl[\hat\Dirac_\Xi\,\bigr]^\su=\imath_*\bigl[\Dirac_{\imath^*\Xi}
\,\bigr]^\uo
\label{DiracSU2eq}\eeq
in $\K^{\rm a}_\su(\R^{2n}\times\C P^1)$ which leads to
\beq
\bigl[\R^{2n}\,,\,\xi\,,\,\imath\bigr]\=\bigl[\R^{2n}\times
\C P^1\,,\,\Xi\,,\,{\rm id}\bigr] \qquad\mbox{with}\qquad
\Xi\=\su\times_\uo\xi
\label{KcycleSU2eq}\eeq
in $\K_\su^{\rm t}(\R^{2n}\times\C P^1)$. The left-hand side of
(\ref{KcycleSU2eq})
corresponds to the class of D$(2n)$ brane-antibrane pairs wrapping
$\R^{2n}$, while the right-hand side corresponds to D$(2n+2)$
brane-antibrane pairs wrapping $\R^{2n}\times\C P^1$. This is just the
equivalence between instantons on $\R^{2n}\times\C P^1$ and vortices
on $\R^{2n}$. We note that in the case $m=1$, the monopole field is
automatically spherically symmetric on $\C P^1$ and one can formulate the
equivalence (\ref{KcycleSU2eq}) using only the requirement of vector bundle
modification in {\it ordinary} topological K-homology~\cite{LPS}, which is
equivalent to Bott periodicity (\ref{Bottper}). In contrast, for $m>1$
one must appeal to an $\su$-equivariant framework and the identification
(\ref{KcycleSU2eq}) of K-cycles is far more intricate. In this case it
is a result of the equivariant excision theorem, and {\it not} of Bott
periodicity in equivariant K-theory. It is this intricacy that leads
to a more complicated brane-antibrane system when $m>1$.

Using the equivariant ABS construction of the previous section, the
K-homology class of the multi-instanton solution
(\ref{ansatz3})--(\ref{ansatz4}) is given by the left-hand side of
(\ref{KcycleSU2eq}) with
\beq
\xi=\bigl[\Delta_E^+\,,\,\Delta_E^-\,;\,\mu^{~}_{N_0,N_1,\dots,N_m}
\bigr] \ ,
\label{ximultiinst}\eeq
where
\beq
\Delta_E^+\=\bigoplus_{i=m_+}^m\,
E_{k_i}\otimes\underline{V}_{\,|m-2i|}
\qquad\mbox{and}\qquad \Delta_E^-\=
\bigoplus_{i=0}^{m_-}\,
E_{k_i}\otimes\underline{V}_{\,m-2i}
\label{twistedEbundles}\eeq
while
\beq
\mu^{~}_{N_0,N_1,\dots,N_m}=\prod\limits_{i=0}^{m_-}\,\bigl(
\mu^{~}_E\circ\Pi^{~}_i\bigr)^{N_i}\,\prod_{j=m_+}^m\bigl(
\mu_E^\dag\circ\Pi_j^{~}\bigr)^{N_j}
\label{muNs}\eeq
with $\mu^{~}_E:\Delta_E^-\to\Delta_E^+$ acting fibrewise as Clifford
multiplication (\ref{CliffmultV},\ref{Cliffmultisot}). The class
(\ref{ximultiinst}) is the K-theory class of the noncommutative
soliton field (\ref{totalTN}). The relation (\ref{KcycleSU2eq})
equates the resulting K-homology class with that defined by
\beq
\Xi=\bigl[\su\times_\uo\Delta_E^+\,,\,\su\times_\uo\Delta_E^-\,;\,
\pi^*\circ\mu^{~}_{N_0,N_1,\dots,N_m}\circ\imath^*\bigr] \ ,
\label{Ximultiinst}\eeq
where the projection $\pi:\R^{2n}\times\C P^1\to\R^{2n}$ is a left inverse to
the inclusion $\imath$, i.e. $\pi\circ\imath={\rm id}$. Through the
standard process of tachyon condensation on the system of D$(2n+2)$
branes and antibranes wrapping $\R^{2n}$, the right-hand side
of (\ref{KcycleSU2eq}) then describes $\sum_{2i<m}\,(m-2i)\,N_i\ $ D2-branes
and
$\sum_{2i>m}\,|m-2i|\,N_i\ $ D2-antibranes. On the left-hand side of
(\ref{KcycleSU2eq}), these are instead D0-branes corresponding to
vortices left over from condensation in the transverse space
$\R^{2n}$.

One can also compute the topological charge in this worldvolume
picture and explicitly demonstrate that the K-theory charges on both
sides of (\ref{KcycleSU2eq}) are the same. The natural charge of
branes defined by elements of equivariant K-theory is given by the
equivariant index ${\rm index}^{~}_\su(\hat\Dirac_\Xi)\in\rep_\su$,
which may be computed by using the $\su$-index theorem~\cite{AS1}
\beq
{\rm index}^{~}_\su~\hat\Dirac_\Xi=-\int_{\R^{2n}\times\C P^1}
\chern^{~}_\su(\Xi)\wedge\widehat{A}\bigl(\R^{2n}\times\C P^1\bigr) \
,
\label{equiindexthm}\eeq
where  $\chern^{~}_\su:\K_\su(\R^{2n}\times\C
P^1)\to\HQ^\bullet_\su(\R^{2n}\times\C P^1;\Q)$ is the equivariant Chern
character taking values in $\su$-equivariant rational
cohomology. Since this index depends only on the equivariant K-homology
class of the Dirac operator on $\R^{2n}\times\C P^1$, we may
explicitly use (\ref{DiracSU2eq}) and perform the dimensional
reduction to write the index (\ref{equiindexthm}) as
\beq
{\rm index}^{~}_\su~\hat\Dirac_\Xi=-\int_{\R^{2n}}\chern^{~}_\su(\xi)
\ .
\label{equiindexred}\eeq

Since the Chern character in (\ref{equiindexred}) is a ring
homomorphism between $\K_\su(\R^{2n})\cong\rep_\su$ and
$\HQ^\bullet(\R^{2n};\Q)\otimes\rep_\su$, upon substitution of
(\ref{ximultiinst},\ref{twistedEbundles}) we can use its additivity
and multiplicativity to compute
\beq
\chern^{~}_\su\bigl(\xi\bigr)\=\chern^{~}_\su\bigl(
\Delta_E^+\ominus\Delta_E^-\bigr)\=\sum_{i=m_+}^m
\,\chern\bigl(E_{k_i}\bigr)\otimes
\chi^{~}_{\underline{V}_{\,|m-2i|}}-\sum_{i=0}^{m_-}
\,\chern\bigl(E_{k_i}\bigr)\otimes\chi^{~}_{\underline{V}_{\,|m-2i|}}
\ ,
\label{chernmultadd}\eeq
where $\chi^{~}_{\underline{V}_{\,|m-2i|}}:\su\to\C$ are the
characters of the $\su$ representations
$\underline{V}_{\,|m-2i|}\cong\C^{|m-2i|}$. This enables us to write
the equivariant index on $\R^{2n}\times\C P^1$ in terms of ordinary
indices on $\R^{2n}$ to get
\beq
{\rm index}^{~}_\su~\hat\Dirac_\Xi=\sum_{i=0}^{m_-}
\,{\rm index}\bigl(\Dirac_{E_{k_i}}\bigr)
\otimes\chi^{~}_{\underline{V}_{\,|m-2i|}}
-\sum_{i=m_+}^m\,{\rm index}\bigl(\Dirac_{E_{k_i}}\bigr)\otimes
\chi^{~}_{\underline{V}_{\,|m-2i|}} \ .
\label{indexord}\eeq
We can turn (\ref{indexord}) into a linear map $\K_\su(\R^{2n}\times\C
P^1)\to\Z$ by composing it with the projection $\pi_0:\rep_\su\to\Z$
onto the trivial representation. Acting on the character ring this
gives
\beq
\pi_0\bigl(\chi^{~}_{\underline{V}_{\,|m-2i|}}\bigr)\=
\chi^{~}_{\underline{V}_{\,|m-2i|}}({\rm id})
\=\dim\underline{V}_{\,|m-2i|}\=|m-2i|
\label{pi0compose}\eeq
and one finally arrives at
\beq
\pi_0\bigl({\rm index}^{~}_\su~\hat\Dirac_\Xi\bigr)=\sum_{i=0}^{m_-}
|m-2i|~{\rm index}\bigl(\Dirac_{E_{k_i}}\bigr)
-\sum_{i=m_+}^m|m-2i|~{\rm index}\bigl(\Dirac_{E_{k_i}}\bigr) \ .
\label{pi0indexfinal}\eeq

Alternatively, one may arrive at the same formula by directly
computing the {\it ordinary} index of the Dirac operator
(\ref{Diracgradeddef}) with $\mphi=0$ using (\ref{DiracS2decomp}) and
(\ref{Diracpkernel})--(\ref{solschargepos}). Since
\beq
{\rm index}~\Dirac_{m-2i}\=\dim\ker\Dirac^+_{m-2i}-
\dim\ker\Dirac^-_{m-2i}\=-(m-2i) \ ,
\label{indexCP1}\eeq
the index of (\ref{Diracgradeddef}) acting on sections of the bundle
(\ref{spinortotgen}) coincides with (\ref{pi0indexfinal}). For a gauge
field configuration appropriate to the K-theory class defined by the
tachyon field (\ref{muNs}), these topological charges coincide
with (\ref{TopchargeYM}).

\bigskip

\section{Vacuum solutions \label{Vacsols}}

The extremal cases for which the Higgs fields have the configurations
$\{\phi_{i+1} =0,\ i=0,1,\dots,m-1\}$ and $\{\partial_\mu\phi_{i+1}=0,\
i=0,1,\dots,m-1\}$, fall
outside of the general scope of the previous analysis and are worth
special consideration. They correspond to vacuum sectors of the
noncommutative gauge theory and are associated with indecomposable
representations of the quiver $\quiverm$ that have no
arrows. Nevertheless, these vacuum sectors admit non-trivial BPS
solutions which signal the presence of stable D-branes attached to
the closed string vacuum after condensation on the brane-antibrane
system. We shall now study them in some detail.

\bigskip

\noindent
{\bf Monopole vacuum.\ }
Let us first look at the case $\partial_\mu\phi^{~}_{i+1}=0,
\ i=0,1,\dots,m-1$. The nonabelian coupled vortex equations
(\ref{ddd1})--(\ref{ddd2}) then imply
\begin{equation}
A^0\=A^1\=\dots\=A^m~=:~A \quad\mbox{and}\quad F^0\=F^1\=\dots\=F^m~=:~F\ ,
\end{equation}
which is only possible in the equal rank case $r=k_0=k_1=\dots=k_m$
corresponding to the gauge symmetry breaking pattern ${\rm
  U}(k)\to{\rm U}(r)^{m+1}$ with $k=(m+1)\,r$. Thus we take
$\phi_{i+1}^{~}=\alpha^{~}_{i+1}~\Idd_r^{~}$ and
$\phi_{i+1}^\dag=\bar\alpha^{~}_{i+1}~\Idd^{~}_r$ with
$i=0,1,\dots,m-1$, where $\a_{i+1}$ are given in (\ref{7.8}). In
quiver gauge theory, the BPS conditions in this sector thus correspond to the
representation of $\quiverm$ which is $r$ copies of the indecomposable
quiver representation
$\simple_{\,0}\oplus\simple_{\,1}\oplus\cdots\oplus\simple_{\,m}$.
They also require
\begin{equation} \label{phi0F}
\de^{a\bb}\,F_{a\bb} \= 0 \qquad\textrm{and}\qquad
F_{\ab\bb} \= 0 \= F_{ab} \ ,
\end{equation}
which are simply the DUY equations on $\rt$.
Note that (\ref{Fyyb}) implies ${\cf}_{y\yb}=0$
in this case, giving the trivial dimensional reduction to $\rt$.
After switching to matrix form via (\ref{X}), we obtain
\begin{equation}\label{mf}
\de^{a\bb}\,\big[X_{a}\,,\,X_{\bb}\big] + \de^{a\bb}\,\th_{a\bb}\=0
\qquad\textrm{and}\qquad \big[X_{\ab}\,,\,X_{\bb}\big]\=0\=
\big[X_a\,,\,X_b\big] \ .
\end{equation}

The obvious solution to (\ref{mf}) is the trivial one with
$X_a=\th_{a\bb}\,{\zb}^{\bb}$, giving $F_{a\bb} = 0$. This sector
can be understood physically as the endpoint of tachyon condensation,
wherein the Higgs fields $\p_{i+1}$ have rolled to their
minima at $\phi_{i+1}=\a_{i+1}~\Idd_r$ and the fluxes have been radiated away
to infinity. Here the D0-branes have been completely dissolved into
the D(2$n$)-branes.

However, non-trivial solutions of the equations (\ref{mf}) also exist.
For this, let us restrict ourselves to the abelian case $r=1$
and simplify matters by taking $\th^a=\th$ for all $a=1,\ldots,n$.
We fix an integer $l\geq1$ and consider the
ansatz~\cite{Kraus}
\begin{equation}\label{ans}
X_a\=\th_{a\cb}\,\Sigma_{l}^\+\,f(\Ncal\,)\, \zb^{\cb}\,\Sigma^{~}_{l}
\qquad\textrm{and}\qquad
X_{\ab}\=\th_{\ab c}\,\Sigma_{l}^\+\,z^c\,f(\Ncal\,)\,\Sigma^{~}_{l}\ ,
\end{equation}
where $f$ is a real function of the ``total number operator''
\begin{equation}\label{tno}
\Ncal:=\frac{1}{2\,\th}\,\sum\limits^n_{a=1} z^a\,{\zb}^{\ab}
\end{equation}
with the property that $f(r)=0$ for $r\le{l}{-}1$.
The shift operator $\Sigma^{~}_{l}$ in (\ref{ans}) is defined to obey
\begin{equation}
\Sigma_{l}^\+\,\Sigma^{~}_{l}\ =\ 1 \qquad\textrm{while}\qquad
\Sigma^{~}_{l}\,\Sigma_{l}^\+\=1-P^{~}_{l}
\end{equation}
with
\begin{equation}\label{projl}
P^{~}_{l}:=
\sum\limits_{|\,\vec r\,|\le{l}-1}|r_1,\ldots,r_n\>\<r_1,\ldots,r_n| \ ,
\end{equation}
where $\vec r=(r_1,\dots,r_n)$ with $|\,\vec r\,|:= r_1+\ldots
+r_n$. Note that
\begin{equation}
\Sigma_{l}^\+\,P^{~}_{l}\=P^{~}_{l}\,\Sigma^{~}_{l}\=0 \qquad\textrm{and}\qquad
f(\Ncal\,)\,P^{~}_{l}\=P^{~}_{l}\,f(\Ncal\,)\=0\ ,
\end{equation}
and $\Sigma_{l}^\+$ projects all states with $|\,\vec r\,|<l$ out of
the Fock space $\cal H$.

One easily sees that (\ref{ans}) fulfills the homogeneous equations in
(\ref{mf}). Remembering that $\th_{a\bb}=-\th_{\bb a}=\frac{1}{2\,\th}\,
\de_{a\bb}$, we also obtain
\begin{align}\nonumber
\big[X_a\,,\, X_{\bb}\big]\ &=\ \th_{a\cb}\,\th_{\bb
  d}\,\Sigma_{l}^\+\,\left\{
f(\Ncal\,)\,{\zb}^{\cb}\,(1{-}P^{~}_{l})\,z^d\,f
(\Ncal\,)-z^d\,f(\Ncal\,)\,(1{-}P^{~}_{l})\,f(\Ncal\,)
\,\zb^{\cb}\right\}\,\Sigma^{~}_{l} \\[4pt]
\label{mf4}
&=\ -\frac{1}{4\,\th^2}\,\de_{a{\cb}}\,\de_{d{\bb}}\,\Sigma_{l}^\+\,\left\{
f^2(\Ncal\,)\,{\zb}^{\cb} z^d-f^2(\Ncal{-}1)\,z^d
{\zb}^{\cb}\right\}\,\Sigma^{~}_{l}
\end{align}
with the help of the identities
$\ {\zb}^{\cb}\,P_{l}=P_{l -1}\,{\zb}^{\cb}\ $ where $P_0 := 0$. We
have also used
\begin{equation}
{{\zb}^{\cb}}\,f(\Ncal\,) \= f(\Ncal{+}1)\,{\zb^{\cb}}
\quad\textrm{and}\quad
{z^d}\,f(\Ncal\,) \= f(\Ncal{-}1)\,{z^d}\ .
\end{equation}
Substituting (\ref{mf4}) into (\ref{mf}), we employ
\begin{equation}
\de_{\cb d}\,z^d\,{\zb}^{\cb} \= 2\,\th\,\Ncal \qquad\textrm{and}\qquad
\de_{\cb d}\,{\zb}^{\cb}\,z^d \= 2\,\th\,(\Ncal+n)
\end{equation}
to find the conditions
\begin{align}\nonumber
0 &\= \de^{a\bb}\,\bigl[X_a\,,\,X_{\bb}\bigr]+
\de^{a\bb}\,\th_{a\bb} \\[4pt] \nonumber
&\=-\frac{1}{2\,\th}\,
\Sigma_{l}^\+\,\biggl\{ f^2(\Ncal\,)\,(\Ncal{+}n)-
f^2(\Ncal{-}1)\,\Ncal\biggr\}\,\Sigma^{~}_{l}+
\frac{n}{2\,\th} \\[4pt] \label{mf5}
&\=\frac{1}{2\,\th}\,\Sigma_{l}^\+\,\biggl\{ \Ncal\,f^2(\Ncal{-}1)-
(\Ncal{+}n)\,f^2(\Ncal\,)+
n\biggr\}\,\Sigma^{~}_{l}
\end{align}
on the operator $f$. With the initial conditions
$f(0)=f(1)=\dots=f(l-1)=0$ and the finite-energy condition $f(r)\to1$
as $r\to\infty$, these recursion relations are solved by
\begin{equation}\label{fsol}
f^2(\Ncal\,)=
\Bigl(1-\frac{Q\ n!}{(\Ncal{+}1)\cdots(\Ncal{+}n)}\Bigr)\,
(1-P_l) \ ,
\end{equation}
where
\begin{equation}\label{Qpm}
Q:=\frac{l\,(l{+}1)\cdots(l{+}n{-}1)}{n!}
\end{equation}
is the number of states in $\Hcal$ with $\Ncal\leq l-1$, i.e. the
number of states removed by the operator $\Sigma_l^\dag$.

We arrive finally at the non-trivial gauge field configuration given
by
\begin{equation}\label{Xsol}
X_a=\frac{1}{2\,\th} \  \Sigma_{l}^\+~\sqrt{1-
\frac{Q\ n!}{(\Ncal{+}1)\cdots(\Ncal{+}n)}}~
(1-P^{~}_l)\,\de_{a\cb}\,{\zb}^{\cb}\,\Sigma^{~}_{l}\ .
\end{equation}
The field strength $F$ on ${\R}^{2n}_{\th}$ obtained from (\ref{Xsol})
has finite $n$-th Chern number $Q$~\cite{Kraus}.
The topological charge $Q$ given by (\ref{Qpm}) is calculated
here via an integral over $\rt$. However, the $(n{+}1)$-th Chern
number for this configuration considered as a gauge field on
$\R_\theta^{2n}\times\C P^1$ with ${\cf}_{y\yb}=0=\cf_{\vt\vp}$
vanishes. Moreover, this configuration has finite energy (\ref{EF})
proportional to the topological charge~\cite{Kraus},
\beq
E^{~}_{\rm BPS}=(2\pi)^{n+1}\,R^2\,n\,(n-1)\,Q \ ,
\label{EBPSm0}\eeq
as usual for a BPS instanton solution.

\bigskip

\noindent
{\bf Higgs vacuum.\ }
The choice $\phi_{i+1} =0$ for all $i=0,1,\dots,m-1$ is somewhat more
interesting since from (\ref{Fyyb}) and (\ref{mupdef}) we then have
${\cf}_{y\yb}\ne 0$ with
\begin{equation}
{\cf}_{y\yb}=-\frac{R^2}{\left(R^2+y\yb\right)^2}~\mup \ .
\end{equation}
This configuration gives the local maximum of the Higgs potential
corresponding to the open string vacuum containing D-branes. In this
case the vortex equations (\ref{ddd1})--(\ref{ddd2}) reduce to
\begin{equation}\label{rduy1}
\de^{a\bb}\,F^i_{a\bb} \= \frac{m-2i}{4\,R^2}  \qquad\textrm{and}\qquad
F^i_{\ab\bb} \= 0 \= F^i_{ab} \ .
\end{equation}
After switching to matrix form via (\ref{X}) we obtain
\begin{equation}\label{mf1}
\de^{a\bb}\,\big[X^i_{a}\,,\,X^i_{\bb}\big] + \de^{a\bb}\,\Bigl(1-
\frac{(m-2i)\,\th}{2n\,R^2}\,\Bigr)\,\th_{a\bb}\=0
\qquad\textrm{and}\qquad \big[X^i_{\ab}\,,\,X^i_{\bb}\big]\=0
\=\big[X^i_a\,,\,X^i_b\big]\ ,
\end{equation}
where we have used the formula $\th_{a\bb}=\frac{1}{2\,\th}\,\de_{a\bb}$.
Recall that there is no summation over the index $i=0,1,\dots,m$ in
the equations (\ref{mf1}).

By comparing  (\ref{mf1}) and (\ref{mf}), we conclude that (\ref{mf1})
can be solved for each $i$ by the same ansatz as for (\ref{mf}). For
this, let us restrict ourselves again to the abelian case for all
nodes $i=0,1,\dots,m$ (so that $k=m+1$), and fix $m+1$ positive integers
$l_0,l_1,\dots,l_m$. We take
\begin{equation}\label{ansa}
X^i_a\=\th_{a\cb}\,\Sigma_{l_i}^\+\,f^{~}_i(\Ncal\,)\,
\zb^{\cb}\,\Sigma^{~}_{l_i}
\qquad\mbox{and}\qquad
X^i_{\ab}\=\th_{\ab
c}\,\Sigma_{l_i}^\+\,z^c\,f^{~}_i(\Ncal\,)\,\Sigma^{~}_{l_i}
\end{equation}
analogously to (\ref{ans})--(\ref{projl}). Producing then the same
calculations as before, we obtain the gauge field configuration
\begin{equation}\label{Xso}
X^i_a=\frac{1}{2\,\th^i} \ \Sigma_{l_i}^\+~\sqrt{1-
\frac{Q^{~}_{i}\ n!}{(\Ncal{+}1)\cdots(\Ncal{+}n)}}~
(1-P^{~}_{l_i})\,\de_{a\cb}\,{\zb}^{\cb}\,\Sigma^{~}_{l_i}\ ,
\end{equation}
where
\begin{equation}\label{thi}
\th^i:=\frac{\th}{\sqrt{1 - \frac{(m-2i)\,\th}{2n\,R^2}}}
\end{equation}
and
\begin{equation}\label{Qi}
Q_i=\frac{l_i\,(l_i +1)\cdots(l_i +n-1)}{n!} \ .
\end{equation}
We have chosen the radius $R$ of the sphere so that
$R^2>\frac{m\,\th}{2n}$.

The solutions (\ref{Xso}) coincide with those given by (\ref{Xsol}) if
one assigns different noncommutativity parameters $\th^i$ to the
worldvolumes of D($2n$)-branes carrying different magnetic fluxes
proportional to $m-2i$. Then the field strength $F^i(\th^i)$ on
$\R^{2n}_{\th^i}$ obtained from (\ref{Xso}) will have finite topological
charge $Q_i$ given by (\ref{Qi}) and corresponding finite BPS
energy analogous to (\ref{EBPSm0}), and the configuration thus
described extends to instantons on $\R^{2n}_\theta\times\C P^1$.
The interesting idea of introducing distinct noncommutativity
parameters on multiple coincident D-branes, generated by different
magnetic fluxes on their worldvolumes~\cite{Dasgupta}, was discussed
in~\cite{Tatar} as a means (among other things) of stabilizing
brane-antibrane systems. This proposal gains support from our Higgs
vacuum BPS solutions (\ref{Xso}) which carry different magnetic fluxes
on different branes.

\bigskip

\section*{Acknowledgments}

We thank O.~Lechtenfeld for collaboration during the early stages of
this project. The work of A.D.P. was supported in part by the Deutsche
Forschungsgemeinschaft~(DFG). The work of R.J.S. was supported in part
by a PPARC Advanced Fellowship, by PPARC Grant PPA/G/S/2002/00478, and
by the EU-RTN Network Grant MRTN-CT-2004-005104.

\bigskip

\end{document}